\documentclass[aps,prb,twocolumn,showpacs,amsmath,amssymb,superscriptaddress,floatfix]{revtex4-1}
\usepackage{graphicx}
\usepackage{color}
\usepackage{braket}
\usepackage{amsfonts}
\usepackage{MnSymbol}
\usepackage{bm}

\usepackage[dvipsnames]{xcolor}
\usepackage{mathrsfs}
\usepackage{mathtools}
\usepackage{subfigure}

\usepackage[colorlinks = true,linkcolor = blue,urlcolor  = blue,     citecolor = blue,anchorcolor = blue]{hyperref}

\usepackage{bbm}
\usepackage[dvipsnames]{xcolor}

\newcommand{\mikpl}[1]{{\color{ForestGreen}#1}}
\date{\today}                  

\begin{document}

\title{Green's functions of quasi-one-dimensional layered systems and their application to Josephson junctions}

\author{Kiryl Piasotski}
\email[Email: ]{kiryl.piasotski@kit.edu}
\affiliation{Institut f\"ur Theorie der Kondensierten Materie, Karlsruher Institut f\"ur Technologie, 76131 Karlsruhe, Germany}
\affiliation{Institut f\"ur Quanten Materialien und Technologien, Karlsruher Institut f\"ur Technologie, 76021 Karlsruhe, Germany}

\author{Mikhail Pletyukhov}
\affiliation{Institut für Theorie der Statistischen Physik, 
RWTH Aachen, 52074 Aachen, Germany}

\author{Alexander Shnirman}
\affiliation{Institut f\"ur Theorie der Kondensierten Materie, Karlsruher Institut f\"ur Technologie, 76131 Karlsruhe, Germany}
\affiliation{Institut f\"ur Quanten Materialien und Technologien, Karlsruher Institut f\"ur Technologie, 76021 Karlsruhe, Germany}

\begin{abstract}

We develop Green's function formalism to describe continuous multi-layered quasi-one-dimensional setups described by piece-wise constant single-particle Hamiltonians. The Hamiltonians of the individual layers are assumed to be quadratic polynomials in the momentum operator with matrix-valued (multichannel) coefficients. This, in particular, allows one to study transport in heterostructures consisting of multichannel conducting, superconducting, or insulating components with band structures of arbitrary complexity. We find a general expression for the single-particle Green's function of the combined setup in terms of the bulk (translationally invariant) Green's functions of its constituents. Furthermore, we provide the expression for the global density of states of the combined system and establish the bound state equation in terms of bulk Green's functions. We apply our formalism to investigate the spectrum and current-phase relations in ordinary and topological Josephson junctions, additionally showing how to account for the effects of static disorder and local Coulomb interaction.

\end{abstract}

\maketitle

\section{Introduction}
\label{Motivation_sec}
\begin{figure*}
                \includegraphics[scale=0.24]{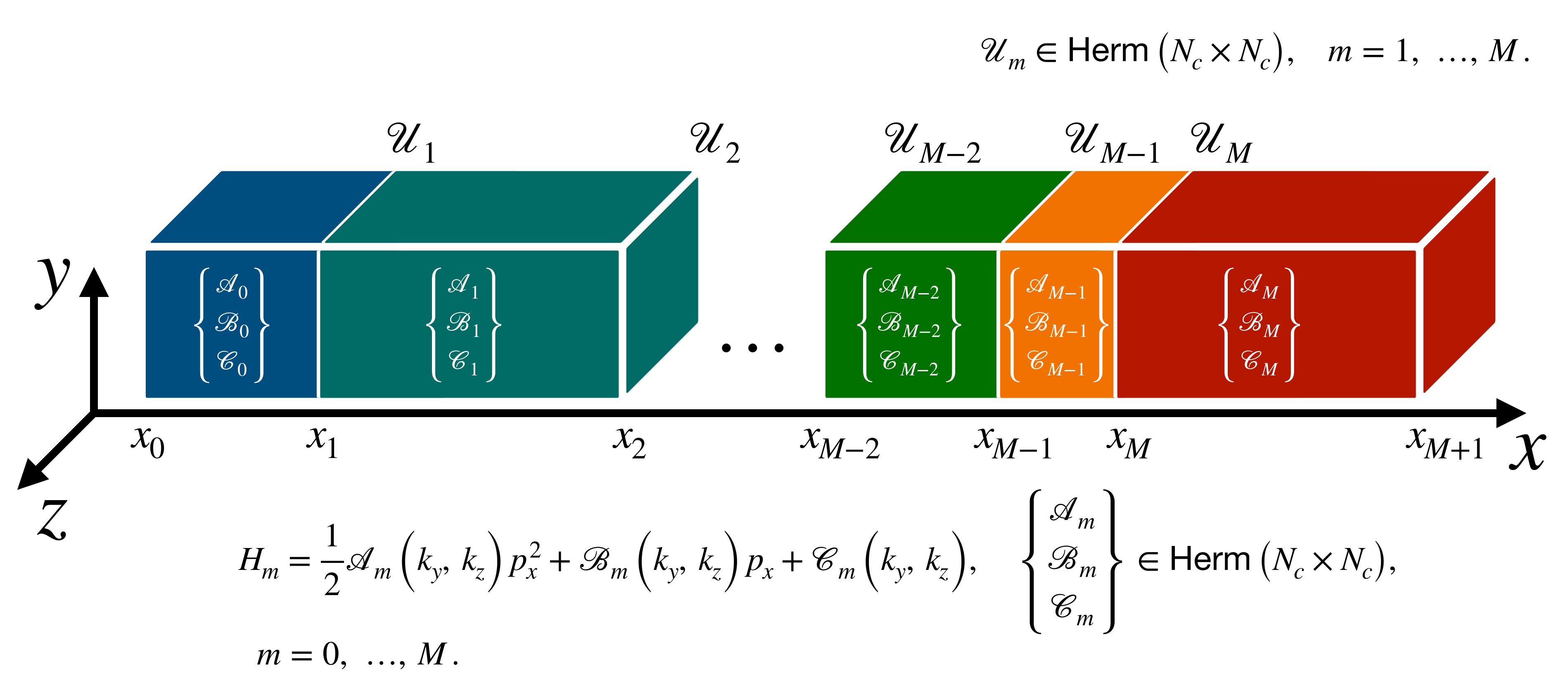}
                \caption{Sketch of the model: A heterostructure consists of $M+1$ layers described by individual matrix-valued Hamiltonians $H_m$, which are consequently coupled with each other across matrix-valued tunneling barriers $\mathcal{U}_m$.}
               \label{Setup_Sketch}
\end{figure*}

Description of equilibrium and transport properties of layered quantum systems is a common problem in the domains of quantum electronics and solid state theory\cite{datta1997electronic,freericks2016transport}. Albeit typically a single-particle quantum mechanical problem, its solution is rarely simple due to the intricate band structures of the  materials forming the layers. Most commonly, nowadays, these systems are analyzed numerically within the tight-binding approximation (like e.g. in Ref.~[\onlinecite{groth2014kwant}]), allowing one to get a good grip on the low-lying excitations, as well as to assess the effects of the static disorder. The main idea behind this approach  consists\cite{PhysRevB.44.8017} in numerical studies of the ballistic conductance in mesoscopic structures  within a lattice model, being expressed in terms of the lattice Green's function. It can be equally well applied to the study of equilibrium properties like Josephson current (JC) in Josephson junctions (JJ) of arbitrary width, see e.g. Refs.~[\onlinecite{PhysRevB.86.134522},\onlinecite{PhysRevB.89.195407}].

An alternative to the tight-binding numerics is the scattering matrix approach \cite{landauer1957spatial,landauer1970electrical} applied to continuum ballistic models. One of its key early-day achievements was the calculation of the two-terminal conductance in terms of the transmission probabilities. The microscopic justification of this method relies either on taking the continuum limit of the wavefunction matching (WFM) of Ref.~[\onlinecite{PhysRevB.44.8017}] or on studying the continuum limit of the atomistic Green's function (AGF) as in Ref.~[\onlinecite{C_Caroli_1971}]. The equivalence of both WMF and AGF approaches has been fully substantiated in Ref.~[\onlinecite{PhysRevB.72.035450}]. The relation between transmission and reflection coefficients of the scattering matrix on the one hand and Green's function, on the other hand, is widely known as the Fisher-Lee relation\cite{PhysRevB.23.6851}. Its mode-resolved generalization has been recently proposed in Ref.~[\onlinecite{Boumrar_2020}].

In application to superconducting systems, the scattering matrix approach has been extended by Blonder, Tinkham, and Klapwijk\cite{PhysRevB.25.4515} (BTK) on the basis of solutions of the Bogoliubov-de-Gennes equations. The BTK theory has been further generalized by C. Beenakker to the multichannel case in Ref.~[\onlinecite{PhysRevLett.67.3836}]. In this work the compact equation for the sub-gap Andreev bound states\cite{andreev1964thermal,andreev1966electron} (ABS) and the expression for the continuous excitation spectrum of the JJ have been established in terms of normal and Andreev scattering matrices. In practice, however, these matrices are often treated in the so-called Andreev limit $\Delta_0  \ll \mu$, where $\Delta_0$ is the absolute value of the superconducting order parameter and $\mu$ is the chemical potential, essentially neglecting the normal reflection at the superconducting interface. 

An alternative theoretical description of the superconducting tunneling and proximity effects has been developed by G. Arnold in Refs.~[\onlinecite{PhysRevB.18.1076},\onlinecite{Arnold1985}] using standard  
nonequilibrium Green's functions. His approach is based on the theory of Feuchtwang\cite{PhysRevB.10.4121,PhysRevB.10.4135,PhysRevB.12.3979}, which does not make use of the tunneling Hamiltonian. This theory shares many common features with the AGF method of Caroli {\it et al} [\onlinecite{C_Caroli_1971}] mentioned above, and these techniques will serve as a starting point for our present consideration. 

Discussing the approaches to interface physics, one cannot but mention that quasiclassical Green's functions treated in terms of the Eilenberger and the Usadel equations are also traditionally used for describing superconducting proximity effects and the JJs, see Refs.~[\onlinecite{RevModPhys.76.411}] and [\onlinecite{doi:10.1098/rsta.2015.0149}] for reviews.

The goal of the present manuscript is to 
generalize and further develop the Green's function techniques of Arnold, Feuchtwang, Caroli \textit{et al}, mentioned above. The structure of the manuscript is twofold.

 First of all, we provide multichannel and multi-interface generalizations of results in Refs.~[\onlinecite{C_Caroli_1971}, \onlinecite{PhysRevB.18.1076},\onlinecite{Arnold1985}, \onlinecite{PhysRevB.10.4121}, \onlinecite{PhysRevB.10.4135}, \onlinecite{PhysRevB.12.3979}], essentially building upon the method of Ref. [\onlinecite{C_Caroli_1971}]. In particular, we derive a compact closed-form expression for the composite  Green's function $G (x,x')$ of a quasi-one-dimensional heterostructure (sketched in Fig. \ref{Setup_Sketch}) given merely in terms of the bulk Green's functions $G^{(0,m)} (x,x')$ of its constituting layers (labeled by $ m=0, \ldots, M$). The knowledge of $G (x,x')$ is beneficial for several reasons: It gives direct access to the system's spectrum, allows for a calculation of various physical observables (like e.g. the density of states (DOS) and the current), may serve as a starting point for low-energy approximations (like e.g. the Andreev limit), as well as provides essential input for a perturbative diagrammatic treatment of disorder and many-body interaction effects. The bulk single-layer Green's functions $G^{(0,m)} (x,x')$ are obtained by means of the standard Fourier transformation and are computationally inexpensive. For few-channel models $G^{(0,m)} (x,x')$ may often be calculated analytically, as demonstrated in our examples below.

We shall point out that ideas of that sort, namely the extraction of the properties of inhomogeneous quantum systems from their bulk counterpart, are traditional in solid-state theory. This subject has a long history, and it often bears different names: the quantum theory of surface states\cite{Garcia_Moliner_1969_I,Garcia_Moliner_1969_II,GARCIAMOLINER199183,GARCIAMOLINER1994332,Rodriguez_Coppola_1990} or the method of embedding\cite{Inglesfield_1971,Inglesfield_1981,PhysRevB.65.165103}, for example.

Secondly, we apply the derived expression for $G (x,x')$ to the two paradigmatic examples: (1) Josephson model of a tunneling barrier between two $s$-wave superconductors, and (2) the model of a JJ\cite{PhysRevB.96.075404,PhysRevB.101.224501} between two semiconducting wires with strong spin-orbit interaction and proximity induced superconductivity, submersed into the parallel magnetic field (each of which is capable of hosting a Majorana zero mode\cite{PhysRevLett.105.077001,PhysRevLett.105.177002}). In particular, we demonstrate that both the excitation spectrum (i.e. the global DOS) as well as the JC can be obtained in terms of a single matrix $d = \begin{pmatrix}
G(0,0) & G (0,W)  \\
G (W,0) & G (W,W) 
\end{pmatrix}^{-1}$ for the finite-width junction (with the two interfaces at $x=0$ and $x=W$), degenerating to $d = [G (0,0) ]^{-1}$ in the short junction limit
(single interface at $x=0$). In particular, the correction to the global DOS due to the tunneling between the layers is given by $\delta \rho = - \frac{1}{\pi} \frac{\partial}{\partial \omega} \text{Im} \, \ln \det d$, which resembles analogous expressions for $\delta \rho$ in terms of the scattering matrix\cite{PhysRevLett.66.76,PhysRevB.65.115307}. A general equation for the bound states acquires a particularly simple form $\det d=0$, which is equally well applicable to all types of heterostructures. The relation of our equations for the bound states and for $\delta \rho$ to their scattering matrix analogs becomes transparent on the basis of the Fisher-Lee relation\cite{PhysRevB.23.6851}: The matrix $d^{-1}$ appears as a common factor in expressions for all components of the scattering matrix, and it can be generally interpreted as a core part of the $T$-matrix in the coordinate representation. The advantage of computing $d$ directly from the Green's function $G (x,x')$ consists, though, in its additive form, combining contributions from adjacent layers (and possibly from a local contact potential at their interface). This property of $d$ is thus analogous to that of self-energies.

Using our formulas, we study the excitation spectrum and the JC in various parametric regimes of the two models discussed above, relaxing the Andreev approximation (which might be essential for one-dimensional wires) and allowing for arbitrary values of the junction's width $W$. The knowledge of the explicit form of $G (x,x')$ allows us to account for the effects of random potential disorder and local Coulomb interaction at the interface of the JJ. We find that the effect of static disorder arises only beyond the Andreev limit. Taking into account the local Coulomb interaction, we predict a crossover between the $0$- and the $\pi$-junctions.

\section{Formalism of interface Green's functions}
\label{sec: formalism}
\subsection{Model and problem formulation}
Let us consider a quasi-one-dimensional system that consists of $M+1$ layers (labeled by $m=0,\ldots,M$) subsequently connected with each other across tunneling barriers (also labeled by $m=1, \ldots,M$) along the  spatial $x$-axis, see Fig. \ref{Setup_Sketch}. We further assume that $m$-th layer extends over the spatial range $(x_m, x_{m+1})$, with $x_1$ and $x_{M}$ being the coordinates of the leftmost and the rightmost interfaces, respectively. The leftmost boundary of the whole system is $x_0$ and the rightmost one is $x_{M+1}$. An important class of models with semi-infinite leads is also included in the present consideration: They are realized by setting $x_0 = - \infty$ and $x_{M+1} = + \infty$. For the $m$-th subsystem we assume the Hamiltonian of the form 
\begin{align}
    H_{m}=\frac{1}{2}\mathcal{A}_{m} p^{2}+\mathcal{B}_{m} p+\mathcal{C}_{m}, \label{hamilt_main_parts}
\end{align}
where $\mathcal{A}_{m},\ \mathcal{B}_{m},\ \mathcal{C}_{m}$ are constant $N_{c}\times N_{c}$ hermitian matrices, parametrically dependent on the conserved transverse (quasi)momenta $k_y$ and $k_z$, with $N_c$ being the number of channels (e.g. spin and/or orbital quantum numbers, etc.), and $p \equiv p_x = -i\partial_{x}$ is the momentum operator in the direction along the system.

The tunneling barriers between the subsystems are modeled by delta-like potentials at the corresponding interfaces:
\begin{align}
    \mathcal{U}(x)=\sum_{m=1}^{M}\mathcal{U}_{m}\delta(x-x_{m}),\label{contact_pot_main}
\end{align}
where $\mathcal{U}_{m}$ are generally considered as $N_{c}\times N_{c}$ hermitian matrices (which, in particular, allows us to treat magnetically active contacts as well). 

The Hamiltonian of the combined system is written as
\begin{align}
    H=\frac{1}{2}p\mathcal{A}(x)p+\frac{1}{2}\left\{\mathcal{B}(x),\ p\right\}+\mathcal{C}(x)+\mathcal{U}(x), \label{Hamiltonian_main_contin}
\end{align}
where
\begin{align}
    \begin{Bmatrix}
    \mathcal{A}(x)\\
    \mathcal{B}(x)\\
    \mathcal{C}(x)
    \end{Bmatrix}=\sum_{m=0}^{M}\Theta(x_m < x < x_{m+1})\begin{Bmatrix}
    \mathcal{A}_{m}\\
    \mathcal{B}_{m}\\
    \mathcal{C}_{m}
    \end{Bmatrix}.\label{params_main}
\end{align}

The eigenvalue problem for the Hamiltonian \eqref{Hamiltonian_main_contin} is complemented by matching conditions for the wave function and its derivative at each interface. While the wavefunction is continuous,
\begin{align}
    \Psi(x_m^{-})=\Psi(x_{m}^{+})\equiv\Psi(x_{m}), \quad x_m^{\pm} = x_m \pm 0^+ ,
    \label{general_matching_condition_1}
\end{align}
its derivative satisfies the following matching condition at the $m$-th interface:
\begin{align}
    \frac{\mathcal{A}_{m}\Psi'(x_{m}^{+})-\mathcal{A}_{m-1}\Psi'(x_{m}^{-})}{2}=\left[\mathcal{U}_{m}- i \frac{\mathcal{B}_{m}-\mathcal{B}_{m-1}}{2}\right]\Psi(x_m).
    \label{general_matching_condition_2}
\end{align}
This condition is traditionally derived by integrating the Schr\"odinger equation $H \Psi (x) = E \Psi (x)$ over the infinitesimally small region $(x_m^- , x_m^+)$ enclosing the contact's coordinate $x_m$. Physically it expresses the conservation of the current density across the interface.

The main goal of our present consideration is to find an expression for the (retarded) Green's function of the whole system\footnote{Note that the  heterostructural Green's function $G(x, x')$ is a highly non-trivial object, containing information about all possible multi-barrier scattering phenomena and interface-localized bound states.}, which satisfies the equation
\begin{align}
    [z - H] G (x,x'; z) = \delta (x-x')
    \label{general_GF_equation}
\end{align}
for arbitrary complex-valued spectral parameter $z$ (with $\text{Im} \, z >0$, in particular for $z=\omega + i 0^+$, in the case of the retarded function). The differential equation \eqref{general_GF_equation} is complemented by the vanishing boundary conditions $G (x_0, x';z) = G (x_{M+1},x'; z) =0$ at the system's ends. By the virtue of the Lehmann representation, the Green's function $G (x,x';z)$ must also obey the matching conditions \eqref{general_matching_condition_1} and \eqref{general_matching_condition_2} in the variable $x$. We note that $G (x,x';z)$ can be alternatively defined as a solution of a reciprocal differential equation with respect to the variable $x'$ equipped with the boundary conditions $G (x, x_0;z) = G (x, x_{M+1}; z) =0$, while the corresponding interface matching conditions are obtained by hermitian conjugating \eqref{general_matching_condition_1} and \eqref{general_matching_condition_2}.

As an input we use the {\it translationally invariant} (aka bulk) Green's functions $G^{(0,m)} (x,x'; z)$ satisfying the equation
\begin{align}
    [z - H_m] G^{(0,m)} (x,x'; z) = \delta (x-x')
    \label{bulk_GF_equation}
\end{align}
on the whole spatial axis. They are evaluated by means of the Fourier transformation
\begin{align}
     G^{(0,m)}(x,x';z)=\int_{-\infty}^{\infty}\frac{dk}{2\pi}  \frac{e^{ i k (x-x')}}{z-h_m (k)}, 
     \label{Lehmann_main_1}
\end{align}
where $h_m (k)$ is the Hamiltonian obtained from $H_m$ in \eqref{hamilt_main_parts} by the substitution $p \to k$. Often the integral in \eqref{Lehmann_main_1} may be evaluated analytically by means of the integration in the complex $k$-plane: To this end, one needs to establish the complex-valued roots $k (z)$ of the secular equation $\det [z-h_m (k)]=0$ with positive imaginary parts, $\text{Im} \, k (z) >0$. Then the result of integration in \eqref{Lehmann_main_1} is represented as a sum of residua at the corresponding poles $k(z)$ (see Appendix \ref{Bulk_propagators} for details). For few-channel models, the equation $\det [z-h_m (k)]=0$ admits analytical solutions (some of them will be demonstrated in the following examples), while in general the roots $k (z)$ have to be determined numerically. As it will be demonstrated later, the root-searching routine is the only numerical part in solving the problem of establishing $G (x,x';z)$, and for this reason, the Green's function approach developed below should provide a considerable speed up in the study of arbitrary heterostructures.

An opening move in establishing $G(x, x'; z)$ is a determination of the set of Green's functions $ G^{m}(x, x'; z)$ describing every isolated layer $m$ on the corresponding spatial interval $(x_m , x_{m+1})$. The Green's function $G^{m} (x, x'; z)$ also satisfies the differential equation \eqref{bulk_GF_equation}, but -- in contrast to  $ G^{(0,m)}(x, x'; z)$ -- it vanishes at the interval's ends, that is $ G^{m}(x_m, x'; z)= G^{m}(x_{m+1}, x'; z )=0$.

Thankfully, the determination of $ G^{m}(x, x')$ (the argument $z$ is omitted for brevity) is conveniently solved by the so-called boundary Green's function technique, allowing one to write a simple relation between the propagators of infinite [$G^{(0,m)}(x,x')$] and bounded [$G^{m}(x, x')$] systems (see Appendix \ref{bound_props_aps}, and references therein, for the summary of the key results). 

Next, we find out how the isolated layers are coupled with each other. 
As such, this coupling is dictated by the matching conditions  \eqref{general_matching_condition_1} and \eqref{general_matching_condition_2}. The key question is then how does one implement these requirements in terms of a local potential? Once this is accomplished, a set of Dyson's equations may be set up, relating the full Green's function $G(x, x')$ to those $G^{m}(x, x')$ of the individual layers. Furthermore, it is expected that the set of these equations admits a closed-form analytical solution, as by the locality of the coupling potential the integral equations are expected to reduce to algebraic ones for a finite number of unknowns.

To give an answer to the key question we underscore the two approaches which we find especially useful in practice.

The first approach, inspired by the earlier ideas of the seminal paper Ref. [\onlinecite{C_Caroli_1971}], is based on discretizing the equation \eqref{general_GF_equation} on a lattice with the spacing $a$, solving the obtained tight-binding counterpart, and taking carefully the continuum limit $a \to 0$.  As in the tight-binding description the space consists of discrete points, the tunneling between the disjoint parts of the system may be defined unambiguously, and hence the above-described program with a formal solution of the corresponding Dyson equations can be successfully executed.  Further evaluation of the limit $a \to 0$ allows us to recover the complete Green's function of the corresponding continuum theory, expressed in terms of boundary Green's functions of the individual layers as well as their spatial derivatives up to the second order.
In this paper, we describe the major steps of implementing this approach for the models of our present interest. 

The second approach is based on (the multichannel generalization of) the Sturm-Liouville theory for second-order differential operators. It constructively exploits the matching conditions \eqref{general_matching_condition_1} and \eqref{general_matching_condition_2} to recover $G (x,x')$ staying within the paradigm of continuum models. The results of its application naturally reproduce those described below in the present paper. Further details of the second approach will be reported elsewhere [\onlinecite{future}].

\subsection{Single barrier}
\label{Sec: single_barrier_theory}

In this subsection, we extrapolate the construction technique of the composite Green's function of a double-layer system of Ref. [\onlinecite{C_Caroli_1971}] to the multichannel case.

Let us start by considering the simplest case of a single barrier separating left $L\ (m=0,\ x\in(x_{0}, 0),\ x_{0}<0)$ and right $R\ (m=1,\ x\in (0,x_{2}),\ x_{2}>0)$ subsystems at $x=x_{1}=0$. 

We introduce the following lattice counterpart of the single-barrier continuum model
\begin{align}
    H &= H_L + H_R  +  | 0 \rangle W_0 \langle 0 | \label{H_junction_disjoint} \\
    & -  |0 \rangle t_L \langle -1| - |-1 \rangle t_L^{\dagger} \langle 0|-  |1 \rangle t_R \langle 0 |- |0 \rangle t_R^{\dagger} \langle 1 | ,
    \label{H_junction_coupling}
\end{align}
where the left and right disjoint subsystems are defined on the lattice sites $n_0 \leq n \leq -1$ and $1 \leq n \leq n_2$, respectively. They are described by the Hamiltonians
\begin{align}
    H_{L} =& - \sum_{n=n_0+1}^{-1} ( |n \rangle  t_L \langle n-1| +  |n-1 \rangle t_L^{\dagger} \langle n| ) \nonumber \\
    &+  \sum_{n=n_0}^{-1} |n \rangle W_L \langle n | , \label{HL_lattice} \\
    H_{R} =&  - \sum_{n=1}^{n_2-1} (|n +1 \rangle  t_R \langle n| + |n \rangle t_R^{\dagger} \langle n+1|) \nonumber \\
    &+  \sum_{n=1}^{n_2} |n \rangle W_R \langle n | , \label{HR_lattice}
\end{align}
with constant, matrix-valued (in the channel space) hopping amplitudes $t_L,t_R$ and onsite potentials $W_L,W_R$.
The central cite $n=0$ is characterized by the onsite potential $W_0$. It is coupled to both subsystems by the same nearest-neighbor hopping amplitudes $t_L$ and $t_R$ as occur in the Hamiltonians \eqref{HL_lattice} and \eqref{HR_lattice}, respectively.

Treating the coupling term \eqref{H_junction_coupling} as a perturbation, we set up the following Dyson equation in the coordinate representation for the Green's function $G = \frac{1}{z-H}$ of the full system
\begin{align}
    G_{n,n'} &= G^{L}_{n,n'}+G^{C}_{n,n'}+G^{R}_{n,n'} \nonumber \\
    &- (G_{n,-1}^{L} t_L^{\dagger} + G_{n,1}^{R} t_R) G_{0,n'} \nonumber \\
    &- \delta_{n,0} G_{0,0}^{C} (t_L G_{-1,n'} + t_R^{\dagger} G_{1,n'} )
    \label{Dyson_eq_junction}
\end{align}
 in terms of the Green's functions $G^L= \frac{1}{z-H_L}$, $G^C = | 0 \rangle \frac{1}{z - W_0} \langle 0 |$, $G^R = \frac{1}{z-H_R}$ of the three disjoint subsystems. The Green's function $G_{n,n'}^L$ is nonzero only for $n,n'\leq -1$, while $G_{n,n'}^R$ is nonzero only for $n,n' \geq 1$. In the corresponding domains, they are expressed according to \eqref{BGF_lattice} in terms of the Green's functions $G^{(L,0)}_{n,n'}$ and $G^{(R,0)}_{n,n'}$ of the two auxiliary models defined on the whole lattice and using the constant parameters from the left and the right subsystems, respectively.
 
Due to the locality of the perturbation, the Dyson equation \eqref{Dyson_eq_junction} admits the explicit solution
\begin{align}
     G_{n,n'} = G_{n,n'}^{L} + G_{n,n'}^{R} +  F_n D^{-1} \bar{F}_{n'} ,
     \label{GF_junction}
\end{align}
where
\begin{align}
    D= z - W_0 -  t_L   G_{-1,-1}^{L} t_L^{\dagger} - t_R^{\dagger}   G_{1,1}^{R} t_R 
    \label{D_def}
\end{align}
and
\begin{align}
    F_n &= -  \delta_{n,0} + G_{n,-1}^{L} t_L^{\dagger} + G_{n,1}^{R} t_R , \label{Fn_1} \\
    \bar{F}_{n'} &= -\delta_{0,n'} + t_L G_{-1,n'}^{L} + t_R^{\dagger}  G_{1,n'}^{R}. \label{Fnp_1}
\end{align}
We particularly note that
\begin{align}
    F_0 & = \bar{F}_0 = -1, \label{FF0} \\
    G_{0,0} &= D^{-1}.
    \label{G00}
\end{align}

On the basis of the solution \eqref{GF_junction} it is straightforward to compute the global DOS. Setting $z=\omega + i 0^+$, we evaluate
\begin{align}
    & \sum_{n=n_0}^{n_2} \text{tr} \left[ G_{n,n} - G_{n,n}^{L} - G_{n,n}^{R} \right] \\
    &= \text{tr} \left[ \frac{1}{D} \right]
    + \sum_{n=n_0}^{-1}  \text{tr} \left[ \frac{1}{D} t_L G_{-1,n}^L G^L_{n,-1} t_L^{\dagger} \right] \nonumber \\
    & + \sum_{n=1}^{n_2}  \text{tr} \left[ t_R^{\dagger} G_{1,n}^R G_{n,1}^R t_R\frac{1}{D} \right],
\end{align}
where the trace operation is performed in the channel space. Using the identities
\begin{align}
    \sum_{n=n_0}^{-1}  G_{-1,n}^L G^L_{n,-1} &= - \frac{\partial G_{-1,-1}^L}{\partial \omega} , \\
    \sum_{n=1}^{n_2} G_{1,n}^R G_{n,1}^R &= - \frac{\partial G_{1,1}^R}{\partial \omega} ,
\end{align}
which are most easily proven in the Lehmann representation, and the Jacobi's formula $\text{tr} \, [D^{-1} \frac{\partial D}{\partial \omega}] = \frac{\partial}{\partial \omega} \ln \det D$, we establish that
\begin{align}
    \rho (\omega) = \rho^L (\omega) + \rho^R (\omega) -\frac{1}{\pi} \text{Im} \, \frac{\partial}{\partial \omega} \ln \det D (\omega + i 0^+).
    \label{DOS_single}
\end{align}
The last term in this expression represents a correction to the global DOS due to the tunneling between the subsystems. It has a form analogous to that of the familiar expression in terms of the scattering matrix\cite{PhysRevB.65.115307}. The terms $\rho^L (\omega)$ and $\rho^R (\omega)$ represent the global DOS of the left and right disjoint subsystems, respectively.

In order to derive the continuum limit of \eqref{GF_junction} we make certain assumptions about the scaling of the Hamiltonian parameters with the lattice constant $a$. In particular, we define for $m=L,R$
\begin{align}
    t_{m} + t_{m}^{\dagger} = \frac{\mathcal{A}_{m}}{a^2}  , \label{A_scale} \\
    i (t_{m} - t_{m}^{\dagger}) = \frac{\mathcal{B}_{m}}{a} , \label{B_scale} \\
    W_{m} =  \frac{\mathcal{A}_{m}}{a^2}   + \mathcal{C}_{m},
\end{align}
where $\mathcal{A}_{m}, \mathcal{B}_{m}, \mathcal{C}_{m} = O (a^0)$ are constant matrices. 

In addition, we choose
\begin{align}
    W_0 &=  \frac{\mathcal{A}_L + \mathcal{A}_R}{2 a^2} + \frac{\mathcal{U}_1}{a} . \label{V_0} 
\end{align}
This choice of the leading $O (\frac{1}{a^2})$ term in \eqref{V_0} is important\cite{PhysRevA.52.1845} for ensuring the fulfillment of the matching conditions \eqref{general_matching_condition_1} and \eqref{general_matching_condition_2} for the Green's function in the continuum limit [see also in the end of the subsection].  In turn, the subleading $O (\frac{1}{a})$ term  induces the impurity  delta-potential \eqref{contact_pot_main}, which is generally matrix-valued. 

It is important to emphasize that our limiting procedure essentially differs from the one frequently used in the wide-band limit treatment of the tunneling regime $t'_{L,R} \ll t_{L,R}$, where the hopping amplitudes $t_{L,R}$ in the leads considerably dominate over the hopping amplitudes $t'_{L,R}$ from the leads onto the quantum dot at the site $n=0$, and the characteristic tunneling rates $\Gamma_{L,R} = \pi (t'_{L,R})^2/t_{L,R}$ giving rise to (the imaginary part of) the dot's self-energy are then much smaller than the bandwidths of the leads $\sim t_{L,R}$. Recall that in our treatment we set $t'_{L,R} = t_{L,R}$.

Defining the continuous variable $x = na$ in the limit $a \to 0$, we recover the continuum analog \eqref{hamilt_main_parts} of the lattice Hamiltonians \eqref{HL_lattice} and \eqref{HR_lattice}. Relating the continuum and the lattice Green's functions via
\begin{align}
    G (x,x') = \lim_{a \to 0} \frac{1}{a} G_{n,n'},
    \label{G_limit}
\end{align}
we derive in Appendix \ref{app:Caroli_approach} the continuum analog of \eqref{GF_junction}. It reads
\begin{align}
\nonumber
    G(x, x')=&G^{L}(x, x')+G^{R}(x, x')\\
    &+F(x) d^{-1} \bar{F}(x'), \label{t_matrix_style}
\end{align}
where $G^{L}(x, x')$ and $G^{R}(x, x')$ are the boundary Green's functions of the corresponding disjoint regions $(x_0 , 0)$ and $(0,x_2)$ which identically vanish outside of them, respectively. The last term in Eq. \eqref{t_matrix_style} contains non-diagonal contributions in the subsystem's basis (i.e. it is generically non-zero for all $x,\ x'\in (x_{0},\ x_{2})$), and thereby it mediates the coupling between the subsystems. It is defined through the following objects
\begin{widetext}
   \begin{align}
     F (x)= \lim_{a \to 0}  F_n = &- \Theta (-x) \left[ G_2^{(0,L)} (x,0^+) + G_2^{L} (x,0^-) - G_2^{(0,L)} (x, 0^-) \right]\frac{\mathcal{A}_{L}}{2} \nonumber \\
    &   + \Theta (x)   \left[ G_2^{(0,R)} (x, 0^-) +G_2^{R} (x, 0^+) -G_2^{(0,R)} (x, 0^+) \right]\frac{\mathcal{A}_{R}}{2}, \label{sing_bar_gf1} \\
     \bar{F} (x')= \lim_{a \to 0}  \bar{F}_{n'}=& -  \Theta (-x')\frac{\mathcal{A}_{L}}{2} \left[ G_1^{(0,L)} (0^+, x') + G_1^{L} (0^-, x') - G_1^{(0,L)} (0^-, x') \right] \nonumber \\
    &   +  \Theta (x') \frac{\mathcal{A}_{R}}{2}\left[ G_1^{(0,R)} (0^-,x') +G_1^{R} (0^+, x') -G_1^{(0,R)} (0^+, x') \right], \label{sing_bar_gf2} \\
    d = \lim_{a \to 0} a \, D =& - \mathcal{U}_{1} - \frac{1}{8} \mathcal{A}_{L}\lim_{x \to 0^-} \frac{d^2}{d x^2} G^{L} (x, x)\mathcal{A}_{L} - \frac{1}{8} \mathcal{A}_{R}\lim_{x \to 0^+} \frac{d^2}{d x^2} G^{R} (x, x)\mathcal{A}_{R}, \label{sing_bar_gf3}
\end{align}
Above we employed a set of compact notations for the coordinate derivatives of Green's functions
\begin{align}
    G^{m}_{1}(x, x')=\frac{\partial G^{m}(x, x')}{\partial x}, \quad G^{m}_{2}(x, x')=\frac{\partial G^{m}(x, x')}{\partial x'},
\end{align}
with analogous abbreviations for the translationally invariant propagators $G^{(0, m)}(x, x')$.
\end{widetext}

The obtained expressions represent a multichannel generalization of Eqs.~(22) and (30) in Ref. [\onlinecite{C_Caroli_1971}]. It is also worth 
mentioning that similar expressions for the two-channel case were previously derived in Refs. [\onlinecite{PhysRevB.18.1076},\onlinecite{Arnold1985}].

The expression \eqref{t_matrix_style} can be interpreted in terms of the $T$-matrix. Formally rewriting it as
\begin{align}
     G (x, x' ) &=  G^{L+R} (x,x') + \langle x | G^{L+R} l^{\dagger} d^{-1} l G^{L+R} | x' \rangle \\
     &= G^{L+R} (x,x') \nonumber \\
     &+ \int d y \int d y'  G^{L+R} (x,y) T (y,y') G^{L+R} (y',x'),
\end{align}
where $G^{L+R} = G^L + G^R$ and $l$ is the formally introduced boundary hermitian operator in terms of 
\begin{align}
    \langle x | G^{L+R} l^{\dagger} | y \rangle &= i \delta (y) F (x), \\
    \langle y' | l G^{L+R} | x' \rangle &= -i \delta (y') \bar{F} (x'), 
\end{align}
we identify $T (y,y') = \langle y| l^{\dagger} d^{-1} l | y'\rangle$ with the $T$-operator in the coordinate representation.

The functions \eqref{sing_bar_gf1} and \eqref{sing_bar_gf2} have the remarkable properties
\begin{align}
    F (0^+) = F (0^-) \equiv F (0) & =-1 , \label{Fcontact} \\
    \bar{F} (0^+) = \bar{F} (0^-) \equiv \bar{F} (0) & =-1 , \label{Fbcontact}
\end{align}
which follow from the standard jump conditions
\begin{align}
[G_{2}^{(0,m)} (0,0^+)- G_2^{(0,m)} (0,0^-)] \frac{\mathcal{A}_m}{2} &= 1, \label{jump1_1st} \\
\frac{\mathcal{A}_m}{2} [G_{1}^{(0,m)} (0^+,0)- G_1^{(0,m)} (0^-,0)] &= 1
\label{jump2_1st}
\end{align} 
for the Green's function derivatives. The expressions \eqref{Fcontact} and \eqref{Fbcontact} thus appear to be consistent with their lattice analogues \eqref{Fn_1} and \eqref{Fnp_1}, respectively.

By the virtue of $G^L (0,0) = G^R (0,0)$ and \eqref{Fcontact}, \eqref{Fbcontact} we recover
\begin{align}
    G (0,0) = d^{-1}.
    \label{G00_d_rel}
\end{align}
This result is also consistent with its lattice analog \eqref{G00}. Replacing $D \to d/a$ in the lattice version of the global DOS \eqref{DOS_single}, we immediately obtain its continuum counterpart
\begin{align}
    \rho (\omega) = \rho^L (\omega) + \rho^R (\omega) -\frac{1}{\pi} \text{Im} \, \frac{\partial}{\partial \omega} \ln \det d (\omega + i 0^+).
    \label{DOS_single_continuum}
\end{align}
The global DOS $\rho^m (\omega)$ of the disjoint subsystem $m$ (here $m=L,R$) is generally given by
\begin{align}
    \rho^{m}(\omega)=-\frac{1}{\pi}\text{Im}\int_{x_m}^{x_{m+1}}{dx}\, \text{tr}\, G^m (x, x;\omega + i 0^+) . \label{SD_disjoint}
\end{align}

Since $d (\omega)$ is hermitian on the real frequency axis, its determinant $\det d (\omega)$ is real-valued. The real-valued roots of the equation
\begin{align}
    \det d (\omega) =0
    \label{BS_equation_basic}
\end{align}
determine energies $E_B$ of bound states induced by the tunneling between the two subsystems. Their contribution to the global DOS \eqref{DOS_single_continuum} naturally appears in the form $\sum_B \delta (\omega - E_B)$.

In the infinite-space model (that is, when both the left and the right subsystems are semi-infinite), the bound states found from the equation \eqref{BS_equation_basic} are localized near the interface of the two subsystems, and their energies $E_B$ reside in the bandgaps of the whole system. It is also remarkable that in this case the  matrix $d$  can be alternatively written as
\begin{align}
    d = [G^{(0,L)} (0,0)]^{-1} - p_0 ,
    \label{d_repr_T}
\end{align}
and hence the bound state equation \eqref{BS_equation_basic} acquires the form
\begin{align}
    \det \left[ 1- G^{(0,L)} (0,0; \omega ) p_0 (\omega) \right] =0 .
\end{align}
Here
\begin{align}
    p_0 &= \mathcal{U}_1 \\
    &+ \frac{1}{8} \lim_{x \to 0^+} \frac{d^2}{d x^2} \left[ \mathcal{A}_R G^R (x,x) \mathcal{A}_R - \mathcal{A}_L G_0^L (x,x) \mathcal{A}_L \right] \nonumber 
\end{align}
is the (generally energy-dependent) term breaking the translational invariance of the auxiliary infinite-space model using the parameters of the left subsystem, i.e. described by $G^{(0,L)} (x,x')$.  In addition, we employed the auxiliary Green's function $G_0^L (x,x')$ which describes the model to the right from the hard-wall potential at $x=0$ but uses the parameters of the left subsystem (see Appendix \ref{bound_props_aps} for its explicit expression as well as for the adopted conventions regarding notations). To establish \eqref{d_repr_T} we used the identity
\begin{align}
    & [G^{(0,L)} (0)]^{-1} \label{G0_inv} \\
    &= -\frac18 \mathcal{A}_L \left[ \lim_{x \to 0^-} \frac{d^2}{d x^2} G^L (x,x) + \lim_{x \to 0^+}  \frac{d^2}{d x^2} G_0^L (x,x) \right] \mathcal{A}_L , \nonumber   
\end{align}
which naturally arises in the translationally invariant case $\mathcal{U}_1 =0$ and $(\mathcal{A}_L ,\mathcal{B}_L , \mathcal{C}_L ) = (\mathcal{A}_R ,\mathcal{B}_R , \mathcal{C}_R )$.

Reminding ourselves that the boundary Green's functions $G^{L}(x,x')$ and $G^{R}(x,x')$, entering Eqs.~\eqref{t_matrix_style}-\eqref{sing_bar_gf3}, admit a simple representation (see Appendix \ref{bound_props_aps}) in terms of  $G^{(0, L)}(x,x')$ and $G^{(0, R)}(x,x')$, respectively, we assert that the above construction completes our program of establishing $G (x,x')$ in the special $M=1$ case. 

To verify that $G (x,x')$ given by \eqref{t_matrix_style} does provide the solution of the equation \eqref{general_GF_equation} with the matching conditions \eqref{general_matching_condition_1} and \eqref{general_matching_condition_2}, we first note that both $G^L (x,x') + G^R (x,x')$ and $F (x)$ satisfy the differential equation \eqref{general_GF_equation}. Second, on the basis of the relations $G^L (0,x') = G^R (0,x')=0$ and \eqref{Fcontact} we establish the continuity of $G (x,x')$ at $x=0$, i.e. the condition \eqref{general_matching_condition_1} is fulfilled.
Third, we express the condition \eqref{general_matching_condition_2} for $G (x,x')$ in the form
\begin{align}
     \frac{\mathcal{A}_R}{2} \left[G_1^R (0^+ , x') + F' (0^+) d^{-1} \bar{F} (x') \right] \nonumber \\
    - \frac{\mathcal{A}_L}{2} \left[G_1^L (0^- , x') + F' (0^-) d^{-1} \bar{F} (x') \right] \nonumber \\
    = - \left[\mathcal{U}_1 - i  \frac{\mathcal{B}_R - \mathcal{B}_L}{2} \right] d^{-1} \bar{F} (x'),
\end{align}
and check whether it is fulfilled for $x' \neq 0$ (i.e. for $x' > 0^+$ and for $x' < 0^-$). Under this condition we identify $\bar{F} (x') =\frac12 \mathcal{A}_R G_1^R (0^+ , x')  - \frac12 \mathcal{A}_L G_1^L (0^- , x')$, and it remains to prove that
\begin{align}
      d  = - \mathcal{U}_1 - \frac{\mathcal{A}_R}{2} F' (0^+)+ \frac{\mathcal{A}_L}{2} F' (0^-) + i  \frac{\mathcal{B}_R - \mathcal{B}_L}{2} .
      \label{d_identity}
\end{align}
The fulfillment of this condition is shown in Appendix \ref{app:proof_match}. Thus it is finally justified that the derived $G (x,x')$ satisfies the both matching conditions.

It is remarkable that the formula \eqref{d_identity} along with the expressions \eqref{Fprime_R} and \eqref{Fprime_L} for $ F' (0^+)$ and $ F' (0^-)$, respectively, represent the simplest way of determining the matrix $d$ in the single-barrier case:
\begin{align}
    d &=  - \mathcal{U}_1 + \mathcal{L}_R  -  \mathcal{L}_L, \label{D_simple_form_1bar} \\
    \mathcal{L}_R &=  \frac{\mathcal{A}_R}{2}  G_1^{(0,R)} (0^+,0) [G^{(0,R)} (0,0)]^{-1} + \frac{i}{2} \mathcal{B}_R, \label{L_R}\\
    \mathcal{L}_L &=  \frac{\mathcal{A}_L}{2} G_1^{(0,L)} (0^-,0) [G^{(0,L)} (0,0)]^{-1} + \frac{i}{2} \mathcal{B}_L. \label{L_L}
\end{align}
In this setting, the relation of $d$ to the matching condition \eqref{general_matching_condition_2}, expressing the conservation of the current density, becomes especially transparent. The importance of the objects $\mathcal{L}_{R/L}$ for the properties of the boundary charge in the half-space models has been recently elucidated in Ref. [\onlinecite{PhysRevB.106.165405}].

\subsection{Multiple barriers}

A direct generalization of the single-barrier result \eqref{t_matrix_style} to the case of multiple barriers is given by the expression (see Appendix \ref{app:mult_junc} for its derivation)
\begin{align}
    G (x,x') &= \sum_{m=0}^M G^m (x,x') \nonumber \\ 
           &+ \sum_{m,m'=1}^M F_m (x) (d^{-1})_{m,m'} \bar{F}_{m'} (x'),
\end{align}
where $d$ is a block tridiagonal matrix, with the additional barrier indices $m,m'$ labeling the blocks. The diagonal blocks
\begin{align}
    d_{m,m} &= - \mathcal{U}_m - \frac18 \mathcal{A}_{m-1} \lim_{x \to x_m^-} \frac{d^2}{d x^2} G^{m-1} (x,x) \mathcal{A}_{m-1} \nonumber \\
    & - \frac18 \mathcal{A}_{m} \lim_{x \to x_m^+} \frac{d^2}{d x^2} G^{m} (x,x) \mathcal{A}_{m}
    \label{multi_bar_gf3_diag}
\end{align}
represent a generalization of \eqref{sing_bar_gf3}: They are defined locally at the interface positions $x_m$. In turn, the off-diagonal blocks
\begin{align}
    d_{m,m+1} &= \frac14 \mathcal{A}_m G_{12}^m (x_m^+ , x_{m+1}^-) \mathcal{A}_m, \label{multi_bar_gf3_above} \\
    d_{m+1,m} &= \frac14 \mathcal{A}_m G_{12}^m (x_{m+1}^- , x_{m}^+) \mathcal{A}_m \label{multi_bar_gf3_below}
\end{align}
describe the propagation between the two adjacent barriers $x_m$ and $x_{m+1}$ across the $m$-th subsystem. These blocks thereby account for the quantum interference effects.

The functions
\begin{widetext}
\begin{align}
    F_m (x) &= -\Theta (x_{m-1} < x < x_m) \left[ G_2^{(0,m-1)} (x, x_m^+ ) + G_2^{m-1} (x,x_m^-) - G_2^{(0,m-1)} (x,x_m^-)\right] \frac{\mathcal{A}_{m-1}}{2} \nonumber \\
            &+ \Theta (x_m < x < x_{m+1}) \left[  G_2^{(0,m)} (x, x_m^- ) + G_2^{m} (x,x_m^+) - G_2^{(0,m)} (x,x_m^+)\right] \frac{\mathcal{A}_{m}}{2} , \label{multi_bar_gf1} \\
    \bar{F}_{m'} (x') &= -\Theta (x_{m'-1} < x' < x_{m'}) \frac{\mathcal{A}_{m'-1}}{2}  \left[  G_1^{(0,m'-1)} (x_{m'}^+ ,x' ) + G_1^{m'-1} (x_{m'}^- , x') - G_1^{(0,m'-1)} (x_{m'}^- , x') \right] \nonumber \\
            &+ \Theta (x_{m'} < x' < x_{m'+1}) \frac{\mathcal{A}_{m'}}{2}  \left[  G_1^{(0,m')} (x_{m'}^- ,x' ) + G_1^{m'} (x_{m'}^+ , x') - G_1^{(0,m')} (x_{m'}^+ , x') \right]  \label{multi_bar_gf2}
\end{align}
\end{widetext}
generalize the expressions \eqref{sing_bar_gf1} and \eqref{sing_bar_gf2}, respectively. They also possess the properties
\begin{align}
    F_m (x_{m'}) = - \delta_{m,m'} , \\
    \bar{F}_{m'} (x_{m}) = - \delta_{m',m} ,
\end{align}
analogous to \eqref{FF0} in the single-barrier case. On their basis  we establish that
\begin{align}
    G (x_m , x_{m'}) = (d^{-1})_{m,m'} ,
\end{align}
and thus reveal the physical meaning of the matrix $d$: It is the inverse of the propagator between the contacts $x_m$ and $x_{m'}$.

The global single-barrier DOS \eqref{DOS_single_continuum} has a straightforward multi-barrier generalization (see Appendix \ref{app:mult_junc})
\begin{align}
    \rho (\omega)=&\sum_{m=0}^{M}\rho^{m}(\omega)-\frac{1}{\pi}\text{Im}\frac{\partial}{\partial\omega}\text{ln}\det d(\omega+i 0^+), \label{sd_general}
\end{align}
where the determinant of $d (\omega+i 0^+)$ is now evaluated for the $N_c M \times N_c M$ matrix.

The bound state equation \eqref{BS_equation_basic} retains its form in the multi-barrier case.

In the double-barrier case $M=2$ we can further simplify the matrix $d$. 
We assign the values $x_1 =0$ and $x_2 = W$ to the contact coordinates, such that $W$ is the width of the central ($C$) region. We also
assume that both the left ($L$) and the right ($R$) subsystems are semi-infinite, and in the following we use the labeling of the regions  $m=L,C,R$ instead of $m=0,1,2$. Owing to the expressions \eqref{two_bar_12}, \eqref{two_bar_21}, \eqref{two_bar_11}, \eqref{two_bar_22} derived in Appendix \ref{app:finite_simplifications} we state
 \begin{align}
     d &\equiv \left( \begin{array}{cc} G (0,0) & G (0,W) \\ G (W,0) & G (W,W)  \end{array}\right)^{-1} \nonumber \\
     &= \left( \begin{array}{cc} G^{(C,0)} (0,0) & G^{(C,0)} (0,W) \\ G^{(C,0)} (W,0) & G^{(C,0)} (W,W)  \end{array}\right)^{-1} - \left( \begin{array}{cc} p_0 & 0 \\ 0 & p_W  \end{array}\right).
     \label{d_two_barrier}
 \end{align}
 The last term can be interpreted as the self-energy term breaking the translational invariance of the auxiliary infinite-space model, which uses the parameters of the central subsystem. It is expressed in terms of
the energy-dependent matrices
    \begin{align}
& p_{0} =  \mathcal{U}_{1} \label{simp_form_main_2b_1}  \\
&+ \frac18 \lim_{x \to 0^-}   \left[ \mathcal{A}_L \frac{d^2}{d x^2} G^{L} (x,x) \mathcal{A}_L  -  \mathcal{A}_C  \frac{d^2}{d x^2} G^{C}_{0} (x, x) \mathcal{A}_C \right], \nonumber \\
& p_{W} =  \mathcal{U}_{2} \label{simp_form_main_2b_2} \\
&+ \frac{1}{8}  \lim_{x \to W^+}  \left[ \mathcal{A}_R  \frac{d^2}{d x^2} G^{R} (x, x)  \mathcal{A}_R - \mathcal{A}_C  \frac{d^2}{d x^2} G^{C}_{W} (x, x) \mathcal{A}_C \right].  \nonumber
\end{align}
Hereby we  introduced the auxiliary Green's functions $G^{C}_{0}(x, x')$ and $G^{C}_{W}(x,x')$ describing the models to the left from the hard-wall potential at $x=0$ and to the right from the hard-wall potential at $x=W$, respectively, and both using the parameters of the central subsystem (see Appendix \ref{bound_props_aps} for their explicit expressions).

Applying the formulas \eqref{R_id_pr} and \eqref{L_id_pr}, we further simplify \eqref{simp_form_main_2b_1} and \eqref{simp_form_main_2b_2} to
\begin{align}
    p_0 &=  \mathcal{U}_{1} + \mathcal{L}_L - \mathcal{L}_{L \to C},\label{p0simple} \\
    p_W &=  \mathcal{U}_{2} - \mathcal{L}_R + \mathcal{L}_{R \to C},\label{pWsimple}
\end{align}
where $\mathcal{L}_R$ and $\mathcal{L}_L$ are defined in \eqref{L_R} and \eqref{L_L}, and
\begin{align}
     \mathcal{L}_{L\to C} &=  \frac{\mathcal{A}_C}{2} G_1^{(0,C)} (0^-,0) [G^{(0,C)} (0,0)]^{-1} + \frac{i}{2} \mathcal{B}_C,\label{L_LtoC} \\
    \mathcal{L}_{R \to C} &=  \frac{\mathcal{A}_C}{2}  G_1^{(0,C)} (0^+,0) [G^{(0,C)} (0,0)]^{-1} + \frac{i}{2} \mathcal{B}_C \label{L_RtoC} \\
    &= \mathcal{L}_{L\to C} + [G^{(0,C)} (0,0)]^{-1} . \nonumber 
\end{align}
These formulas allow one to express the matrix $d$ in the double-barrier case given by \eqref{d_two_barrier} in terms of the objects $\mathcal{L}$ containing only first derivatives of the translationally invariant Green's functions $G^{(0,m)}$.

\section{Applications to Josephson systems}
\label{sec: Applications}
To showcase our formalism, we find it instructive to consider a number of standard-issue problems in the theory of Josephson junctions (JJ). 

By a JJ, one commonly understands a weak link between a pair of superconductors. When two BCS condensates (labeled by 1 and 2) are brought together and the tunneling of Cooper pairs between them is then switched on, the combined system finds a new ground state, in which the difference $\varphi=\varphi_{1}-\varphi_{2}$ of the phases $\varphi_{1},\ \varphi_{2}$ of the corresponding superconducting order parameters $\Delta_{1,2}=|\Delta_{1,2}|e^{i\varphi_{1,2}}$ adjusts itself to a particular value $\varphi =\varphi_{\text{min}}$. In particular, the most common cases are $\varphi_{\text{min}}=0$ and $\varphi_{\text{min}}=\pi$, defining the so-called $0$- and $\pi$-junctions. This result implies that the ground state energy of the combined system is a function of the aforementioned phase difference, with a minimum at $\varphi=\varphi_{\text{min}}$. In this regard, it is important to study the spectral flow of Josephson systems with the externally varied phase difference\footnote{In real experiments, the phase difference is typically varied by closing the system into a circular geometry far away from the tunneling region, and varying the magnetic flux threading the resulting ring.}, and this defines the first type of problems for showcasing our formalism. Specifically, in one of the following examples, we shall see how local Coulomb interaction at the contact between the condensates may lead to a crossover between the above-mentioned $\varphi_{\text{min}}=0$ and $\varphi_{\text{min}}=\pi$ ground states.

As Josephson systems feature, by construction, superconducting components, local charge conservation is violated resulting in the non-zero persistent current (known as the Josephson current (JC)) between the condensates comprising the junction. It turns out\cite{abrikosov2017fundamentals} that such a current is also a $\varphi$-dependent quantity that, quite generally, may be shown to be the $\varphi$-derivative of the aforesaid ground state energy. The study of the experimentally measurable JC defines the second problem, which we consider for demonstrating the potential of Green's function formalism.

\subsection{Basic definitions}
\subsubsection{Model Hamiltonian}
\label{sec: SC_models}
Before proceeding with concrete examples, we first specify notations of the model Hamiltonians. 

In what follows we restrict our consideration to spin-$\frac{1}{2}$ $s$-wave superconductors, although -- as is apparent from Section \ref{sec: formalism} -- our formalism allows including arbitrary matrix structure and momentum dependence of the order parameters up to $O (p^2)$. 

We consider the following second quantized Hamiltonian 
\begin{align}
\nonumber
    \mathcal{H}=&\int_{-\infty}^{\infty}{dx} \hat{\psi}^{\dagger}(x)\left[ p\frac{1}{2m(x)}p+\frac{1}{2}\{A(x),\ p\}+V(x)\right]\hat{\psi} (x)\\
    &+\frac{1}{2}\int_{-\infty}^{\infty}{dx}\left[ \hat{\psi}^{\dagger}(x)\hat{\Delta}(x)(\hat{\psi}^{\dagger}(x))^{T}+\text{h.c.}\right], \label{Hamiltonian_Josephson_noNambu}
\end{align}
where $\hat{\psi} (x) = (\hat{\psi}_{\uparrow} (x) , \hat{\psi}_{\downarrow} (x))^T$ is a two-component spinor, whose spin components $\sigma = \uparrow, \downarrow$ are the field operators obeying the standard fermionic anticommutation relations
\begin{align}
    \{ \hat{\psi}_{\sigma}(x),\ \hat{\psi}_{\sigma'}(x')\}=&0,\\ 
    \{\hat{\psi}_{\sigma}(x),\ \hat{\psi}_{\sigma'}^{\dagger}(x')\}=&\delta_{\sigma, \sigma'}\delta(x-x').
\end{align}
The effective mass $m(x)$ is a piece-wise constant scalar function of $x$; $A(x)=A^{\dagger}(x), V(x)=V^{\dagger}(x)$ are piece-wise constant $2\times 2$ hermitian matrices, and $\hat{\Delta} (x)=-\hat{\Delta}^{T}(x)=\Delta(x)i\sigma_{y}$ is the antisymmetric $2\times 2$ $s$-wave paring matrix expressed in terms of the spatially dependent scalar order parameter $\Delta (x)$. We assign the following spatial dependence to these objects: 
\begin{align}
    \begin{Bmatrix}
    m(x)\\
    A(x)\\
    V(x)\\
    \Delta(x)
    \end{Bmatrix}=&\Theta(-x)\begin{Bmatrix}
    m_{L}\\
    A_{L}\\
    V_{L}\\
    \Delta_{L}e^{i\varphi_{L}}
    \end{Bmatrix} \\
    +& \Theta(W > x > 0)\begin{Bmatrix}
    m_{C}\\
    A_{C}\\
    V_{C}\\
    0
    \end{Bmatrix}+\Theta(x-W)\begin{Bmatrix}
    m_{R}\\
    A_{R}\\
    V_{R}\\
    \Delta_{R}e^{i\varphi_{R}}
    \end{Bmatrix}. \nonumber 
\end{align}
Note that the left ($- \infty < x<0$) and the right ($+ \infty > x> W$) regions host superconductors, while the central -- tunneling -- region $W > x >0$ is normal ($|\Delta_C|=0$). Introducing the labelling of the regions in terms of the index $\lambda = L,C,R$ (note that in the previous section it corresponds to the index $m$, but here and below we use $\lambda$ to avoid confusion with the mass notation), we further specify that $m_{\lambda}$ and $\Delta_{L,R}$ take positive real values, while the hermitian matrices $A_{\lambda} = A_{\lambda}^{(0)} + \sum_{j=x,y,z} A_{\lambda}^{(j)} \sigma_j \equiv  A_{\lambda}^{(0)} + \vec{A}_{\lambda} \cdot \vec{\sigma}$ and $V_{\lambda}= V_{\lambda}^{(0)} + \sum_{j=x,y,z} V_{\lambda}^{(j)} \sigma_j \equiv V_{\lambda}^{(0)} + \vec{V}_{\lambda} \cdot \vec{\sigma}$ are spanned by the Pauli matrices $\sigma_j$ and the identity matrix with the real-valued coefficients $A_{\lambda}^{(0,j)}$ and $V_{\lambda}^{(0,j)}$.

Introducing the extended Nambu spinor
\begin{align}
    \hat{\Psi} (x)=\begin{pmatrix} \hat{\psi} (x) \\ i\sigma_{y}(\hat{\psi}^{\dagger}(x))^{T}\end{pmatrix},
    \label{Nambu_extended}
\end{align}
we rewrite the Hamiltonian \eqref{Hamiltonian_Josephson_noNambu} in the form
\begin{align}
    \mathcal{H}=\frac{1}{2}\int_{-\infty}^{\infty}{dx}\, \hat{\Psi}^{\dagger}(x)H \hat{\Psi} (x),
\end{align}
where
\begin{align}
    H=&\begin{pmatrix} h^{(0)} & \Delta(x) \\ \Delta^{*}(x) &  -\sigma_{y}h^{(0)*} \sigma_{y} \end{pmatrix} \label{Hamiltonian_nambu_matrix},\\
    h^{(0)}=&p\frac{1}{2m(x)}p+\frac{1}{2}\{A(x),\ p\}+V(x). \label{Hamiltonian_nambu_matrix_2dline}
\end{align}
As per common practice, we find it convenient to define a new set of Pauli matrices $\tau_{x},\ \tau_{y},\ \tau_{z}$, along with an identity $\tau_{0}$, acting on the space of particles (upper two components of $\Psi(x)$) and holes (lower two components of $\Psi(x)$). The Hamiltonian \eqref{Hamiltonian_nambu_matrix} now may be written as
\begin{align}
    H=&\tau_{z}\left[ p\frac{1}{2m(x)}p+\frac{1}{2}\{\vec{A}(x)\cdot \vec{\sigma},\ p\}+V^{(0)}(x)\right] \label{Hamiltonian_nambu_matrix_taus_1}\\
    &+\tau_{0}\left[\frac{1}{2}\{A^{(0)}(x),\ p\}+\vec{V}(x)\cdot\vec{\sigma}\right] \label{Hamiltonian_nambu_matrix_taus_2}\\
    &+\Delta(x)\tau_{+}+\Delta^{*}(x)\tau_{-}. \label{Hamiltonian_nambu_matrix_taus_3}
\end{align}
In this decomposition of $H$, the terms  \eqref{Hamiltonian_nambu_matrix_taus_1}, that is the kinetic energy, the spin-orbit interaction, and the scalar potential, as well as the pairing potential \eqref{Hamiltonian_nambu_matrix_taus_3} are even under the standard time-reversal operation $\hat{T} = i \sigma_y K$, with $K$ denoting the complex conjugation. In turn, the terms collected in \eqref{Hamiltonian_nambu_matrix_taus_2}, that is the vector potential and the Zeeman field, are odd under the time-reversal operation.

\subsubsection{Observables}
\label{sec_obs_JJ}
It is well-known (see e.g. in Ref.~[\onlinecite{PhysRevB.81.224515}]) that introducing the extended Nambu representation artificially doubles the Hilbert space assigned to the quantum system. This redundancy of the description has to be removed in the calculation of observable quantities by enforcing a pseudo-reality constraint on the Nambu field operators.
Eventually this results in removing the hole-like part of the spectrum residing at negative energies $\omega<0$. It follows that the excitation spectrum of the system may be directly inferred from the Eqs.~\eqref{sd_general}, \eqref{d_two_barrier}, \eqref{p0simple}-\eqref{L_RtoC} by restricting the spectral range to $\omega>0$. For the JJ models, we choose the parameters 
\begin{align}
    \mathcal{A}_{\lambda}=\frac{\tau_{z}}{m_{\lambda}}, \quad \mathcal{U}_{1,2}=0.
\end{align}
Note that the contact potential strengths $\mathcal{U}_{1,2}$ are neglected, since the physical effect of the tunneling barrier in the finite-$W$ setup is accounted for by an appropriate tuning of $V_{C}$.

In the short junction limit, that is when the junction width $W$ is much smaller than all other physical length scales in the system, we impose the scaling  $V_C = \frac{\mathcal{U}_1}{W}$  on the potential of the central region, while all other terms in $H_C$ are supposed to be of $O (W^0)$. Then  $H_C = [\mathcal{U}_1 +O (W)] \, \delta_W (x)$, with the nascent $\delta$-function $\delta_{W}(x)=\frac{1}{W}\Theta(W > x > 0)$. In the limit $W \to 0$, we get the delta-distribution $\delta (x) = \lim_{W \to 0} \delta_{W}(x)$ and neglect the $O (W)$ terms accompanying $\mathcal{U}_1$. The resulting model has the single barrier at $x=0$ with the contact potential $\mathcal{U}_1 \delta (x)$, and the corresponding formulas for the spectral density \eqref{DOS_single_continuum}, \eqref{D_simple_form_1bar}-\eqref{L_L} become applicable.

In the expressions \eqref{DOS_single_continuum} and \eqref{sd_general} for the spectral density $\rho (\omega)$ of the composite system there are terms $\rho^m (\omega)$ expressing the spectral density of the isolated subsystems and the term expressed via $d$. This last term represents the correction due to the tunneling between the subsystems. It is the only term containing the dependence on the phase difference $\varphi$, which we indicate explicitly:
\begin{align}
     \delta\rho(\omega, \varphi)=-\frac{1}{\pi}\text{Im}\frac{\partial}{\partial\omega} \ln \det \, d(\omega + i 0^+, \varphi). \label{spec_d_short_intermed}
\end{align}

The Josephson current may be expressed as the derivative of the Gibbs free energy $F$ with respect to the phase difference across the superconducting leads, $J (\varphi) = \frac{2e}{\hbar} \frac{d F}{d \varphi}$. It is then given in terms of \eqref{spec_d_short_intermed} by
\begin{align}
    J(\varphi)=-\frac{2e}{\hbar \beta}\int_{0}^{\infty}d\omega\ln\left(2\cosh \frac{\beta\omega}{2} \right)\frac{\partial}{\partial \varphi} \delta \rho(\omega, \varphi), \label{finite_temp_rep1}
\end{align}
where  $\beta=\frac{1}{k_{B}T}$ is the inverse temperature. Integrating by parts we reveal an alternative representation
\begin{align}
    J(\varphi)=-\frac{1}{\Phi_{0}}\text{Im}\int_{0}^{\infty}d\omega \,  \tanh \frac{\beta\omega}{2}    \frac{\partial}{\partial \varphi} \ln \det d (\omega + i 0^+, \varphi), \label{key_current formula_1}
\end{align}
where $\Phi_{0}=\frac{h}{2e}$ is the superconducting magnetic flux quantum.

With the help of \eqref{key_current formula_1}, the identity
\begin{align}
     \frac{\partial}{\partial \varphi} \ln \det d (z, \varphi) = \text{tr}\left\{\left[d(z, \varphi)\right]^{-1}\frac{\partial d(z, \varphi)}{\partial \varphi}\right\} 
\end{align}
and the granted particle-hole symmetry, one can also convert the integral over the real frequency axis into the (quickly convergent) Matsubara sum
\begin{align}
    J(\varphi)=&-\frac{\pi }{\Phi_{0}\beta}\sum_{i\omega_{n}}\text{tr}\left\{\left[d(i\omega_{n}, \varphi)\right]^{-1}\frac{\partial d(i\omega_{n}, \varphi)}{\partial \varphi}\right\} \label{key_current formula_2}\\
    \stackrel{T=0}{\rightarrow}&-\frac{1}{2 \Phi_{0}}\int_{-\infty}^{\infty} d\omega \, \text{tr}\left\{\left[d(i \omega, \varphi)\right]^{-1}\frac{\partial d(i \omega, \varphi)}{\partial \varphi}\right\} \label{key_current formula_3},
\end{align}
which  in the zero-temperature limit goes over into the integral \eqref{key_current formula_3} over the imaginary frequency axis. These imaginary frequency representations appear very efficient in numerical calculations of $J (\varphi)$.

\subsection{Single barrier examples}
\label{sec: 1_barrier_examples}
First, we study the short junction case, in which the tunneling region is described by the contact potential $\propto \delta (x)$, and the results of Section \ref{Sec: single_barrier_theory} are then employed.

\subsubsection{Benchmark example: The Josephson model}
\label{Old_josephson_main}
Let us start by considering the paradigmatic problem of a pair of $s$-wave superconductors tunnel-coupled with each other across a barrier at $x=0$. In this case, the extended Nambu representation \eqref{Nambu_extended} is not needed, and the matrix Hamiltonians are expressed in the reduced Nambu basis $\hat{\Psi} (x) =(\hat{\psi}_{\uparrow} (x), \hat{\psi}_{\downarrow}^{\dagger} (x))^T$ as
\begin{align}
    H_{\lambda}=\begin{pmatrix} \frac{p^{2}}{2m}-\mu & \Delta_{\lambda} \\   \Delta_{\lambda}^{*} & -\frac{p^{2}}{2m}+\mu \end{pmatrix}, \quad \lambda =R,L ,
    \label{first_example_hamster}
\end{align}
where the order parameter is assumed to be homogeneous in magnitude throughout the sample $|\Delta_{\lambda}|=\Delta_{0} > 0$. Choosing the symmetric gauge, we express the phases of $\Delta_{\lambda}$ via the externally induced phase difference $\varphi=\varphi_{R}-\varphi_{L}$ such that 
\begin{align}
    \Delta_{\lambda}=\Delta_{0}e^{i\varphi_{\lambda}},\quad \varphi_{\lambda}=\lambda\frac{\varphi}{2}.
\end{align}
Hereby we conveniently re-labeled right $\lambda=+$ and left $\lambda=-$ subsystems. As is traditionally done, the nature of the barrier is modeled by the contact potential 
\begin{align}
    \mathcal{U} (x)=\delta(x)V_{0}\tau_{z}. \label{contact_pot_1bar_SC}
\end{align}

\begin{figure}
                \includegraphics[scale=0.53]{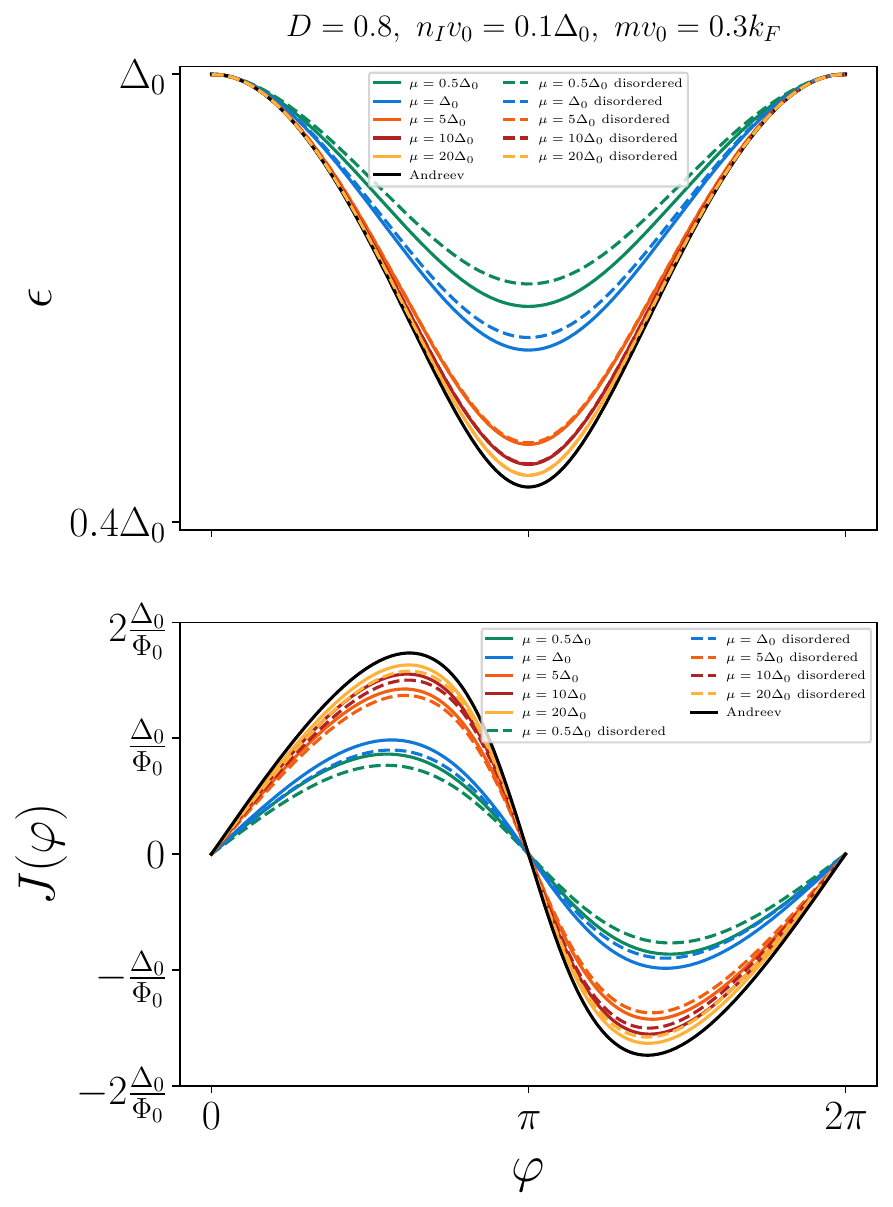}
                \caption{The phase-dispersion of the ABS (top panel) and zero-temperature JC (bottom panel). The results for a clean system are shown in solid lines and are compared to their disordered counterparts plotted in dashed lines. Various colors correspond to different values of the chemical potential, as indicated in the legends. To achieve the Andreev limit universality, we fix the values (see above the top panel) of  the barrier transparency $D$ as well as of the energy  $n_{I}v_{0}$ and momentum $mv_{0}$ scales associated with the impurity scattering.}
               \label{Andreev_disorder}
\end{figure}

The $d$-matrix for this model is calculated in Appendix \ref{ap_old_Josephson}. For $\text{Im} \, z >0$ it reads
\begin{align}
    d (z,\varphi) = -  V_{0} \tau_{z} + i \frac{k^{(+)}}{m}  \tau_{z}      + i \frac{ k^{(-)}}{m} \frac{z \mikpl{-}   \Delta_0 \cos \frac{\varphi}{2} \tau_x}{z \sqrt{1-(\frac{\Delta_0}{z})^2}}  , \label{spectral_matrix_example_1}
\end{align}
where 
\begin{align}
    k^{(\pm)} = k_F \frac{\sqrt{1 + \frac{z}{\mu} \sqrt{1 -(\frac{\Delta_0}{z})^2 }} \mp \sqrt{1 - \frac{z}{\mu} \sqrt{1-(\frac{\Delta_0}{z})^2 }} }{2}, \label{some_expr_wave_vec}
\end{align}
and $k_{F} =\sqrt{2m\mu}$. The general bound state equation \eqref{BS_equation_basic} yields
    \begin{align}
    & 0 = \frac{m^2}{k_F^2} \det d (\omega + i 0^+, \varphi) \label{Andreev_dispersion_clean_general} \\
    &= \left( \frac{k^{(+)}}{ik_{F}}  + \frac{m V_0}{k_F} \right)^2+\frac{k^{(-)2}}{k_{F}^{2}} \left( 1-   \frac{\Delta_{0}^2}{\Delta_{0}^2 -\omega^2} \sin^2 \frac{\varphi}{2} \right), \nonumber
    \end{align}
where  $D = 1/[1+ (m V_0 /k_F)^2]$ is the transparency of the tunneling barrier. It can be satisfied in the gap $|\omega | < \Delta_0$, where it holds
\begin{align}
    z \sqrt{1-\left(\frac{\Delta_0}{z} \right)^2} \bigg|_{z = \omega + i 0^+} = i \sqrt{\Delta_0^2 - \omega^2}, 
\end{align}
and the corresponding replacements are to be made in $k^{(\pm)}$. It follows that equation \eqref{Andreev_dispersion_clean_general} defines the exact energy-phase relation, 
valid for all values of $\mu$ and $D$.

Evaluating the JC on the basis of \eqref{key_current formula_1} [additionally multiplying it by the factor 2 to account the two copies of \eqref{spectral_matrix_example_1} needed to reproduce the extended Nambu representation] we note that the sub-gap contribution appears as the residuum 
\begin{align}
    J_{\text{ABS}} (\varphi) &= \frac{2\pi}{\Phi_0} \tanh \frac{\beta \omega_A }{2} \frac{\partial_{\varphi} \det d (\omega_A , \varphi) }{\partial_{\omega} \det d (\omega_A , \varphi)}  \\
    &= - \frac{2\pi}{\Phi_0} \tanh \frac{\beta \omega_A (\varphi)}{2} \frac{d \omega_A (\varphi)}{d \varphi}
    \label{J_ABS}
\end{align}
at the pole $\omega = \omega_A$, given by the ABS energy found from the equation \eqref{Andreev_dispersion_clean_general}. In turn, the continuum contribution to \eqref{key_current formula_1} equals
\begin{align}
     J_{\text{cont}} (\varphi) =& -\frac{ \sin \varphi}{ \Phi_{0}}\int_{\Delta_0}^{\infty}d\omega \,  \tanh \frac{\beta\omega}{2}  \frac{\Delta_{0}^2}{\omega^2-\Delta_{0}^2}  \nonumber \\
    & \times \text{Im} \frac{ k^{(-)2}   }{m^2 \,\det d (\omega + i 0^+, \varphi)} . \label{JC_cont}
\end{align}

To obtain the known ABS expression 
\begin{align}
    \omega_{A}(\varphi) = \Delta_{0} \sqrt{1- D \sin^2 \frac{\varphi}{2}}, \label{Josephs_dispers_LET}
\end{align}
we invoke the so-called Andreev approximation relying on  $\mu \gg \Delta_0$. In this limit we find that $k^{(+)}\to 0$ and $k^{(-)}\to k_{F}$, simplifying 
\begin{align}
    d (z, \varphi) \approx - V_0 \tau_z + \frac{i k_F}{m} \frac{z- \Delta_0 \cos \frac{\varphi}{2} \tau_x}{z \sqrt{1- (\frac{\Delta_0}{z})^2}},
    \label{d_Andreev}
\end{align}
as well as the equation \eqref{Andreev_dispersion_clean_general} to the form which admits the solution \eqref{Josephs_dispers_LET}. In addition, we notice that the JC is completely mediated by the ABS, since $k^{(-)}$ and $\det d (\omega + i 0^+)$ in \eqref{JC_cont} become purely real, and therefore $J_{\text{cont}} (\varphi)$ vanishes in the Andreev limit. Thus, Eq.~\eqref{J_ABS} yields the following universal relation\cite{https://doi.org/10.1002/pssa.2210470266,PhysRevLett.67.3836}
\begin{align}
    J(\varphi) = &\frac{\pi \Delta_{0}}{2 \Phi_{0}} \frac{D \tanh\left[\frac{\beta\Delta_{0} \sqrt{1- D \sin^2 \frac{\varphi}{2}}}{2}\right]}{\sqrt{1- D \sin^2 \frac{\varphi}{2}}}\sin \varphi. \label{Josephs_current_LET}
\end{align}

In Fig. \ref{Andreev_disorder}, in solid lines, we show how the energy-phase and the corresponding current-phase relations approach the universal results of Eqs. \eqref{Josephs_dispers_LET}, \eqref{Josephs_current_LET} upon an increase in the chemical potential $\mu$.

It is also worth mentioning that the expression \eqref{d_Andreev}  multiplied by $-1$ is analogous by virtue of \eqref{G00_d_rel} to the  self-energy  of a quantum dot coupled to superconducting leads in the weak tunneling regime which is further approximated in the wide-band limit, see e.g. Ref. [\onlinecite{Zonda2015}] for that self-energy expression. One has to replace the tunneling rate $\Gamma$ of that model with $\frac{k_F}{m}$ to gain the formal analogy with \eqref{d_Andreev}. In our treatment, however, an effective $\Gamma$ is itself of the bandwidth's order of magnitude, that is  $\sim \mu$ [cf. the general discussion in Section \ref{Sec: single_barrier_theory}].

\begin{figure*}
                \includegraphics[scale=0.28]{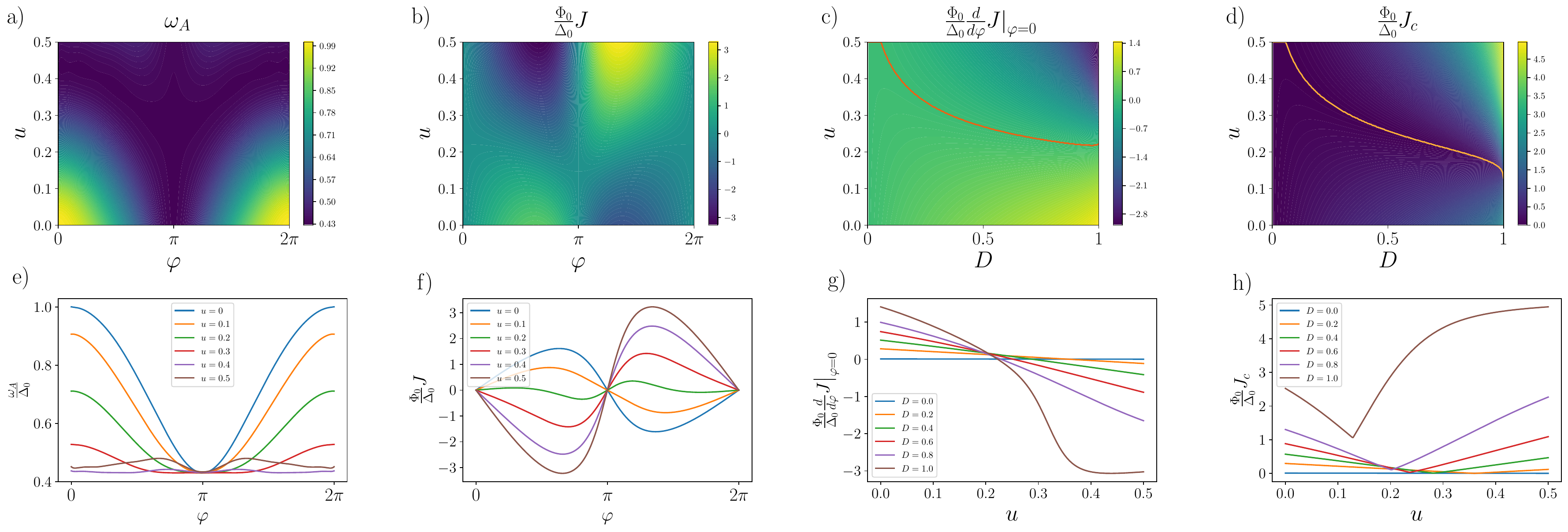}
                \caption{Panels a) and e): Dispersion of sub-gap states as a function of $\varphi$ and $u$ for a system with $\mu=3\Delta_{0},\ D=0.9$. Panels b) and f): Josephson current as a function of $\varphi$ and $u$ for $\mu=3\Delta_{0},\ D=0.9$. Panels c) and g): derivative of Josephson current with respect to phase difference $\varphi$ at $\varphi=0$ for a system with $\mu=3\Delta_{0}$. The orange line in panel c) indicates the sign change of $\frac{d}{d\varphi}J(\varphi=0)$. Panels d) and h): The critical current as a function of $u$ and $D$ for $\mu=3\Delta_{0}$. The orange line in panel d) indicates the location of the cusp in the critical current [explicitly seen in panel h)] as a function of Coulomb interaction strength $u$.}
               \label{Andreev_Coulomb}
\end{figure*}

One may further ask whether the JC value beyond the Andreev limit in our model is robust against static disorder, for example. Our approach allows one to get analytical insights into such questions at a modest expense. In particular, we may ignore the effects of disorder on the barrier tunneling, assuming that its imperfection is completely accounted for by the contact potential \eqref{contact_pot_1bar_SC}. Under such an assumption, we put the self-energy insertions into the bulk propagators $G^{(0, R/L)}$ alone.

Let us consider random non-magnetic impurities, characterized by the strength $v_0$ of the short-range impurity potential  and the impurity density $n_I$. Employing the standard $T$-matrix approximation [\onlinecite{altland2010condensed}], we obtain the following disorder-induced self-energy
\begin{align}
    \Sigma^{\lambda}(z)=& U_{\lambda} \Sigma (z) U_{\lambda}^{\dagger} ,  \quad U_{\lambda} = e^{\frac{i}{4} \tau_z \lambda \varphi} , \label{self_dis1} \\
    \Sigma (z)=& n_{I}v_{0}\tau_{z}\frac{1}{1-G^{(0)}(0,0;z) v_{0}\tau_{z}}+O(n_{I}^{2}), \label{self_energy_impurity_main}
\end{align}
dressing the momentum-space propagators of the bulk superconductors:
\begin{align}
     G^{(0, \lambda)}_{k}(z) &=  U_{\lambda} G^{(0)}_{k}(z)  U_{\lambda}^{\dagger} 
    \rightarrow  \tilde{G}^{(0, \lambda)}_{k}(z) = U_{\lambda} \tilde{G}^{(0)}_{k}(z)  U_{\lambda}^{\dagger} , \nonumber \\
    \tilde{G}^{(0)}_{k}(z)  &=  \frac{1}{[G^{(0)}_{k}(z)]^{-1}-\Sigma (z)} , \label{GF_impurity_main}
\end{align}
where $G^{(0)}_{k}(z)$ is given by \eqref{S_wave_appendix}. Representing
\begin{align}
    \tilde{G}^{(0)}_{k}(z) = \frac{1}{\tilde{z} (z) - \tau_z [\frac{k^2}{2m} - \tilde{\mu} (z)] - \tau_x \tilde{\Delta}_0 (z)}
\end{align}
and comparing it with \eqref{GF_impurity_main}, we find that the effect of the random disorder consists in the following energy-dependent renormalization of the model parameters
\begin{widetext}
    \begin{align}
    z\rightarrow& \tilde{z}(z)=z\left(1-n_{I}v_{0}k_{1,+} k_{2,+}\frac{i m v_{0}k^{(-)}  }{(m v_{0}k^{(-)})^{2}+(k_{1,+} k_{2,+}+  i m v_{0}k^{(+)})^{2}}\frac{1}{z\sqrt{1-(\frac{\Delta_0}{z})^2 }}\right),\\
    \Delta_{0}\rightarrow& \tilde{\Delta}_{0}(z)=\Delta_{0}\left(1-n_{I}v_{0}k_{1,+} k_{2,+}\frac{i m v_{0}k^{(-)}}{(m v_{0}k^{(-)})^{2}+(k_{1,+} k_{2,+}+  i m v_{0}k^{(+)})^{2}}\frac{1}{z\sqrt{1-(\frac{\Delta_0}{z})^2} }  \right),\\
    \mu\rightarrow&\tilde{\mu}(z)=\mu-n_{I}v_{0}k_{1,+} k_{2,+}\frac{k_{1,+} k_{2,+}+  i m v_{0}k^{(+)}}{(m v_{0}k^{(-)})^{2}+(k_{1,+} k_{2,+}+  i m v_{0}k^{(+)})^{2}},
\end{align}
\end{widetext}
where $k_{1,+}$ and $k_{2,+}$ are given in \eqref{roots_k1} and \eqref{roots_k2}. 

Since the $d$-function is eventually expressed via the parameters of the bulk systems, it is sufficient to make the above replacements in \eqref{spectral_matrix_example_1} in order to extract spectral information of the composite system subject to random potential disorder. Using the invariance 
\begin{align}
    \frac{z}{\Delta_0}=\frac{\tilde{z}(z)}{\tilde{\Delta}_0(z)}, \label{Anderson_scale_invariance}
\end{align}
one finds that the dispersion of Andreev levels in Eq. \eqref{Andreev_dispersion_clean_general} is modified by the following replacement 
\begin{align}
    & \frac{k^{(\pm)}}{k_F} \to  \frac{\tilde{k}^{(\pm)}}{k_F} = \sqrt{\frac{\tilde{\mu} (z)}{\mu}}  \\
    & \times \frac{\sqrt{1 + \frac{\tilde{z} (z)}{\tilde{\mu}(z)} \sqrt{1 -(\frac{\Delta_0}{z})^2 }} \mp \sqrt{1 - \frac{\tilde{z}(z)}{\tilde{\mu} (z)} \sqrt{1-(\frac{\Delta_0}{z})^2 }} }{2}. \nonumber
\end{align}
Returning back to the Andreev approximation $\tilde{\mu}(z)\approx \mu\gg\Delta_{0}$, we find that the results \eqref{Josephs_dispers_LET} and \eqref{Josephs_current_LET} remain intact.

In Fig. \ref{Andreev_disorder}, in dashed lines, we show how the energy-phase and the corresponding current-phase relations approach the universal results of Eqs. \eqref{Josephs_dispers_LET}, \eqref{Josephs_current_LET} upon an increase in the chemical potential $\mu$. In addition, Fig. \ref{Andreev_disorder} showcases the effect of disorder on the energy- and current-phase relations away from the universal Andreev limit, showing that static disorder tends to push the Andreev states into the continuum decreasing the critical current, similar to the effect of decreasing the barrier's transparency.

Next, we discuss the effect of the local Coulomb interaction at the point contact of the two superconductors. We model it as the Coulomb repulsion between fluctuations of local spin-up and spin-down densities,
\begin{align}
    V_{U}=&U \left\{ \hat{\psi}^{\dagger}_{\uparrow} (0) \hat{\psi}_{\uparrow}(0) - \langle n_{\uparrow} (0) \rangle \right\} \left\{ \hat{\psi}^{\dagger}_{\downarrow}(0)\hat{\psi}_{\downarrow}(0) - \langle n_{\downarrow} (0) \rangle \right\}.
    \label{V_U}
\end{align}
With this term our model resembles the Anderson-Josephson model of a quantum dot with the on-site Coulomb interaction tunnel-coupled to two superconducting leads (see the review about various studies of this model in Ref. [\onlinecite{Meden_2019}]). However, in our case, there is no well-defined quantum dot, since we treat the contact far beyond the tunneling regime.

We also note that the subtraction of the density averages in Eq. \eqref{V_U} eventually leads to the energy renormalization in the Cooper channel alone (see Eq. \eqref{HF_Wick} below, derived under the assumption of no spontaneous spin-symmetry breaking). Performing this subtraction, we get rid of an uninteresting renormalization of the diagonal contact-potential component in the particle-hole basis, which is largely dominated by high-energy normal-state contributions.  One can alternatively envisage this subtraction as a result of combining the particle-hole diagonal contribution to the (restricted) Hartree-Fock self-energy with the bare contact potential, which leads to an effective contact potential and defines the contact's physical transparency renormalized by the local Coulomb interaction.

The Hartree-Fock approximation to the local Coulomb interaction \eqref{V_U} is found on the basis of the Wick's theorem
\begin{align}
V_U 
&\simeq U \langle  \hat{\psi}_{\downarrow} (0)  \hat{\psi}_{\uparrow} (0) \rangle  \hat{\psi}_{\uparrow}^{\dagger} (0)  \hat{\psi}_{\downarrow}^{\dagger} (0)  \nonumber \\ 
&+ U \langle \hat{\psi}_{\uparrow}^{\dagger} (0)  \hat{\psi}_{\downarrow}^{\dagger} (0) \rangle  \hat{\psi}_{\downarrow} (0)  \hat{\psi}_{\uparrow} (0) \nonumber \\
&-  U \langle \hat{\psi}_{\uparrow}^{\dagger} (0)  \hat{\psi}_{\downarrow}^{\dagger} (0) \rangle \langle \hat{\psi}_{\downarrow} (0)  \hat{\psi}_{\uparrow} (0)  \rangle \nonumber \\
&= \Delta_{loc} \{  \hat{\psi}_{\uparrow}^{\dagger} (0)  \hat{\psi}_{\downarrow}^{\dagger} (0) +\hat{\psi}_{\downarrow} (0)  \hat{\psi}_{\uparrow} (0) \}  - \frac{\Delta_{loc}^2}{U}, \label{HF_Wick}
\end{align}
where $\Delta_{loc} = U \langle  \hat{\psi}_{\downarrow} (0)  \hat{\psi}_{\uparrow} (0) \rangle = U \langle \hat{\psi}_{\uparrow}^{\dagger} (0)  \hat{\psi}_{\downarrow}^{\dagger} (0) \rangle$ is a local superconducting order parameter.

The local Green's function $G_U (0,0)$ in the Hartree-Fock approximation results from the Dyson equation
\begin{align}
    d_U &= [G_U (0,0)]^{-1} \\
    &= [G (0,0)]^{-1} - \Delta_{loc} \tau_x = d - \Delta_{loc} \tau_x .
\end{align}
Using it in the expression \eqref{key_current formula_2} [times factor 2 because of the present usage of the reduced Nambu basis] for the JC and accounting for the correction $ - \frac{\Delta_{loc}^2}{U}$ to the free energy occurring in \eqref{HF_Wick}, we establish the mean-field expression for the JC
\begin{align}
J (\varphi) =& -\frac{2 \pi }{\Phi_{0}\beta}\sum_{i\omega_{n}}\text{tr}\left\{\left[d_U (i\omega_{n}, \varphi)\right]^{-1}\frac{\partial d_U (i\omega_{n}, \varphi)}{\partial \varphi}\right\} \nonumber \\
&- \frac{2 \pi}{\Phi_0} \frac{d}{d \varphi}\frac{\Delta_{loc}^2}{U} .
\label{current_U}
\end{align}
With the help of the self-consistency equation (see in Appendix \ref{app:coul_corr})
\begin{align}
    \Delta_{loc} = \frac{U}{2 \beta} \sum_{i \omega_n} \text{tr} \left\{ [d_U (i \omega_n , \varphi) ]^{-1} \tau_x\right\}
\end{align}
the expression \eqref{current_U} is modified  to the form
\begin{align}
J (\varphi) =& -\frac{2\pi }{\Phi_{0}\beta}\sum_{i\omega_{n}}\text{tr}\left\{\left[d_U (i\omega_{n}, \varphi)\right]^{-1}\frac{\partial d (i\omega_{n}, \varphi)}{\partial \varphi}\right\} .
\label{current_U}
\end{align}

A further application of the self-consistent Hartree-Fock approach to the present model reaches its limitation, emerging in the form of an ultraviolet divergence which is caused by the ultra-local form of the Coulomb interaction (see discussion in Appendix \ref{app:coul_corr}), and one has to refine the method. 

A similar problem arises in the bare perturbation theory in the Andreev limit $\mu \gg \Delta_0$: The local superconducting order parameter $\Delta_{loc}$ features the logarithmic behavior $\propto \frac{k_F}{m} \ln \frac{\mu}{\Delta_0}$, which hints at the nontrivial competition between the superconductivity and the Kondo effect.\cite{Meden_2019}

To give a conservative quantitative estimate of the interaction effects on the JC and the Andreev spectrum, we provide below the results of the first-order perturbation theory for $\mu \gtrsim \Delta_0$ and the dimensionless interaction parameter $u = \frac{U m^2 \Delta_0}{ \pi k_F^2} \ll 1$. In particular, we approximate the local self-energy term $\Delta_{loc} (\varphi) = \frac{k_F}{m} \bar{\Delta}_{loc} (\varphi)$ by its leading $O (u)$ contribution given in \eqref{Delta_loc_PT}. The resulting $d_U$ is used for the JC evaluation by means of \eqref{current_U} [or \eqref{current_U_zero_T} at zero temperature] as well as in the bound state equation
\begin{align}
    \det d_U (\omega , \varphi) =0.
\end{align}

Numerical data for the ABS and JC for a junction with $\mu=3\Delta_{0}$ are shown in Fig. \ref{Andreev_Coulomb}. The panels a) and e) demonstrate the energy-phase relation of the sub-gap states for various values of the interaction parameter $u\in[0,\ 0.5]$ and contact transparency $D=0.9$, while the panels b) and f) show the phase-dispersion of the JC at these parameters. We observe that upon increasing $u$ above $\sim 0.2$ the JC derivative [with respect to $\varphi$] changes its sign from positive to negative, indicating the $0-\pi$ phase transition. It is very analogous to the phase transition which is well-established in superconductor-ferromagnet-superconductor junctions [\onlinecite{robinson2007zero}, \onlinecite{chtchelkatchev2001pi}]. The panels c) and g) show the JC derivative  at $\varphi=0$ as a function of $u$ and $D$. In particular, we observe that the phase transition occurs at smaller values of $u$ in more transparent junctions with $D\to 1$. The panels d) and h) demonstrate the critical current $J_{c}=\max_{\varphi}|J(\varphi)|$ as a function of $u$ and $D$. We reveal that around the phase transition, the critical current displays a non-differentiable cusp in the dependence on the interaction parameter $u$.

\subsubsection{Junction of two Majorana wires: spectral properties}
\label{sec: Short_majorana_junction}
Let us now apply our formalism to the famous Majorana wire problem [\onlinecite{PhysRevLett.105.077001},\onlinecite{PhysRevLett.105.177002}]. In particular, we consider a pair of two semiconducting wires with the strong spin-orbit interaction $\alpha$ and induced superconducting correlations, which are additionally submersed into the external magnetic field $B$ pointing in the wires' direction. In the extended Nambu basis, introduced in Section \ref{sec: SC_models}, they are described by the following Bogoliubov-de-Gennes Hamiltonians
\begin{align}
    H_{\lambda} =& \left( \begin{array}{cc} h_p^{(0)} + \alpha p \sigma _z - B \sigma_x & \Delta_{\lambda} \\ \Delta_{\lambda}^{*} & -h_p^{(0)} -  \alpha p \sigma_z - B \sigma_x \end{array} \right),  \label{Majorana_wire_hamiltonian} \\ 
    & h_p^{(0)}= \frac{p^{2}}{2m}-\mu ; \quad \lambda =R,L \equiv +, -  . \label{Majorana_wire_hamiltonian_line2}
\end{align}
Assuming the wires to be semi-infinite, we bring them in contact\cite{PhysRevB.96.075404,PhysRevB.101.224501} at $x=0$ enabling the tunneling across the contact potential
\begin{align}
     \mathcal{U} (x) =\delta(x)\, V_{0}\, \tau_{z} \, \sigma_0, 
\end{align}
where $\sigma_0$ is the identity matrix in the spin space.

 As in the previous consideration, we consider again for simplicity the case of the isospectral junction with $|\Delta_{\lambda}|=\Delta_{0} > 0$, with a symmetrically induced phase difference $\Delta_{\lambda}=\Delta_{0} e^{\frac{i}{2 }\lambda\varphi}$ across the interface.

The bulk Green's functions 
\begin{align}
    G^{(0,\lambda)} (x,x') & = U_{\lambda} \, g (x-x') \, U_{\lambda}^{\dagger}, \quad U_{\lambda} = e^{\frac{i}{4} \tau_z \lambda \varphi} , \label{g_x_Maj} 
\end{align}
associated with the Hamiltonians \eqref{Majorana_wire_hamiltonian}, are evaluated in Appendix \ref{Ap_bulk_SC_majorana_model}. This information appears sufficient for establishing the composite Green's function $G (x,x')$ \eqref{t_matrix_style} and the $d$-matrix \eqref{D_simple_form_1bar} containing  spectral properties relevant for the study of the ABS and JC. For the present model we obtain
\begin{align}
    d =&  \frac{\tau_z}{2 m} \left\{- 2 m V_{0}\right. \label{d_Maj_single} \\
    & \left. \qquad + U_+ g' (0^+) g^{-1} (0) U_- - U_-  g' (0^-) g^{-1} (0) U_+ \right\}. \nonumber 
\end{align}

\begin{figure}
                \includegraphics[scale=0.19]{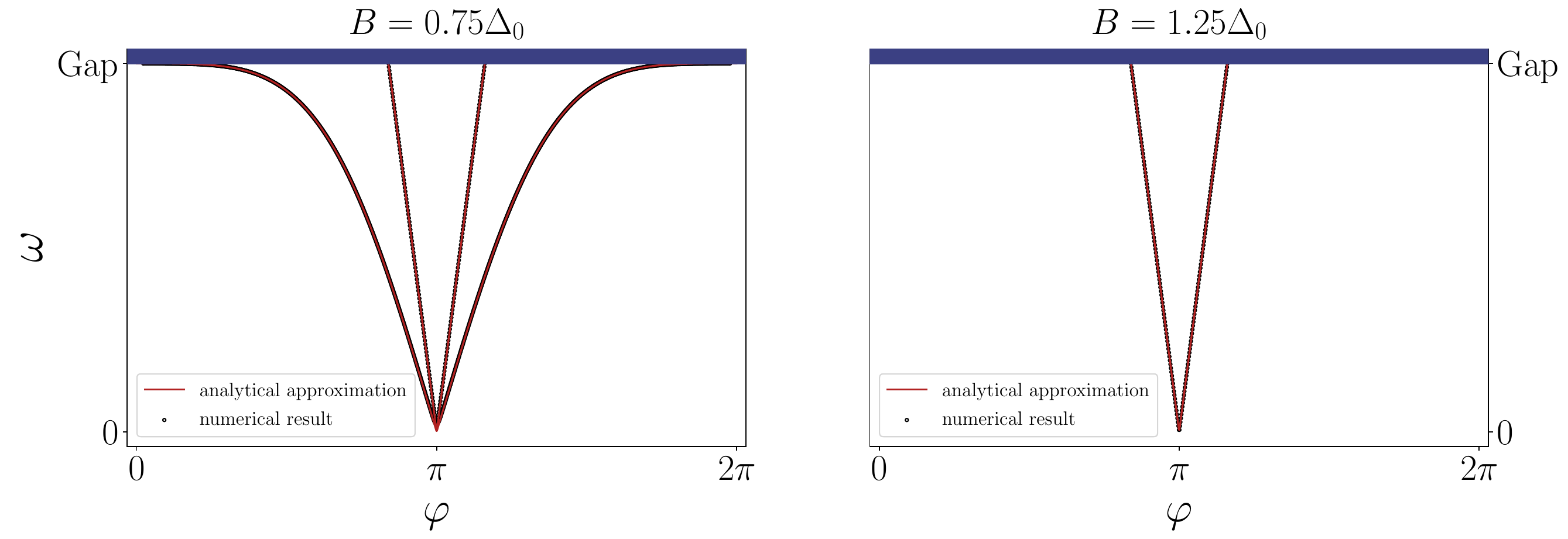}
                \caption{The ABS dispersion in the short junction of the two Majorana wires at zero chemical potential $\mu=0$ and large spin-orbit energy $E_{\text{SO}}=\frac{m\alpha^{2}}{2} = 60 \, \Delta_{0}$ dominating over $\Delta_0$ and $B$. Left and right panels refer to the non-topological ($B= 0.75 \,  \Delta_0 <\Delta_{0}$) and topological ($B = 1.25 \, \Delta_0 >\Delta_{0}$) regimes, respectively. As one can see, the numerical result based on the exact equation $\det d =0$ perfectly agrees with the approximate low-energy result  derived in Appendix \ref{Glazman_result} [see Eqs. \eqref{ABS_Majorana1} and \eqref{ABS_Majorana2}], which coincides with that of Ref. [\onlinecite{PhysRevB.101.224501}].}
               \label{fig:mu0_short_SO_dom}
\end{figure}

The calculation of the JC by means of  Eqs. \eqref{key_current formula_1}-\eqref{key_current formula_3} requires the additional derivative $\partial_{\varphi} d$. Since the $\varphi$-dependence of $d$ in the above formula enters only via the gauging matrices $U_{\pm}$, we note the following useful relation
\begin{align}
    \partial_{\varphi} d =&   \frac{i \tau_z}{8 m} \left\{ U_+ [\tau_z , g' (0^+) g^{-1} (0) ] U_- \right. \nonumber \\
    & \left. \qquad + U_-  [ \tau_z , g' (0^-) g^{-1} (0) ]  U_+ \right\}. \label{Der_spec_mat_short}
\end{align}

On the basis of the exact expressions for $g (0)$ and $g' (0^\pm)$ (see in  Appendix \ref{Ap_bulk_SC_majorana_model}) one can address various limits of the model's parameters. For example, we show in Appendix \ref{Glazman_result} how the recent results of Ref.  [\onlinecite{PhysRevB.101.224501}] for the energy spectrum and the JC in the regime of dominating spin-orbit energy $E_{\text{SO}} \equiv \frac{m \alpha^2}{2} \gg \Delta_0, B , \mu$ can be analytically recovered from our general expression for the $d$-function. We also use it to numerically evaluate the ABS dispersion at large $E_{\text{SO}} = 60 \, \Delta_0$ [see in Fig. \ref{fig:mu0_short_SO_dom}] to benchmark the present exact interface Green's function approach versus the scattering approach of Ref.  [\onlinecite{PhysRevB.101.224501}] relying on low-energy approximations in the spin-orbit dominated regime.

The demonstration presented in Appendix \ref{Glazman_result} serves a more general purpose of explaining how one can derive low-energy approximations for $G (x,x')$ in arbitrary heterostructures. Since $G (x,x')$ can be always expressed according to our present findings in terms of bulk Green's functions, it suffices to make a low-energy approximation for these functions. This approximation typically relies on the bulk spectrum linearization near Fermi points (see also Ref.~[\onlinecite{PhysRevB.106.165405}] for a similar discussion), and the approximate bulk Green's functions are evaluated much easier than their exact counterparts. 

At the same time, it is no longer needed to investigate how the low-energy approximation affects the matching condition \eqref{general_matching_condition_2} involving derivatives of the wavefunctions. This is usually a subtle problem since the spectrum linearization lowers by one the order of the Schr{\"o}dinger differential equation, and this requires relaxing the condition on the first derivatives. On the other hand, the matching condition \eqref{general_matching_condition_2} expresses the current density conservation, which must be somehow accounted for in the construction of the eigenfunctions. Our approach circumvents this problem, since it allows us to make the low-energy approximation directly for $G (x,x')$, skipping any intermediate approximate treatment of Eq. \eqref{general_matching_condition_2}.

The generality of our approach allows one to go beyond the low-energy approximation  relying on the dominance of the spin-orbit interaction energy $E_{\text{SO}}$ in the present model. Thus it enables exploring arbitrary parameter regimes of the Majorana junction model, as  demonstrated in the following. 

\begin{figure}[t!]
                \includegraphics[scale=0.20]{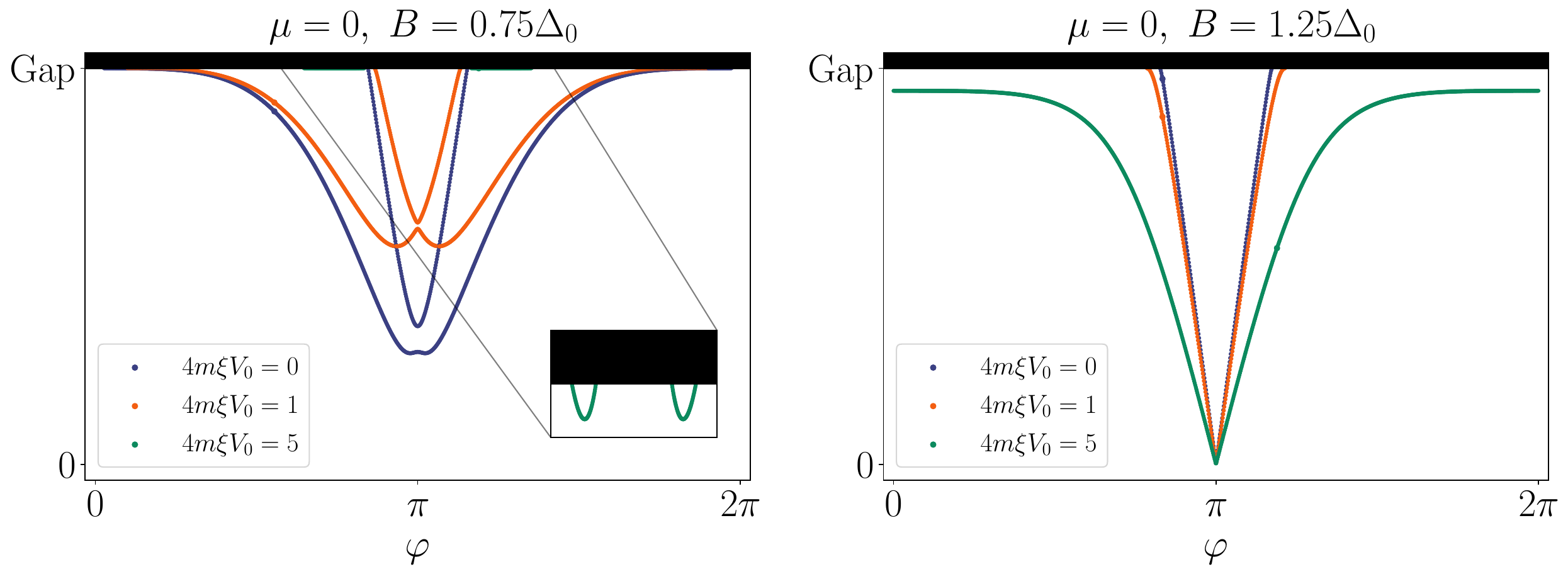}
                \caption{The ABS dispersion in the short Majorana junction at zero chemical potential $\mu=0$ and moderate spin-orbit interaction strength $E_{\text{SO}}=\frac{m\alpha^{2}}{2}=4\Delta_{0}$. Left and right panels refer to the non-topological ($B<\Delta_{0}$) and topological ($B>\Delta_{0}$) regimes, respectively. Blue, orange, and green colors are used to mark three different strengths of the contact potential $4m\xi V_{0}=0$, $4m\xi V_{0}=1$, and $4m\xi V_{0}=5$, respectively, which are expressed in terms of the  Ginzburg–Landau coherence length $\xi  =1/\sqrt{2m\Delta_{0}}$.}
               \label{fig:mu0_short}
\end{figure}

In particular, in Fig. \ref{fig:mu0_short} we display the dispersion of the sub-gap states for the moderate spin-orbit  energy $E_{\text{SO}}=4\, \Delta_{0}$, at zero chemical potential $\mu=0$. We note that in the non-topological regime $B<\Delta_{0}$, addressed in the left panel of Fig. \ref{fig:mu0_short}, Andreev levels are being pushed into the continuum upon an increase of the contact potential strength $V_{0}$, akin to  the behavior in ordinary JJs. In the topological regime, shown in the right panel of Fig. \ref{fig:mu0_short}, we find that the increase of $V_0$ results in the unwrapping of the Andreev mode, such that its tails near $\varphi=0,\ 2\pi$ are pushed towards zero energy.

\begin{figure}[t!]
               \includegraphics[scale=0.20]{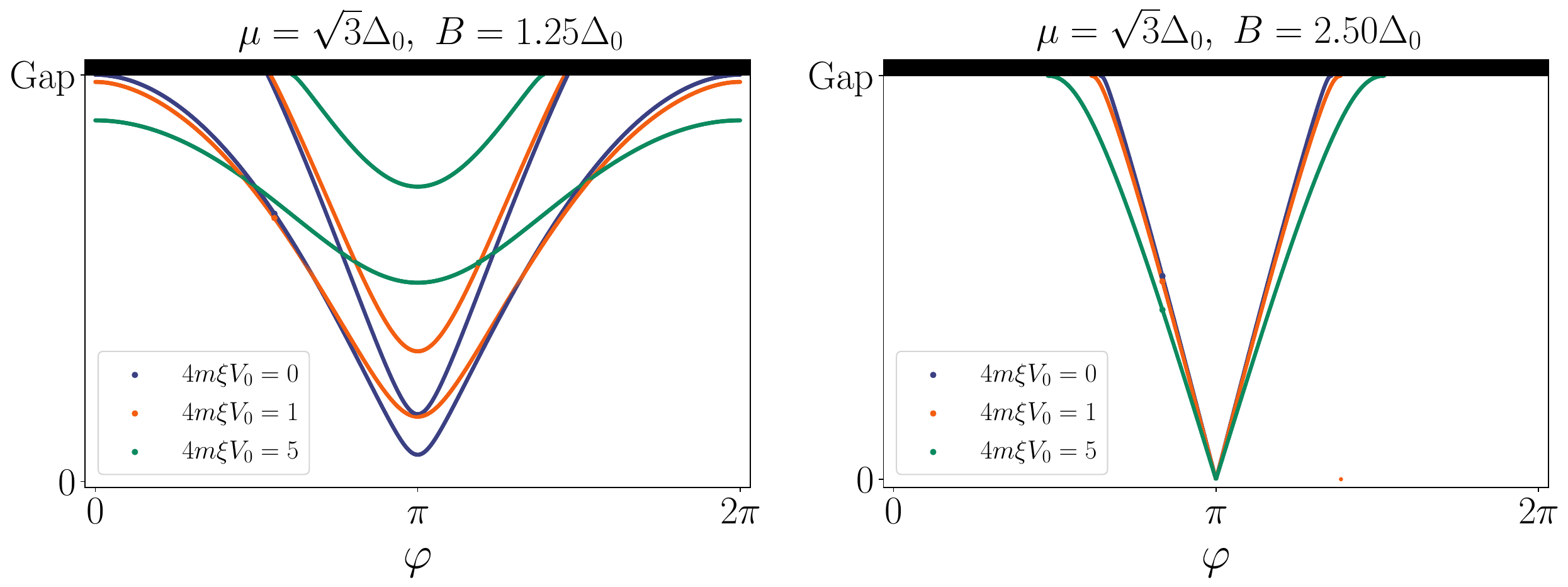}
                \caption{The ABS dispersion in the short Majorana junction at finite chemical potential $\mu=\sqrt{3}\Delta_{0}$ and moderate spin-orbit interaction strength $E_{\text{SO}}=\frac{m\alpha^{2}}{2}=4\Delta_{0}$. Left and right panels refer to the non-topological ($B<\sqrt{\Delta_{0}^{2}+\mu^{2}}$) and topological ($B>\sqrt{\Delta_{0}^{2}+\mu^{2}}$) regimes, respectively. Blue, orange, and green colors are used to mark three different strengths of the contact potential $4m\xi V_{0}=0$, $4m\xi V_{0}=1$, and $4m\xi V_{0}=5$, respectively. }
               \label{fig:mus3_short}
\end{figure}

The results for a yet different parameter regime with $E_{\text{SO}}=4\Delta_{0}$ and  -- more essentially -- nonzero $\mu=\sqrt{3}\Delta_{0}$  are shown in Fig. \ref{fig:mus3_short}. As before, the left and right panels refer to the non-topological and topological regimes, respectively. The nonzero $\mu$ replaces the  boundary value $B$ of the topological phase transition from $\Delta_{0}$ to the larger value $\sqrt{\Delta_{0}^{2}+\mu^{2}}$ ($= 2\Delta_{0}$ in our specific example). 
Apart form that, we observe the apparent increase in the value of $V_0$ required to push the Andreev states into and away from the continuum of the scattering states in the non-topological and topological phases, respectively. 

\subsubsection{Junction of two Majorana wires: Josephson current}

\begin{figure}[t!]
                \includegraphics[scale=0.20]{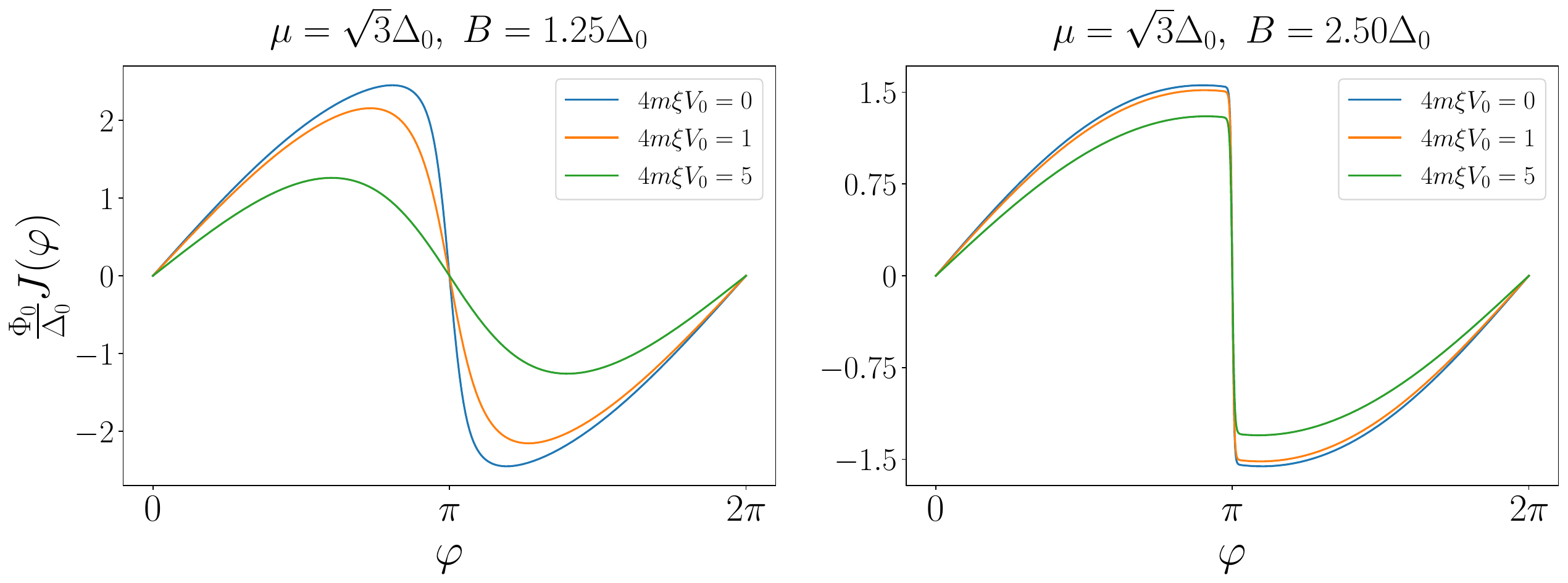}
                \caption{The zero-temperature ($T=0$) limit of the JC calculated at finite chemical potential $\mu=\sqrt{3}\Delta_{0}$ and relatively small spin-orbit energy $E_{\text{SO}}=\frac{m\alpha^{2}}{2}=4\Delta_{0}$. Left and right panels refer to the non-topological ($B=1.25\Delta_{0}<\sqrt{\Delta_{0}^{2}+\mu^{2}}=2\Delta_{0}$) and topological phases ($B=2.50\Delta_{0}>2\Delta_{0}$), respectively.}
               \label{fig:Current_short_T0}
\end{figure}

\begin{figure}[t!]
                \includegraphics[scale=0.20]{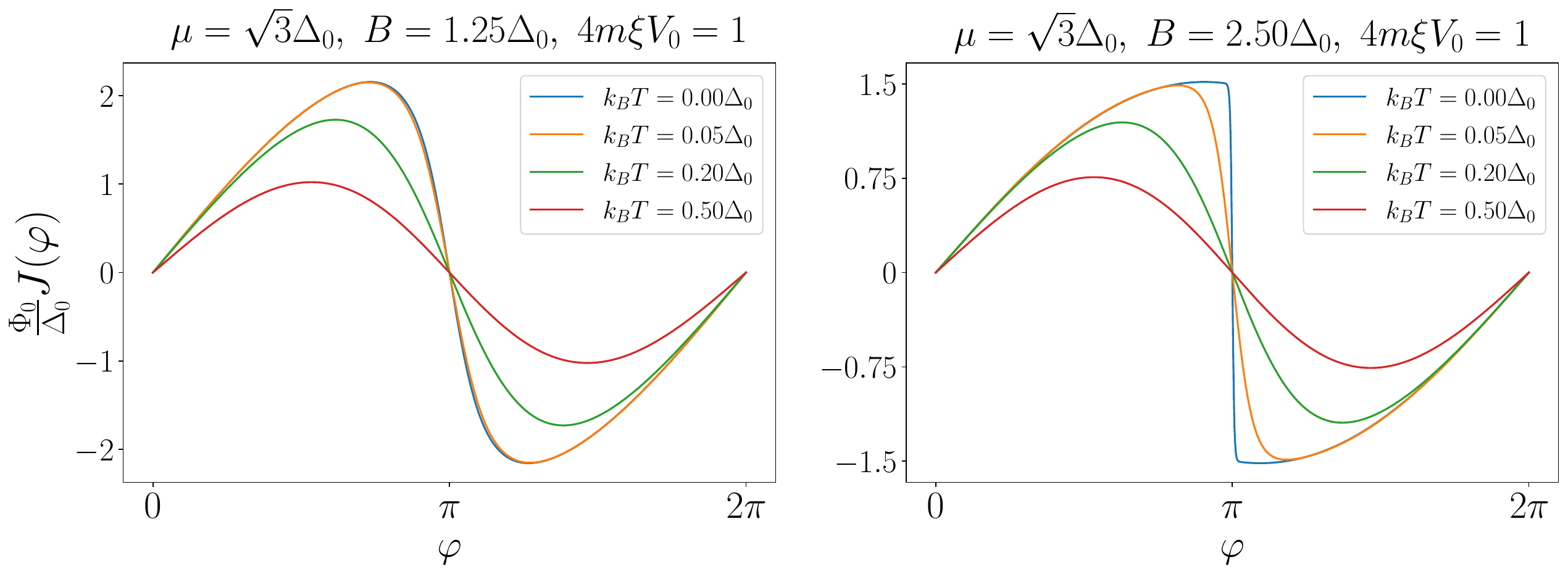}
                \caption{The effect of the finite temperature on the JC, calculated at finite chemical potential $\mu=\sqrt{3}\Delta_{0}$ and relatively small spin-orbit energy $E_{\text{SO}}=\frac{m\alpha^{2}}{2}=4\Delta_{0}$. Left and right panels refer to the non-topological ($B=1.25\Delta_{0}<\sqrt{\Delta_{0}^{2}+\mu^{2}}=2\Delta_{0}$) and topological phases ($B=2.50\Delta_{0}>2\Delta_{0}$), respectively.}
              \label{fig:Current_short_Tf}
\end{figure}

An example of the zero-temperature JC computed for the model parameters $E_{\text{SO}}=4\Delta_{0}$, $\mu=\sqrt{3}\Delta_{0}$ is shown in Fig. \ref{fig:Current_short_T0}, with the left and right panels referring to the non-topological and topological phases, respectively. In the non-topological phase, the JC demonstrates a sinusoidal behavior with a tilt towards the high-symmetry point $\varphi=\pi$. On the contrary, we see that the JC exhibits a sharp step-like discontinuity at the high-symmetry point $\varphi=\pi$ in the topological phase (see the left panel of Fig. \ref{fig:Current_short_T0}). This property relates to the discontinuity in the derivative of the ground state energy (see Fig. \ref{fig:mus3_short}) with respect to the phase difference $\partial_{\varphi} E_{\text{GS}}(\varphi)\propto\text{sgn}(\varphi-\pi),\ \varphi\approx\pi$. Such a result is nonphysical and arises from the non-commutativity of the zero-energy and zero-temperature limits, as is most easily seen in the representation \eqref{key_current formula_1}:
\begin{align}
    \lim_{T\rightarrow 0^{+}}\lim_{\omega\rightarrow 0^{+}}\tanh\left[\frac{\omega}{2k_{B}T}\right]=0,\\
    \lim_{\omega\rightarrow 0^{+}}\lim_{T\rightarrow 0^{+}}\tanh\left[\frac{\omega}{2k_{B}T}\right]=1.
\end{align}
This implies that any arbitrarily small but finite temperature will smear the sharp step (see Fig. \ref{fig:Current_short_Tf}). Additionally, Fig. \ref{fig:Current_short_T0} demonstrates the effect of the barrier imperfection $V_{0}\neq0$ on the JC, quite conventionally implying the reduction of the critical current $J_{c}=\max_{\varphi}|J(\varphi)|$ with the increase in the back-scattering strength $V_{0}$. 

The finite-temperature effects are shown in Fig. \ref{fig:Current_short_Tf}, with the left and right panels referring to the non-topological and topological phases, as before. In the non-topological phase, the key effects of thermodynamic fluctuations are the simultaneous decrease in the critical current and demolition of the aforementioned $\pi$-tilt of the current-phase relation. As is mentioned in the previous paragraph, in the topological phase the non-zero temperature destroys the sharp step in the current profile, further affecting the current-phase relation in a manner typical for the non-topological phase.

\subsection{Double barrier example: Long junction of two Majorana wires}
\label{sec:majorana_thick}

We continue by studying a long Josephson junction between the two Majorana wires. Now we assume that our system consists of the three layers: the right ($R$) and the left ($L$) superconducting leads, and the central ($C$) normal region which is free of superconducting correlations, as is outlined in Section \ref{sec: SC_models}. The Hamiltonians of the right and left superconducting leads are given by Eq. \eqref{Majorana_wire_hamiltonian}, while the Hamiltonian of the central region is
\begin{align}
    H_{C} =& \tau_{z}h_p^{(0, C)} + \tau_{z}\alpha_{C} p \sigma _z - B_{C} \sigma_x,\label{central_long_wire_hamiltonian}
\end{align}
with $h^{(0, C)}_{p}$ defined as in Eq. \eqref{Majorana_wire_hamiltonian_line2} with $\mu\rightarrow\mu_{C}$ and $m_{C}=m$.

The bound state equation \eqref{BS_equation_basic}, using \eqref{d_two_barrier} in the two-barrier case,
can be conveniently written as
\begin{align}
    \det \left( \begin{array}{cc} 1-  \hat{p}_0 \, g^C (0) & - \hat{p}_0 \, g^C (-W) e^{\frac{i}{2} \tau_z \varphi} \\  - \hat{p}_W \, g^C (W) e^{-\frac{i}{2} \tau_z \varphi} & 1- \hat{p}_W \, g^C (0)\end{array} \right) =0 , 
\end{align}
where the translationally invariant Green's function $g^C (x-x')$ for the normal central region is evaluated in Appendix \ref{Ap_bulk_Norm_majorana_model}. We also shifted the phase dependence from $p_0 = U_- \, \hat{p}_0 \, U_+$ and $p_W = U_+ \, \hat{p}_W \, U_-$ to the off-diagonal elements, responsible for the quantum coherence in the normal section, in order to elucidate its importance for supporting the JC. In the present model, the non-decaying terms $g^C (\pm W)$ are achieved due to the gapless spectrum of the Hamiltonian \eqref{central_long_wire_hamiltonian} (in particular, its outer branches do not gap out due to $\Delta_0 =0$). The matrices 
\begin{align}
    \hat{p}_{0,W} &= \pm \frac{\tau_z}{2m} g'(0^{\mp})  [g (0)]^{-1} \mp \frac{\tau_z}{2m}  g^{C \, '} (0^{\mp})  [g^C (0)]^{-1} \nonumber \\
    & \pm \frac{i}{2} (\alpha - \alpha_C ) \tau_z \sigma_z  
\end{align}
do not carry the $\varphi$-dependence and expressed with the help of the Green's function $g (x-x')$ of the bulk superconductors evaluated in Appendix \ref{Ap_bulk_SC_majorana_model}.

In Fig. \ref{fig: Spec_long_es} we show the phase-dependence  of the junction-localised bound states and of the JC at zero and finite temperatures. We focus on the case of the  isospectral junction with $B_{C}=B,\ \mu_{C}=\mu=\sqrt{3}\Delta_{0},\ \alpha_{C}=\alpha=4\Delta_{0}\xi$ varying the junction's width $W =0.5 \, \xi,\ 2.5 \xi ,\ 5 \xi ,\ 10 \xi$ [in the units of the coherence length $\xi = 1/\sqrt{2 m \Delta_0}$] at the two values of the magnetic field $B=1.25\Delta_{0},\ 2.5\Delta_{0}$. 

\begin{figure*}
    \centering
                \includegraphics[scale=0.29]{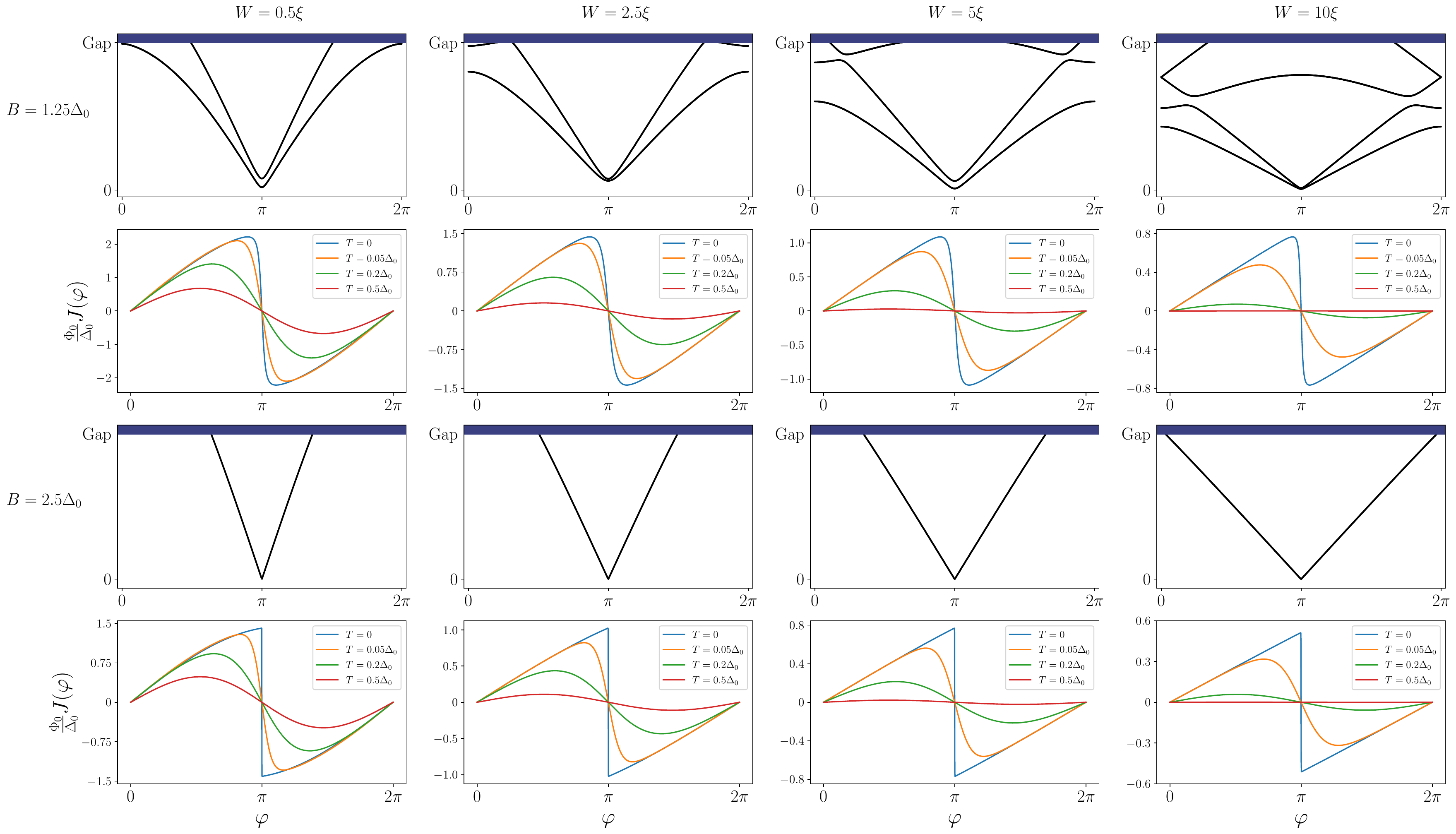}
                \caption{The phase dependence of the ABS and JC for long 
                JJs with $\alpha_{C}=\alpha=4\Delta_{0}\xi,\ \mu_{C}=\mu=\sqrt{3}\Delta_{0}$, and $B_{C}=B$, for a variety of the junction's widths $W=0.5\xi,\ 2.5\xi,\ 5\xi$, and $W=10\xi$ (four different columns). The two upper rows correspond to the non-topological regime $B=1.25\Delta_{0}<\sqrt{\Delta^{2}_{0}+\mu^{2}}=2\Delta_{0}$, while the two lower rows correspond to the topological phase $B=2.5\Delta_{0} > 2 \Delta_0$. In the second and fourth rows, different colors are used to mark different temperatures $T=0,\ 0.05\Delta_{0},\ 0.2\Delta_{0}$, and $0.5\Delta_{0}$.}
               \label{fig: Spec_long_es}
\end{figure*}

Specifically, the first and third rows demonstrate the evolution of the energy dispersion with an increase in the width of the central region. We see that in the non-topological phase ($B=1.25 \, \Delta_0<\sqrt{\Delta_{0}^{2}+\mu^{2}}=2\Delta_{0}$) the addition of the normal segment  between the two Majorana wires has the effect of lowering the energy of the sub-gap states, as well as increasing their number. In turn, in the topological phase ($B=2.5 \, \Delta_0 > 2 \Delta_0$) the only bound state mode tends to spread over the whole phase interval featuring the piece-wise linear branches of its phase dispersion.

As for the JC shown in the second and fourth rows of the same figure, we find that it features qualitatively similar behavior to its short-junction counterpart, with the major effect of the finite width being the decrease in the critical current.

\begin{figure*}
    \centering
                \includegraphics[scale=0.24]{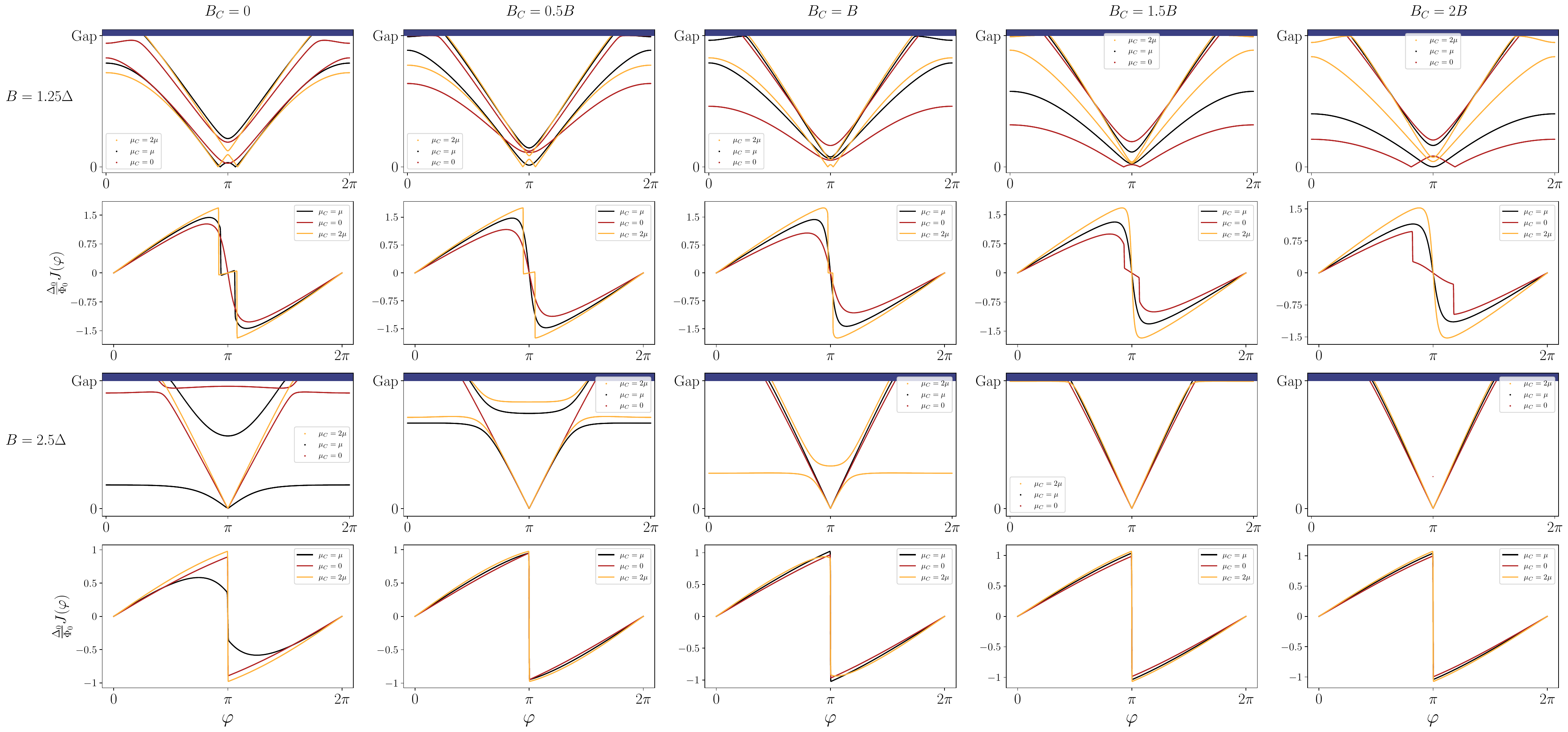}
                \caption{The phase dependence of the ABS and JC for long 
                JJs for the long ($W=2.5\xi$) JJ with $\alpha_{C}=\alpha=4\Delta_{0}\xi$. The two upper rows correspond to the non-topological phase $B=1.25\Delta_{0}$, while the two lower rows describe the topological regime $B=2.5\Delta_{0}$. The five different columns correspond to  five different values $B_{C}=0,\ 0.5B,\ B,\ 1.5B$, and $B_{C}=2B$ of the Zeeman field in the central region. Three distinct colors mark three different values $\mu_{C}=0,\ \mu,\ 2\mu$ of the chemical potential in the central region. }
               \label{fig: Spec_long_nes}
\end{figure*}

Next, we consider an example of the model's realization, in which some of the parameters of the central region are distinct from the corresponding ones in the superconducting leads. For instance, let us study the effect of modifying the parameters of Abelian ($\mu_{C}$) and non-Abelian ($B_{C}$) parts of the scalar potential. Experimentally this may be achieved by applying an appropriate gate voltage (to affect $\mu_{C}$) and by bringing the wire in proximity with a ferromagnet (to affect $B_{C}$).

In Fig. \ref{fig: Spec_long_nes}, we present the results for the  ABS and  the zero temperature JC in the model with $\alpha_{C}=\alpha=4\xi\Delta_{0},\ W=2.5\xi$, and $\mu=\sqrt{3}\Delta_{0}$. The five different columns correspond to five different values of  $B_{C}=0,\ 0.5B,\ B,\ 1.5B,$ and $2B$. Like in Fig. \ref{fig: Spec_long_es}, the first and third rows of Fig. \ref{fig: Spec_long_nes} show the phase-dispersion of the ABS in the non-topological ($B=1.25\Delta_{0}$) and topological ($B=2.5\Delta_{0}$) regimes of the Majorana wires, respectively, while the second and fourth rows give the corresponding zero-temperature current-phase relations. Three distinct colors are used to indicate three different values of the chemical potential in the normal region: $\mu_{C}=0$ (red),   $\mu_{C}=\mu$ (black), and  $\mu_{C}=2\mu$ (yellow). 

In the non-topological phase ($B=1.25\Delta_{0}$, first and second rows),  we observe that, by varying the parameters of the central region, it is possible to push the ABS energies downwards as much as needed for creating the crossing points at zero energy. They are similar to the one observed in the topological phase in the short junction. However, these crossing points are not topological in nature, and arise from the crossing of two particle-like and hole-like Andreev bands, that are symmetric around the $\varphi=\pi$ point, and hence come in pairs \footnote{Note that one can envisage the situation in which the two bands touch one another at exactly zero energy. In this case, however, the states around zero energy are expected to disperse quadratically, as opposed to the linear dispersion found in the topological regime.}. In turn, the crossing at zero energy in the topological phase arises from the intersection of two $\varphi=\pi$-asymmetric bands, belonging to two distinct parity branches, and is topologically protected [\onlinecite{PhysRevB.86.134522}]. We note that an analogous effect was also observed in superconductor-ferromagnet-superconductor heterostructures in the presence of spin-orbit coupling in Ref. [\onlinecite{PhysRevB.89.195407}]. 

As for the zero-temperature JC in the non-topological phase, we reveal that the appearance of the zero-energy-touching Andreev branches results in the discontinuities of the current-phase relations, akin to the topological regime. Like the touching points themselves, the step-like discontinuities also come in pairs and are notably falling behind the critical current in size. The latter effect has to do with the presence of additional strongly dispersing Andreev states, which significantly contribute to the current. 

In the topological phase ($B=2.5\Delta_{0}$, third and fourth rows), we see that independently of the parameter values inside the central region, there is always only one bound state touching the zero energy at $\varphi=\pi$. In addition, for the cases of the equal ($\mu_{C}=\mu$, black curve) and enhanced ($\mu_{C}=2\mu_{C}$, yellow curve) chemical potentials of the central segment  we observe that an additional bound state appears for $B_{C}<B$. As in the previous consideration of the topological regime, we find that the current-phase relation again features sharp step-like discontinuity at the high-symmetry point $\varphi=\pi$. We note that for most parameter values the size of this discontinuity is inappreciably smaller than the critical current.

\section{Conclusions and Outlook}
In this paper, we provided a detailed account of the interface Green's function technique regarding a large class of quasi-one-dimensional models of heterostructures. In our analysis, we assumed the Hamiltonians of the individual layers to be $N_{c}\times N_{c}$ matrix-valued quadratic polynomials in momentum operator conjugate to the composition axis of the structure. Such Hamiltonians may be seen as arising from the expansion of microscopic ones around the respective Fermi surfaces, allowing one to model a broad range of realistic physical systems (multi-band systems, systems with superconducting correlations, etc). For the considered class of models, we established the representation of the position space Green's function of the entire system in terms of bulk position space Green's functions of its sub-parts. By demonstrating that the calculation of latter objects is uninvolved, we opened up a simple pathway to study the multi-barrier scattering phenomena and the interface-localized bound states in systems of interest. Furthermore, our work lays one of the first stones into the analysis of many-body and disorder effects in layered systems, requiring knowledge of Green's functions as input. By further extracting the global DOS of the system from its Green's function, we reveal that the spectral information may, largely, be drawn from a single matrix $d$. For a system with $M$ tunneling barriers, we show that $d$ is an $N_{c} M\times  N_{c}M$ block matrix, comprised of 
$N_{c}\times N_{c}$-sized blocks admitting for a simple representation in terms of bulk Green's functions of the layers. As the spectral density relates to the logarithmic derivative of $\det \, d$, we also establish that the bound state energies of the system lie at its zeroes. 

We shall point out that our formalism explicitly deals with ballistic contact problems (beyond the tunneling regime), commonly studied in modern-day experiments [\onlinecite{PhysRevLett.128.197702}]. As nowadays the veil of theoretical secrecy over the ballistic regime is just being lifted, we hope our method finds its extensive use in this regard.

The method is further exemplified on models of normal and topological Josephson junctions. As a prototypical example, we considered a junction between two conventional superconductors with an induced phase difference across the interface. We derived the exact energy- and current-phase relations for the considered model, showing how they reduce to the commonly known formulas in the Andreev approximation, for which the chemical potential in superconducting leads is assumed to be the largest energy scale in the model. Further, we studied whether the universal Josephson relations, arising in the Andreev limit, are robust against static disorder. By disorder-dressing the bulk Green's functions of superconductors in the $T$-matrix approximation and using those to assemble the Green's function of the composite system, we demonstrated the stability of the Josephson relations in the Andreev approximation, revealing only small corrections to the ABS energy beyond it. Next, we studied the effect of the local Coulomb interaction at the contact point between superconductors, showing how it leads to the $0-\pi$ phase transition, detectable through the change in the current-phase relation of the system. 

As for the topological Josephson systems, we considered models of short and long junctions between two Majorana wires. We demonstrated how to calculate energy-phase and current-phase relations for the considered models beyond the low-energy approximation, reducing the numerical computation to root-searching problems.

In addition, we indicate that our method may serve as a convenient starting point to develop low-energy approximations for considered models. When projecting the Hamiltonians of different subsystems onto the low-energy window of interest, one is typically challenged to identify and further implement the correct matching conditions on the distinct low-energy degrees of freedom. Our method allows us to elegantly circumvent this problem by developing the low-energy approximations of bulk single-layer Green's functions, as in Ref. [\onlinecite{PhysRevB.106.165405}]. In this paper, we demonstrated how this strategy is realized in  the Majorana junction model (see Appendix \ref{Glazman_result}), reproducing the low-energy results of Ref. [\onlinecite{PhysRevB.101.224501}].

In our future work, we are going to generalize our developments to the class of position-dependent single-layer Hamiltonians, which are important to account for the effects of inhomogeneous external fields.

\section*{Acknowledgments}

KP thanks P. Ostrovsky for valuable comments. MP is grateful to P. Khomyakov, V. Meden, H. Schoeller, and A. Svetogorov for useful discussions.  KP and AS acknowledge funding from DFG Projects SH 81/6-1 and SH 81/7-1.

\begin{appendix}

\section{Evaluation of the Green's functions}
\subsection{Green's functions of translationally invariant systems}
\label{Bulk_propagators}
Let us consider the infinite-space model described by the Hamiltonian \eqref{hamilt_main_parts} and evaluate its Green's function \eqref{Lehmann_main_1} in the position representation. It can be conveniently rewritten as
\begin{align}
    G^{(0, m)}(x,x';z)=\int_{-\infty}^{\infty}\frac{dk}{2\pi}\frac{e^{ik(x-x')}}{Q (k; z)} P(k; z), \label{Lehman_ap_1}
\end{align}
where $P (k; z)= \text{adj} [z-h_m (k)]$ is the adjugate matrix, and $Q (k; z)=\det [z-h_m (k)]$ is a $2N_c$-order polynomial of $k$, admitting the representation 
\begin{align}
    Q (k; z)= \det ( \textstyle{\frac12} \mathcal{A}_m) \prod_{s=1}^{2N_{c}}(k-k_{s}(z)) \label{ploy_rep_1_ap}
\end{align}
in terms of the roots $k_s (z)$.

By the hermiticity of $h_m (k)$ the determinant $Q (k; \omega)$ on the real frequency axis may only have either a pair of complex conjugated roots or purely real roots. Upon the shift $\omega \to \omega + i 0^+$ the latter acquire infinitesimal imaginary parts, whose signs are given by the signs of $\frac{d k_s (\omega)}{d \omega}$, or equivalently by the signs of the dispersion slope (group velocity) $\frac{d \omega_k}{d k} \big|_{k=k_s}$. For the Hamiltonian \eqref{hamilt_main_parts}, a number of dispersion branches pointing up (down) at negative $k$ is the same as a number of dispersion branches pointing up (down) at positive $k$. This implies that for every $\omega$ the number of real roots with the positive dispersion slope equals the number of real roots with the negative dispersion slope. Therefore we can generally classify all roots into the two equally sized sets $\{k_{s,+} (\omega + i 0^+)\}_{s=1}^{N_c}$ and $\{k_{s,-} (\omega + i 0^+)\}_{s=1}^{N_c}$ according to
\begin{align}
    \text{Im}\left[ k_{s,+}(\omega+i 0^+)\right] >0, \quad 
    \text{Im}\left[ k_{s,-}(\omega+i 0^+)\right] <0. \label{post_selection_criteria}
\end{align}

Evaluating \eqref{Lehman_ap_1} at $z=\omega + i 0^+$ we close the integration contour in the upper and lower halves of the complex momentum plane for $x>x'$ and $x<x'$, respectively, to correspondingly pick up the residues at the isolated singularities $k_{s,+}$ or $k_{s,-}$. This results in 
\begin{widetext}
\begin{align}
    G^{(0,m)}(x,x'; \omega + i 0^+)=&\sum_{s=1}^{N_{c}}\frac{i\Theta(x-x') e^{i k_{s,+} |x-x'|}}{\det (\frac12 \mathcal{A}_m) \prod_{s'\neq s} (k_{s,+}-k_{s',+}) \prod_{s'}(k_{s,+}-k_{s',-})} P (k_{s,+}; \omega ) \nonumber \\
    &-\sum_{s=1}^{N_{c}} \frac{i\Theta(x'-x) e^{-i k_{s,-} |x-x'|}}{\det (\frac12 \mathcal{A}_m) \prod_{s'}(k_{s,-}-k_{s',+}) \prod_{s' \neq s}(k_{s,-}-k_{s',-})} P (k_{s,-}; \omega ).
    \label{G0m_app}
\end{align}    
\end{widetext}

Thus the evaluation of \eqref{Lehmann_main_1} reduces to a problem of the polynomial factorization \eqref{ploy_rep_1_ap} with a root post-selection criteria \eqref{post_selection_criteria}. In many practical applications, these tasks may be executed analytically, otherwise one can resort to numerical methods, such as the NumPy's built-in routine $\text{numpy.roots}$, for example [\onlinecite{harris2020array}].

\subsection{Boundary Green's functions}
\label{bound_props_aps}
For completeness of the exposition, we shall quote some basic results on the propagators of bounded systems. For a detailed account of the subject, we refer an interested reader to Refs. [\onlinecite{wimmer2009quantum}, \onlinecite{piasotskiuniversal}], while for various interesting applications to transport and solid state phenomena see Refs. [\onlinecite{PhysRevB.106.165405, zazunov2018josephson, PhysRevB.101.094511, PhysRevB.100.081106, PhysRevB.101.115405, PhysRevB.96.155103, PhysRevB.96.024516, PhysRevB.95.235143, PhysRevB.94.014502, PhysRevB.80.014425, muller2021universal}].

Consider a model described by Green's function $G^{(0,m)}(x,x')$. Inserting an ultra-local infinite-height potential at $x=X$, which creates a hard-wall boundary, we separate our initial system into two disjoint subsystems $x>X$ and $x<X$. The Green's functions of the both of them 
are given by the expression
\begin{align}
    &G^{m}_{X}(x, x')=G^{(0,m)}(x,x')  \nonumber\\
    &-G^{(0,m)}(x,X)\left[G^{(0,m)}(X,X)\right]^{-1}G^{(0,m)}(X,x'),
    \label{BGF_continuum}
\end{align}    
satisfying the boundary conditions $G^{m}_{X}(X, x')=G^{m}_{X}(x, X)=0$. It also holds that $G^{m}_{X}(x, x') $ identically vanishes when either $x>X,x'< X$ or $x<X,x'>X$.

The expression \eqref{BGF_continuum} also has the  lattice analog  (see e.g. in Ref. [\onlinecite{muller2021universal}])
\begin{align}
    G_{n,n'}^{m} = G^{(0,m )}_{n,n'} -  G^{(0,m )}_{n,0}  [G_{0,0}^{(0,m )} ]^{-1}  G_{0,n'}^{(0,m )} 
    \label{BGF_lattice}
\end{align}
for a nearest-neighbor tight-binding model with the infinite-height potential at the site $n=0$.

Adding the second hard-wall boundary at $x=Y$, we obtain the Green's function
\begin{align}
&G^{m}_{X, Y}(x, x')=G^{m}_{X}(x,x') \nonumber \\
&-G^{m}_{X}(x, Y)\left[G^{m}_{X}(Y,Y)\right]^{-1}G^{m}_{X}(Y, x') \label{BGF_non_symmetric_1} \\
&=G^{m}_{Y}(x,x') \nonumber \\
&-G^{m}_{Y}(x, X)\left[G^{m}_{Y}(X,X)\right]^{-1}G^{m}_{Y}(X, x'), \label{BGF_non_symmetric_2}
\end{align}  
satisfying the boundary conditions $G^{m}_{X, Y}(X, x')=G^{m}_{X, Y}(x, X)=G^{m}_{X, Y}(Y, x')=G^{m}_{X, Y}(x, Y)=0$. It can be also represented in the symmetrized form
\begin{align}
&G^{m}_{X, Y}(x,x') = G^{(0,m)} (x,x')  \nonumber \\
 \nonumber
&- G^{(0,m)}(x,X)\left[G^{m}_{Y}(X, X)\right]^{-1} G^{m}_{Y}(X, x') \\
&- G^{(0,m)}(x,Y)\left[G^{m}_{X}(Y, Y)\right]^{-1} G^{m}_{X}(Y, x') .
\label{BGF_symmetric}
\end{align}
Assuming that $X<Y$, we observe that the Green's function $G^{m}_{X, Y}(x, x')$ identically vanishes when $x$ and $x'$ belong to different intervals 
(smaller than $X$, between $X$ and $Y$, larger than $Y$). In addition, the following identities hold:
\begin{align}
    G^{m}_{X, Y}(x, x') = G^{m}_{X}(x, x') \quad x,x' <X , \\
    G^{m}_{X, Y}(x, x') = G^{m}_{Y}(x, x') \quad x,x' > Y .
\end{align}

With these definitions, we obtain the following expression for the boundary Green's functions introduced in the main text:
\begin{align}
G^{m}(x, x')=\Theta(x_{m+1}>x>x_{m}) G^{m}_{x_{m}, x_{m+1}}(x, x'),
\label{proj_bound}
\end{align}  
where we assumed that neither $x_{0}=-\infty$ nor $x_{M+1}=\infty$. In these special cases, we get simpler expressions on the basis of \eqref{BGF_continuum}
\begin{align}
G^0(x, x') &= \Theta(x_1-x) G^{0}_{x_1} (x, x'), \quad x_{0}=-\infty, \\
G^M (x, x') &= \Theta(x-x_M) G^{M}_{x_M}(x, x'), \quad x_{M+1}=\infty.
\end{align}

\section{Derivation of the continuum limit in the single-barrier model}
\label{app:Caroli_approach}

To derive the continuum limit $a \to 0$ of the lattice Green's functions \eqref{t_matrix_style} we expand $G^L_{-1,-1}$ and $G^R_{1,1}$ by two orders higher than the leading one: This is necessary to cancel the leading $O (\frac{1}{a^2})$ term in \eqref{D_def} and to correctly extract the subleading $O (\frac{1}{a})$ contribution.  This expansion is facilitated by the properties $G^L_{-1,0}=G^L_{0,-1}=G^L_{0,0}=0$ and $G^R_{1,0}=G^R_{0,1}=G^R_{0,0}=0$ which are inherent to the boundary Green's functions \eqref{BGF_lattice}. Using these identities as well as $G^{(0,L )}_{-1,-1} = G^{(0,L )}_{0,0}$ we have
\begin{align}
    G^L_{-1,-1} &=  G^{(0,L )}_{0,0} - G^{(0,L )}_{-1,0} - G^{(0,L )}_{0,-1}  + G^{(0,L )}_{0,0}  \nonumber \\
    & -  [G^{(0,L)}_{-1,0} - G^{(0,L)}_{0,0} ] [G_{0,0}^{(0,L )} ]^{-1} [G_{0,-1}^{(0,L)}-G_{0,0}^{(0,L)}] \nonumber  \\
    & \approx   a^2 G_1^{(0,L)} (0^-,0) - \frac{a^3}{2} G_{11}^{(0,L)} (0^-,0) \nonumber  \\
    & - a^2 G_1^{(0,L)} (0^+,0) - \frac{a^3}{2} G_{22}^{(0,L)} (0,0^-) \nonumber \\
    & - a^3 G_1^{(0,L )} (0^-,0)] [G^{(0,L )} (0,0)]^{-1} G_2^{(0,L )} (0,0^-). \label{cont_limit_GL11}
\end{align}
Hereby we identify $G_{n,n'}^{(0,L)}$ with $G^{(0,L)} (x,x')$ via \eqref{G_limit} and use the subindices 1 and 2 to indicate partial derivatives of the latter function with respect to the corresponding arguments.

Observing that $t_m = \frac{\mathcal{A}_m - i a \mathcal{B}_m}{2 a^2}$ and
\begin{align}
    & \lim_{x \to 0^-} \frac{d^2}{d x^2} G^L (x,x) = - G_{11}^{(0,L)} 
    (0^-,0) - G_{22}^{(0,L)} (0,0^-) \nonumber  \\
    &- 2 G_1^{(0,L)} (0^-,0) [G^{(0,L)} (0,0)]^{-1} G_2^{(0,L)} (0, 0^-) ,
\end{align}
and using  the jump conditions \eqref{jump1_1st},\eqref{jump2_1st}, we obtain
\begin{align}
   & t_L G^L_{-1,-1} t_L^{\dagger} = \frac{\mathcal{A}_L -i a \mathcal{B}_L}{2a^2} G^L_{-1,-1} \frac{\mathcal{A}_L + i a \mathcal{B}_L}{2 a^2} \nonumber \\
   &= -\frac{\mathcal{A}_L}{2 a^2} + \frac{1}{8a} \mathcal{A}_L \lim_{x \to 0^-} \frac{d^2}{d x^2} G^L (x,x) \mathcal{A}_L + O (a^0).
\end{align}

Analogously we find that
\begin{align}
    & t_R^{\dagger} G^R_{1,1} t_R \nonumber \\
    &= -\frac{\mathcal{A}_R}{2 a^2} + \frac{1}{8a} \mathcal{A}_R \lim_{x \to 0^+} \frac{d^2}{d x^2} G^R (x,x) \mathcal{A}_R + O (a^0).
\end{align}

Collecting together the terms defining $D$ in \eqref{D_def} we observe the cancellation of the $O (\frac{1}{a^2})$ terms thanks to the choice of the $a$-scaling in \eqref{V_0}. The remaining $O (\frac{1}{a})$ terms give rise to \eqref{sing_bar_gf3} defining the matrix $d$. Note that the linear $z$-dependence is suppressed in this limit since it appears in the next-to-subleading order of the $a$-expansion.

To approximate the remaining terms $G^L_{n,-1}$, $G^R_{n,1}$, $G^L_{-1,n'}$, and $G^R_{1,n'}$ in \eqref{GF_junction}, it suffices to expand them up to the leading order which turns out to be $O (a^2)$:
\begin{align}
     & G^L_{n,-1} = G^L_{n,-1} - G^L_{n,0} \approx 
     - a^2 G_2^{(0,L)} (x,0^+)\nonumber \\
    &  + a^2 G^{(0,L)} (x) [G^{(0,L)} (0,0)]^{-1} G_2^{(0,L)} (0,0^-) \nonumber \\
    & = - a^2 \left[ G_2^{(0,L)} (x,0^+) + G_2^L (x,0^-) - G_2^{(L,0)} (x,0^-) \right], \label{FL_lat_cont} \\
    & G^R_{n,1} = G^R_{n,1} - G^R_{n,0} \approx 
     a^2 G_2^{(0,R)} (x,0^-) \nonumber \\
    & - a^2 G^{(0,R)} (x,0) [G^{(0,R)} (0,0)]^{-1} G_2^{(0,R)} (0,0^+) \nonumber  \\
    & = a^2 \left[ G_2^{(0,R)} (x,0^-) + G_2^R (x,0^+) - G_2^{(0,R)} (x,0^+)\right], \label{FR_lat_cont} \\
    & G^L_{-1,n'} = G^L_{-1,n'} - G^L_{0,n'} \approx 
     -a^2 G_1^{(0,L)} (0^+,x') \nonumber \\
    & + a^2 G_1^{(0,L)} (0^-,0) [G^{(0,L)} (0,0)]^{-1}  G^{(0,L)} (0,x') \nonumber \\
    & = -a^2 \left[ G_1^{(0,L)} (0^+,x') + G_1^L (0^-, x')  - G_1^{(0,L)} (0^-, x') \right], \label{FbL_lat_cont} \\
    & G^R_{1,n'} = G^R_{1,n'} - G^R_{0,n'} \approx a^2 G_1^{(0,R)} (0^-,x') \nonumber \\
    & -a^2 G_1^{(0,R)} (0^+,0) [G^{(0,R)} (0,0)]^{-1}  G^{(0,R)} (0,x') \nonumber \\
    & = a^2 \left[ G_1^{(0,R)} (0^-,x') + G_1^R (0^+,x') - G_1^{(0,R)} (0^+,x') \right]. \label{FbR_lat_cont} 
\end{align}
Multiplying them with  $t_L^{\dagger}$, $t_R$, $t_L$, $t_R^{\dagger}$, respectively, cancels the factor $a^2$, and we eventually obtain \eqref{sing_bar_gf1} and \eqref{sing_bar_gf2}.

\section{Proof of the relation \eqref{d_identity}}
\label{app:proof_match}

To prove \eqref{d_identity} it suffices to show that
\begin{align}
  \mathcal{A}_R F' (0^+) - i  B_R =   \frac{1}{4} \mathcal{A}_{R}\lim_{x \to 0^+} \frac{d^2}{d x^2} G^{R} (x, x)\mathcal{A}_{R} , \label{R_id_pr} \\
  \mathcal{A}_L F' (0^-) - i B_L = - \frac{1}{4} \mathcal{A}_{L}\lim_{x \to 0^-} \frac{d^2}{d x^2} G^{L} (x, x)\mathcal{A}_{L}. \label{L_id_pr}
\end{align}

First we evaluate
\begin{align}
& F' (0^+) = [G_{12}^{(0,R)} (0^+,0^-) \nonumber \\
&- G_1^{(0,R)} (0^+,0) [G^{(0,R)} (0,0)]^{-1} G_2^{(0,R)} (0,0^+)] \frac{\mathcal{A}_R}{2}, \\
& F' (0^-) = -[G_{12}^{(0,L)} (0^-,0^+) \nonumber \\
&- G_1^{(0,L)} (0^-,0) [G^{(0,L)} (0,0)]^{-1} G_2^{(0,L)} (0,0^-)] \frac{\mathcal{A}_L}{2}.
\end{align}
Using the identities
\begin{align}
&  G_{12}^{(0,R)} (0^+,0^-) \nonumber \\
&= G_1^{(0,R)} (0^+,0) [G^{(0,R)} (0,0)]^{-1} G_2^{(0,R)} (0,0^-), \\
& G_{12}^{(0,L)} (0^-,0^+) \nonumber \\
&= G_1^{(0,L)} (0^-,0) [G^{(0,L)} (0,0)]^{-1} G_2^{(0,L)} (0,0^+),
\end{align}
previously derived in Ref. [\onlinecite{PhysRevB.106.165405}] [see Eqs. (B2) and (B3) therein], as well as the jump conditions \eqref{jump1_1st} and \eqref{jump2_1st}, we simplify
\begin{align}
    F' (0^+) &=  - G_1^{(0,R)} (0^+,0) [G^{(0,R)} (0,0)]^{-1} , \label{Fprime_R} \\
    F' (0^-) &= - G_1^{(0,L)} (0^-,0) [G^{(0,L)} (0,0)]^{-1} . 
    \label{Fprime_L}
\end{align}

Next, we evaluate
\begin{align}
    &- \frac18 \mathcal{A}_R \left[\lim_{x \to 0^+} \frac{d^2}{d x^2} G^{R} (x, x)-\lim_{x \to 0^-} \frac{d^2}{d x^2} G_0^{R} (x, x) \right] \mathcal{A}_R \nonumber \\ 
    &= \frac14   \mathcal{A}_R G_1^{(0,R)} (0^+,0) [G^{(0,R)} (0,0)]^{-1} G_2^{(0,R)} (0,0^+)  \mathcal{A}_R  \nonumber \\
    & - \frac14  \mathcal{A}_R G_1^{(0,R)} (0^-,0) [G^{(0,R)} (0,0)]^{-1} G_2^{(0,R)} (0,0^-)  \mathcal{A}_R \nonumber \\
    & = \frac14   \mathcal{A}_R [ G_{12}^{(0,R)} (0^+,0^-) -G_{12}^{(0,R)} (0^-,0^+)]  \mathcal{A}_R \nonumber \\
    & - \mathcal{A}_R F' (0^+) - [G^{(0,R)} (0,0)]^{-1} 
\end{align}
and
\begin{align}
     & - \frac18 \mathcal{A}_L \left[ \lim_{x \to 0^-} \frac{d^2}{d x^2} G^{L} (x, x)  -\lim_{x \to 0^+} \frac{d^2}{d x^2} G_0^{L} (x, x) \right] \mathcal{A}_L  \nonumber \\
    & = \frac14 \mathcal{A}_L  G_1^{(0,L)} (0^-,0) [G^{(0,L)} (0,0)]^{-1} G_2^{(0,L)} (0,0^-) \mathcal{A}_L \nonumber \\
    & -\frac14 \mathcal{A}_L G_1^{(0,L)} (0^+,0) [G^{(0,L)} (0,0)]^{-1} G_2^{(0,L)} (0,0^+) \mathcal{A}_L \nonumber \\
    &= \frac14   \mathcal{A}_L [ G_{12}^{(0,L)} (0^-,0^+) -G_{12}^{(0,L)} (0^+,0^-)]  \mathcal{A}_L \nonumber \\
    & + \mathcal{A}_L  F' (0^-) - [G^{(0,L)} (0,0)]^{-1}.
\end{align}

Further using \eqref{G0_inv} we establish that
\begin{align}
    & \mathcal{A}_R F' (0^+) - \frac14 \mathcal{A}_R \lim_{x \to 0^+} \frac{d^2}{d x^2} G^{R} (x, x) \mathcal{A}_R   \nonumber \\ 
    & =  \frac14   \mathcal{A}_R [ G_{12}^{(0,R)} (0^+,0^-) -G_{12}^{(0,R)} (0^-,0^+)]  \mathcal{A}_R  
\end{align}
and
\begin{align}
     &   \mathcal{A}_L  F' (0^-) + \frac14 \mathcal{A}_L \lim_{x \to 0^-} \frac{d^2}{d x^2} G^{L} (x, x)   \mathcal{A}_L   \nonumber \\
    &= \frac14   \mathcal{A}_L [ G_{12}^{(0,L)} (0^+,0^-) -G_{12}^{(0,L)} (0^-,0^+)]  \mathcal{A}_L .
\end{align}
The relations \eqref{R_id_pr} and \eqref{L_id_pr} follow from these by virtue of the jump  condition in the mixed derivatives
\begin{align}
    \frac14 \mathcal{A}_{m} [G^{(0, m)}_{12}(0^{+},0^-)-G^{(0, m)}_{12}(0^-,0^+)] \mathcal{A}_{m}= i\mathcal{B}_{m},
\end{align}
 which is a generalization of the equation (18) in Ref. [\onlinecite{PhysRevB.106.165405}].

\section{Green's function derivation for the multi-barrier model}
\label{app:mult_junc}

The Green's function computation in the single-barrier model presented in Section \ref{Sec: single_barrier_theory}  can be straightforwardly generalized to the case of arbitrary $M$ barriers. Defining their positions at sites $n=n_1, \ldots , n_M$ and assigning the local onsite potentials $W_{n_m}$ to each of them, we also introduce $M+1$ disjoint subsystems defined on the intervals $n_m + 1 \leq n \leq n_{m+1} -1$ and describe them by the Green's functions $G^m$.

Switching on the coupling $\sum_{m=1}^{M} v_m$ between the disjoint subsystems via the hoppings through the barrier sites $n_m$,
\begin{align}
     v_m = &-  |n_m \rangle t_{n_m -1} \langle n_m-1| - |n_m-1 \rangle t_{n_m-1}^{\dagger} \langle n_m|\nonumber \\
     &-  |n_m +1 \rangle t_{n_m} \langle n_m |- |n_m \rangle t_{n_m}^{\dagger} \langle n_m + 1 |,
\end{align}
we set up the Dyson equation treating $v_m$'s as a perturbation. It has the form
\begin{align}
 & G_{n,n'} = G^{\{ m\}}_{n,n'}  \label{Dyson_eq_multiple} \\
    &- \sum_{m'=1}^M (G_{n,n_{m'}-1}^{m'-1} t_{n_{m'}-1}^{\dagger} + G_{n,n_{m'}+1}^{m'} t_{n_{m'}}) G_{n_{m'},n'} \nonumber \\
    &- \sum_{m'=1}^M  \delta_{n,n_{m'}} g_{n_{m'},n_{m'}}^{m'} (t_{n_{m'}-1} G_{n_{m'}-1,n'} + t_{n_{m'}}^{\dagger} G_{n_{m'} + 1,n'} )
    \nonumber
\end{align}
generalizing \eqref{Dyson_eq_junction}. Hereby $G^{\{ m\}}_{n,n'}$ contains not only all Green's functions $G^m$ of the finite-ranged disjoint subsystems, but also those $g_{n,n'}^m = \frac{\delta_{n,n_m} \delta_{n',n_m}}{z - W_{n_m}}$ of the barrier sites $n_m$. 

Below we find the solution of \eqref{Dyson_eq_multiple} following the same routine as in Section \ref{Sec: single_barrier_theory}. 

Introducing the $M$-component vectors
\begin{align}
\bar{\mathcal{G}}_{m'}^{(\pm)} &=  G_{n_{m'}\pm 1,n'}, \label{barGpm} \\
    \bar{\mathcal{G}}_{m'}^{(0)} &=  G_{n_{m'},n'} ,  \label{barG0}
\end{align}
we first derive the equation
\begin{align}
    \left( \begin{array}{c} \bar{\mathcal{G}}^{(+)}  \\ \bar{\mathcal{G}}^{(-)}  \\ \bar{\mathcal{G}}^{(0)}  \end{array} \right) = 
    \left( \begin{array}{ccc} 
    1  & 0 & -T^{+,0} \\  
    0 &  1  & - T^{-,0} \\
     -T^{0,+}   & -T^{0,-} & 1 \\
    \end{array}\right)^{-1}
    \left( \begin{array}{c} \bar{\mathcal{G}}_{disj}^{(+)}  \\ \bar{\mathcal{G}}_{disj}^{(-)}  \\ \bar{\mathcal{G}}_{disj}^{(0)}  \end{array} \right),
\end{align}
where 
\begin{align}
\bar{\mathcal{G}}_{disj,m'}^{(+)} &=G_{n_{m'}+ 1,n'}^{m'}, \\
\bar{\mathcal{G}}_{disj,m'}^{(-)} &=G_{n_{m'}- 1,n'}^{m'-1}, \\ 
\bar{\mathcal{G}}_{disj,m'}^{(0)} &=  g_{n_{m'},n_{m'}}^{m'} \delta_{n_{m'},n'},
\end{align} 
and
\begin{align}
    T_{m,m'}^{0,+} &= - \delta_{m,m'} g^{m'}_{n_{m'},n_{m'}} t^{\dagger}_{n_{m'}}, \\
    T_{m,m'}^{0,-} &= - \delta_{m,m'} g^{m'}_{n_{m'},n_{m'}} t_{n_{m'}-1}, \\
    T_{m,m'}^{+,0} &= -\delta_{m,m'-1} G_{n_{m'-1} + 1,n_{m'}-1}^{n_{m'}-1} t_{n_{m'}-1}^{\dagger} 
    \nonumber \\
    & - \delta_{m,m'} G_{n_{m'} +1,n_{m'}+1}^{m'} t_{n_{m'}}, \\
    T_{m,m'}^{-,0} &= - \delta_{m,m'} G_{n_{m'}-1,n_{m'}-1}^{m'-1} t_{n_{m'}-1}^{\dagger} \nonumber \\
    & - \delta_{m,m'+1} G_{n_{m'+1} -1,n_{m'}+1}^{m'} t_{n_{m'}}.
\end{align}

Finding explicitly the solutions  for \eqref{barGpm}, \eqref{barG0} 
\begin{align}
    & G_{n_m , n'} = -\sum_{m'=1}^M (D^{-1})_{m,m'} \bar{F}_{m',n'} , \\
    &  g^{m}_{n_{m},n_{m}} \left( t_{n_m-1} G_{n_{m}- 1,n'} +t^{\dagger}_{n_m} G_{n_{m}+ 1,n'} \right) \nonumber \\
    &=  g_{n_{m},n_{m}}^{m} \delta_{n_{m},n'} + \sum_{m'=1}^M (D^{-1})_{m,m'}
    \bar{F}_{m',n'}   , 
\end{align}
we insert them into \eqref{Dyson_eq_multiple}. Thereby we establish the composite Green's function of the whole system
\begin{align}
    G_{n,n'} = \sum_{m=0}^{M} G_{n,n'}^{m} + \sum_{m,m'=1}^M F_{n,m}  (D^{-1})_{m,m'} \bar{F}_{m',n'},
    \label{GF_junction_multi}
\end{align}
where
\begin{align}
    & D_{m,m'} = \delta_{m,m'} \, (z - W_{n_m} \nonumber \\
    &- t_{n_m-1} G^{m-1}_{n_m-1,n_m-1} t^{\dagger}_{n_m-1} 
    -t_{n_m}^{\dagger} G^m_{n_m +1 , n_m+1} t_{n_m} )  \nonumber \\
    &- \delta_{m,m'+1} t_{n_m-1} G^{m-1}_{n_m -1, n_{m-1}+1} t_{n_{m-1}} \nonumber \\
    &- \delta_{m,m'-1} t^{\dagger}_{n_m} G^m_{n_m +1 , n_{m+1} -1} t_{n_{m+1}-1}^{\dagger},
    \label{Dmm}
\end{align}
and
\begin{align}
    F_{n,m} &=- \delta_{n,n_m}+G_{n,n_{m}-1}^{m-1} t_{n_{m}-1}^{\dagger} + G_{n,n_{m}+1}^{m} t_{n_{m}}, \label{F_multi} \\
    \bar{F}_{m',n'} &= -  \delta_{n_{m'},n'} + t_{n_{m'}-1}  G_{n_{m'}- 1,n'}^{m'-1} +  t^{\dagger}_{n_{m'}} G_{n_{m'}+ 1,n'}^{m'}.
    \label{Fb_multi}
\end{align}

The formula \eqref{GF_junction_multi} is a multi-barrier generalization of the single-barrier formula \eqref{GF_junction}. Observing that
\begin{align}
    \sum_{n} \bar{F}_{m',n} F_{n,m} &= \frac{\partial}{\partial \omega} D_{m',m} ,
\end{align}
we also express the global DOS in the multi-barrier setup by the analogy with the single-barrier formula \eqref{DOS_single} as
\begin{align}
    \rho (\omega)= \sum_{m=0}^M \rho^m (\omega) -\frac{1}{\pi} \text{Im} \, \frac{\partial}{\partial \omega} \ln \det D (\omega + i 0^+).
    \label{DOS_multi}
\end{align}

To take the continuum limit of $D_{m,m'}$ we approximate in \eqref{Dmm} [cf. Eq. \eqref{V_0}]
\begin{align}
   W_{n_m} \approx \frac{\mathcal{A}_{m-1} + \mathcal{A}_m}{2 a^2} + \frac{\mathcal{U}_m}{a} 
\end{align}
and
\begin{align}
   & t_{n_m-1} G^{m-1}_{n_m-1,n_m-1} t^{\dagger}_{n_m-1} \nonumber \\
   & \approx -\frac{\mathcal{A}_{m-1}}{2 a^2} + \frac{1}{8a} \mathcal{A}_{m-1} \lim_{x \to x_m^-} \frac{d^2}{d x^2} G^{m-1} (x,x) \mathcal{A}_{m-1},  \\
   & t_{n_m}^{\dagger} G^m_{n_m +1 , n_m+1} t_{n_m} \nonumber \\
   & \approx -\frac{\mathcal{A}_m}{2 a^2} + \frac{1}{8a} \mathcal{A}_m \lim_{x \to x_m^+} \frac{d^2}{d x^2} G^m (x,x) \mathcal{A}_m, 
\end{align}
analogously to the single-barrier case discussed in Appendix \ref{app:Caroli_approach}. In addition, there are two new objects in \eqref{Dmm}, which are approximated by
\begin{align}
    & t_{n_m-1} G^{m-1}_{n_m -1, n_{m-1}+1} t_{n_{m-1}} \nonumber \\
    & \approx -\frac{1}{4a} \mathcal{A}_{m-1} G_{12}^{m-1} (x_m^- , x_{m-1}^+)  \mathcal{A}_{m-1},  \\
    & t^{\dagger}_{n_m} G^m_{n_m +1 , n_{m+1} -1} t_{n_{m+1} -1}^{\dagger} 
    \nonumber \\
    & \approx -\frac{1}{4a} \mathcal{A}_m G^m_{12} (x_m^+ , x_{m+1}^-) \mathcal{A}_m . 
\end{align}
Altogether these terms provide the continuum limit of $D$ \eqref{Dmm} giving rise to the expressions \eqref{multi_bar_gf3_diag}-\eqref{multi_bar_gf3_below} defining the matrix $d = \lim_{a \to 0} a \, D$ in the multi-barrier case.

A derivation of the continuum limit of \eqref{F_multi} and \eqref{Fb_multi} employs the same type of the $a$-expansion as in \eqref{FL_lat_cont}-\eqref{FbR_lat_cont}. It leads us to the final result \eqref{multi_bar_gf1}, \eqref{multi_bar_gf2} analogous to the result \eqref{sing_bar_gf1}, \eqref{sing_bar_gf2} of the single-barrier case.

\section{Simplification of the Green's function in the double-barrier case}
\label{app:finite_simplifications}

On the basis of expressions \eqref{BGF_continuum} and \eqref{BGF_symmetric} we obtain
\begin{align}
    G_{0,W}^C (x,x') &= g (x-x')\label{G_two_imp} \\
& - g (x) \left\{  g (0)-   g (-W) [g (0) ]^{-1} g (W) \right\}^{-1}  \nonumber \\
& \times \left\{  g (-x')  -  g (-W) [ g (0)]^{-1} g (W-x') \right\} \nonumber \\
&- g (x-W)  \left\{ g (0) - g (W) [g (0)]^{-1} g (-W) \right\}^{-1}  \nonumber \\
& \times \left\{  g (W -x')  -  g (W) [ g (0)]^{-1}  g (-x')\right\},  \nonumber 
\end{align}
in terms of the translationally invariant counterpart $  G^{(C,0)} (x,0) \equiv g (x)$. According to \eqref{proj_bound}, the boundary Green's function of the central region is then given by $G^C (x,x') = \Theta (W>x>0) G_{0,W}^C (x,x')$.

Taking in account that $G_{0,W}^C (x,x')$ identically vanishes when $x$ and $x'$ belong to different spatial regions [either $(-\infty, 0)$ or $(0,W)$ or $(W,\infty)$], we get the identities
\begin{align}
    \frac{\partial^2}{ \partial x \partial x'} G_{0,W}^C (0^- , W^+) =\frac{\partial^2}{ \partial x \partial x'} G_{0,W}^C (0^- , W^-)\\
    =\frac{\partial^2}{ \partial x \partial x'} G_{0,W}^C (0^+ , W^+) =0  .
\end{align}
They allow us to replace
\begin{align}
    &  \frac14 \mathcal{A}_C G_{12}^C (0^+ , W^-) \mathcal{A}_C  \\
    =&  \frac14 \mathcal{A}_C \left[ \frac{\partial^2}{ \partial x \partial x'} G_{0,W}^C (0^+ , W^-) - \frac{\partial^2}{ \partial x \partial x'} G_{0,W}^C (0^- , W^-)\right.  \nonumber \\ 
    & \quad \left. + \frac{\partial^2}{ \partial x \partial x'} G_{0,W}^C (0^- , W^+)- \frac{\partial^2}{ \partial x \partial x'} G_{0,W}^C (0^+ , W^+) \right] \mathcal{A}_C . \nonumber
\end{align}
After the lengthy calculation using \eqref{G_two_imp} and \eqref{jump1_1st}, \eqref{jump2_1st} we obtain 
\begin{align}
    & \frac14 \mathcal{A}_C G_{12}^C (0^+ , W^-) \mathcal{A}_C \nonumber \\
    &=  \left\{  g (W) - g (0) [ g (-W) ]^{-1} g (0)   \right\}^{-1}.
    \label{two_bar_12} 
\end{align}
Analogously we deduce
\begin{align}
    & \frac14 \mathcal{A}_C G_{12}^C (W^- , 0^+) \mathcal{A}_C \nonumber \\
    &= \left\{  g (-W) - g (0) [g (W)]^{-1} g (0)     \right\}^{-1}  .
    \label{two_bar_21}
\end{align}

To derive 
\begin{align}
    & - \frac18 \mathcal{A}_C \left[ \lim_{x \to 0^-} \frac{d^2}{d x^2} G_0^C (x,x) + \lim_{x \to 0^+} \frac{d^2}{d x^2} G^C (x,x)\right] \mathcal{A}_C \nonumber \\
    &= \left\{ g (0) - g (-W) [g (0)]^{-1}   g (W)  \right\}^{-1}, 
    \label{two_bar_11} 
\end{align}
we use the identities
\begin{align}
    \frac{\partial^2}{\partial x \partial x'} G_{0,W}^{C} (0^+ , 0^-) = \frac{\partial^2}{\partial x \partial x'} G_{0,W}^{C} (0^- , 0^+) =0
\end{align}
to replace
\begin{align}
    & - \frac18 \mathcal{A}_C \left[ \lim_{x \to 0^-} \frac{d^2}{d x^2} G_0^C (x,x) + \lim_{x \to 0^+} \frac{d^2}{d x^2} G^C (x,x)\right] \mathcal{A}_C  \\
    = & - \frac18 \mathcal{A}_C \left[ \lim_{x \to 0^-} \frac{d^2}{d x^2} G_{0,W}^C (x,x) + \lim_{x \to 0^+} \frac{d^2}{d x^2} G_{0,W}^C (x,x)\right. \nonumber \\
    & \left. -2 \frac{\partial^2}{\partial x \partial x'} G_{0,W}^{C} (0^+ , 0^-) - 2 \frac{\partial^2}{\partial x \partial x'} G_{0,W}^{C} (0^+ , 0^-) \right] \mathcal{A}_C. \nonumber
\end{align}
After the lengthy calculation using \eqref{G_two_imp} and \eqref{jump1_1st}, \eqref{jump2_1st} we obtain \eqref{two_bar_11}. 

Analogously we deduce
\begin{align}
    & - \frac18 \mathcal{A}_C \left[ \lim_{x \to W^+} \frac{d^2}{d x^2} G_W^C (x,x) + \lim_{x \to W^-} \frac{d^2}{d x^2} G^C (x,x)\right] \mathcal{A}_C \nonumber \\
    &= \left\{ g (0) -  g (W)  [g (0)]^{-1} g (-W)  \right\}^{-1}.
    \label{two_bar_22}
\end{align}

\section{Josephson system}
\subsection{Position-space Green's functions of bulk $s$-wave superconductors}
\label{ap_old_Josephson}
The momentum-space Green's functions of the bulk $s$-wave superconductors  are equal at $\varphi=0$ to
\begin{align}
    G_k^{(0)} (z) =& \frac{1}{z^2 - (h_k^{(0)})^2 - \Delta^2 } \left( \begin{array}{cc} z + h_k^{(0)} & \Delta_0  \\ \Delta_0  &  z - h_k^{(0)} \end{array} \right), \label{S_wave_appendix}
\end{align}
where $h_k^{(0)}=\frac{k^{2}}{2m}-\mu$.

For the evaluation of
\begin{align}
    G^{(0)} (x,x') = \int_{-\infty}^{\infty} \frac{d k}{2 \pi} e^{i k (x-x')} G_k^{(0)} (z)
\end{align}
we note the following poles of $G_k^{(0)} (z)$
\begin{align}
    k_{1, \pm} = \pm  \sqrt{2m \mu + 2m z \sqrt{1 - \left( \frac{\Delta_0}{z} \right)^2}} , \label{roots_k1} \\
    k_{2,\pm} = \mp  \sqrt{2m \mu - 2m z \sqrt{1 - \left( \frac{\Delta_0}{z} \right)^2}} . \label{roots_k2}
\end{align}
Considering $\text{Im} (z) >0$ and deforming appropriately the integration contour in the complex $k$-plane, we obtain by means of the residue theorem
\begin{align}
    G^{(0)} (x,x') =& -  \frac{ i m e^{i k_{1,+} |x-x'|}}{2 k_{1,+} }  \left[ e^{\tau_x \chi} + \tau_z \right] \\
    &+  \frac{ i m  e^{i k_{2,+} |x-x'|}}{2 k_{2,+}   }   \left[ e^{\tau_x \chi} - \tau_z \right] ,
\end{align}
where $e^{\tau_x \chi}=\frac{1 + \tau_x \tanh \chi}{\sqrt{1-\tanh^2 \chi}}$ and $\tanh\chi=\frac{\Delta_0}{z}$. 

For $\text{Im} (z) <0$ it is necessary to replace $k_{1,+} \to k_{1,-}$ and $k_{2,+} \to k_{2,-}$ in the above equations.

Next, we evaluate
\begin{align}
    G^{(0)} (0,0) &=  -  \frac{ i m }{ k_{1,+} k_{2,+}  }  \left[ k^{(+)}  \tau_z - k^{(-)}  e^{\tau_x \chi}  \right] , \\
 \, [ G^{(0)} (0,0)]^{-1} &= -  \frac{ k^{(+)}  \tau_z + k^{(-)}  e^{-\tau_x \chi} }{ i m } ,
\end{align}
where
\begin{align}
    k^{(\pm)} = \frac{k_{1,+} \pm k_{2,+}}{2} \, \text{sgn} \, \text{Im} (z ),
\end{align}
as well as
\begin{align}
    G_1^{(0)} (0^{\pm},0) = G_2^{(0)} (0,0^{\pm})= \pm   m \tau_z .
\end{align}
With this in hands and noting
\begin{align}
    G^{(0, \lambda)} (x,x') = U_{\lambda} G^{(0)} (x,x') U_{\lambda}^{\dagger}, \quad U_{\lambda} =e^{\frac{i}{4} \tau_z \lambda \varphi}, 
\end{align}
one can show on the basis of Eqs.~\eqref{D_simple_form_1bar}, \eqref{L_R}, and \eqref{L_L}  that the matrix $d$ of the Josephson system is given by Eq. \eqref{spectral_matrix_example_1}.

\subsection{Local Coulomb interaction in the Hartree-Fock approximation}
\label{app:coul_corr}

To compute the expectation values $\langle \psi_{\downarrow} (0)\psi_{\uparrow} (0)\rangle$ and $\langle \psi_{\uparrow}^{\dagger} (0)\psi_{\downarrow}^{\dagger} (0) \rangle$ with respect to the Hartree-Fock Hamiltonian, we define the following imaginary-time Matsubara Green's functions
\begin{align}
    G_{\gamma \delta; U}^{\text{Mat}} (x,x'; \tau) =& - \langle T_{\tau} \hat{\Psi}_{\gamma} (x, \tau) \hat{\Psi}_{\delta}^{\dagger} (x',0) \rangle  \label{G_Mats1} \\
    =& \frac{1}{\beta} \sum_{i\omega_n} e^{-i \omega_n \tau} G_{\gamma \delta; U} (x,x'; i \omega_n ) ,
    \label{G_Mats2}
\end{align}
where $\hat{\Psi}_{1} (x) = \hat{\psi}_{\uparrow} (x)$ and $\hat{\Psi}_{2} (x) = \hat{\psi}_{\downarrow}^{\dagger} (x)$. In this terms
\begin{align}
    \langle \psi_{\downarrow} (0)\psi_{\uparrow} (0)\rangle &=  G^{\text{Mat}}_{12,U} (0,0;\tau=0^-) \nonumber \\
    &= \frac{1}{2 \beta} \sum_{i\omega_n} e^{i \omega_n 0^+} \text{tr} \, \left\{[d_U (i \omega_n )]^{-1} \tau_x \right\}, \\
    \langle \psi_{\uparrow}^{\dagger} (0) \psi_{\downarrow}^{\dagger} (0)\rangle & =  G^{\text{Mat}}_{21,U} (0,0;\tau=0^-) \nonumber \\
    &= \frac{1}{2 \beta} \sum_{i\omega_n} e^{i \omega_n 0^+} \text{tr} \, \left\{[d_U (i \omega_n )]^{-1} \tau_x \right\} .
    \label{Matsubara_offdiag}
\end{align}

The function $[d_U (z)]^{-1} = [d (z) - \Delta_{loc} \tau_x]^{-1}$ has the following off-diagonal components
\begin{align}
    & \frac12 \text{tr} \, \left\{ [d_U (z)]^{-1} \tau_x \right\} = -\frac{m}{k_F}  \left(   \frac{  \frac{ k^{(-)}}{k_F}   \frac{i \Delta_0}{z} \cos \frac{\varphi}{2} }{ \sqrt{1-(\frac{\Delta_0}{z})^2}} + \bar{\Delta}_{loc}  \right) \label{offdiag_integrand} \\
    & \times \frac{ 1  }{\frac{\frac{ k^{(-)\, 2}}{k_F^2}}{ 1-(\frac{\Delta_0}{z})^2} + \left( \frac{  \frac{ k^{(-)}}{k_F}   \frac{i \Delta_0}{z} \cos \frac{\varphi}{2} }{ \sqrt{1-(\frac{\Delta_0}{z})^2}} + \bar{\Delta}_{loc} \right)^2 +(  \bar{V}_0 + \frac{k^{(+)}}{ik_F} )^2}, \nonumber
\end{align}
where $\bar{V}_0 = \frac{m V_{0}}{k_F}$ and $\bar{\Delta}_{loc} = \frac{m \Delta_{loc}}{k_F}$.

At zero temperature the Matsubara sum in \eqref{Matsubara_offdiag} turns into the imaginary-frequency integral. The integrand $\propto \Delta_0 \cos \frac{\varphi}{2}$ in \eqref{offdiag_integrand} decays sufficiently fast at high frequencies, so that the convergence factor $e^{i \omega 0^+}$ for it may be omitted. The other integrand $\propto \bar{\Delta}_{loc}$ does require the convergence factor, which is needed to handle the terms $\sim \frac{1}{z}$ at high frequencies.  Collecting all contributions we obtain the following self-consistency equation
\begin{align}
    & \bar{\Delta}_{loc} = - u \int_{0}^{\infty} d \bar{\omega} \frac{\bar{k}^{(-)}  \cos \frac{\varphi}{2} }{ \sqrt{\bar{\omega}^2+1}}  R (\bar{\omega}), \\
    & \qquad -  u  \bar{\Delta}_{loc} \int_{0}^{\infty} d \bar{\omega}   [ R (\bar{\omega}) - R_{0} (\bar{\omega})] \\
    & \qquad  -  u  \bar{\Delta}_{loc} \int_{0}^{\infty} d \bar{\omega} R_0 (\bar{\omega}) \cos (\bar{\omega} 0^+), \label{diverg_delta_loc}
\end{align}
where
\begin{align}
    u = \frac{U m^2 \Delta_0}{ \pi k_F^2}
\end{align}
is a dimensionless interaction parameter, and
\begin{align}
     R (\bar{\omega}) &= \frac{  1 }{\frac{\bar{k}^{(-)\, 2} \bar{\omega}^2}{ \bar{\omega}^2+ 1} + \left( \frac{  \bar{k}^{(-)}      \cos \frac{\varphi}{2} }{ \sqrt{\bar{\omega}^2+1}} + \bar{\Delta}_{loc} \right)^2 +\left(  \bar{V}_0 - i \bar{k}^{(+)} \right)^2} , \\
     \bar{k}^{(\pm)} &= \frac{\sqrt{1 +i \frac{\Delta_0}{\mu} \sqrt{\bar{\omega}^2 +1} } \mp \sqrt{1-i \frac{\Delta_0}{\mu} \sqrt{\bar{\omega}^2 +1} }}{2} , \\
     R_0 (\bar{\omega}) &= \frac{ 1 }{ \bar{k}_0^{(-)\, 2} + \left( \bar{V}_0 - i \bar{k}_0^{(+)} \right)^2   }, \\
    \bar{k}_0^{(\pm)} &= \frac{\sqrt{1 +i \frac{\Delta_0}{\mu} \bar{\omega} } \mp \sqrt{1-i \frac{\Delta_0}{\mu} \bar{\omega} }}{2} .
\end{align}

The perform the integral in \eqref{diverg_delta_loc} we rotate the integration contour to the real axis
\begin{align}
    & u  \bar{\Delta}_{loc} \, i \int_{i0}^{i\infty} d  \bar{z} R_0 (-i \bar{z}) \frac{e^{ \bar{z} 0^+} + e^{ - \bar{z} 0^+}}{2} \nonumber \\
    =& - \frac{i}{2}  u  \bar{\Delta}_{loc}  \int_0^{\infty} d \bar{\omega} \, [ R_0 (i \bar{\omega} +  0^+) -  R_0 (-i \bar{\omega} + 0^+)] \\
    =& -   u  \bar{\Delta}_{loc}  \int_{\mu/\Delta_0}^{\infty}  \frac{ d \bar{\omega} \, e^{-\bar{\omega} 0^+} \, \sqrt{ \frac{\Delta_0}{\mu} \bar{\omega}+1}}{\left(\sqrt{\frac{\Delta_0}{\mu} \bar{\omega}-1} + \bar{V}_0 \right) \left(\frac{\Delta_0}{\mu} \bar{\omega} +1 + \bar{V}_0^2 \right)}.
\end{align}
The resulting integral has the ultraviolet logarithmic divergence which is not cut off at the scale $\mu$. The origin of this divergence is rooted in the ultra-local form of the Coulomb interaction in our model, and for the integral's regularization it is necessary to introduce the cut-off scale $\omega_c \gg \mu$, whose inverse $v_F/\omega_c$  captures the microscopic length scale of the interaction. Thus \eqref{diverg_delta_loc} is estimated by $- u  \bar{\Delta}_{loc} \frac{\mu}{\Delta_0} \ln \frac{\omega_c}{\Delta_0}$. This behaviour signals the necessity to refine the model's consideration, e.g. by resorting to renormalization group methods.\cite{Meden_2019}

In turn, in the bare perturbation theory the local superconducting order parameter is given by the convergent integral
\begin{align}
    & \bar{\Delta}_{loc} \approx - u \int_{0}^{\infty} d \bar{\omega}   \frac{  \bar{k}^{(-)}    \cos \frac{\varphi}{2} }{ \sqrt{\bar{\omega}^2+1}}  R (\bar{\omega}) \bigg|_{\bar{\Delta}_{loc} =0} .
    \label{Delta_loc_PT}
\end{align}
It is remarkable that in the Andreev limit $\mu \gg \Delta_0$ it also features the logarithmic dependence of the high energy scale $\mu$:
\begin{align}
\bar{\Delta}_{loc} \approx - D u  \cos \frac{\varphi}{2} \ln \frac{\mu}{\Delta_0}, \quad D = \frac{1}{1+ \bar{V}_0^2} .
\label{Delta_loc_PT_appr}
\end{align}

In addition,  we provide the zero temperature limit of the JC formula \eqref{current_U} 
\begin{align}
 J (\varphi) &= \frac{2 \Delta_0 \sin \frac{\varphi}{2} }{ \Phi_{0}} 
 \label{current_U_zero_T} \\
 & \times  \int_0^{\infty} d \bar{\omega}  \left(  \frac{  \bar{k}^{(-)}    \cos \frac{\varphi}{2} }{ \sqrt{\bar{\omega}^2+1}} + \bar{\Delta}_{loc} \right)  \frac{R (\bar{\omega}) \bar{k}^{(-)}}{\sqrt{\bar{\omega}^2+1}}.
 \nonumber 
\end{align}

\section{Majorana junction}
In this Appendix, we evaluate bulk Green's functions and their spatial derivatives which are needed for constructing the $d$-matrix in the Majorana junction model (both in the short $W \to 0$ and finite $W$ cases). The derived expressions are used for plotting energy- and current-phase relations in the numerical examples presented in the main text. 

\subsection{Short junction}
\label{Ap_bulk_SC_majorana_model}

The bulk Green's function $g(x)$ occurring in \eqref{g_x_Maj} is defined by
\begin{align}
    g (x)= \int\frac{dk}{2\pi} e^{ik x} g_k  , \label{some_intermediate_formula_some_appendix}
\end{align}
where
\begin{align}
    g_{k}=& \begin{pmatrix} a_k^{(+)} & -\Delta_{0} \\ -\Delta_{0} & a_k^{(-)} \end{pmatrix}^{-1} , \label{g_k_Maj} \\
    a_k^{(\pm)} &= z \mp  h_k^{(0)} \mp \alpha k \sigma _z + B \sigma_x, \\
    h_k^{(0)}=&\frac{k^{2}}{2m}-\mu.
\end{align}
The inverse in \eqref{g_k_Maj}  may be written as
\begin{align}
g_k = \frac{P_k}{Q_k} = \frac{1}{Q_k}  \left(\begin{array}{cc} a_k^{(-)} A_k^{(+)} & \Delta_{0} A_k^{(-)} \\ \Delta_{0}   A_k^{(+)} & a_k^{(+)} A_k^{(-)}   \end{array} \right) , \label{another_intermediate_formula_some_appendix}
\end{align}
where 
\begin{align}
    A_k^{(+)} &= \text{adj} \, [a_k^{(+)}  a_k^{(-)} -\Delta^{2}_{0} ] \nonumber  \\
    &= z^2 -\Delta_0^2+ B^2 -( h_k^{(0)})^2 -\alpha^2 k^2 \nonumber \\
    &- 2  z B \sigma_x +2 \alpha k  h_k^{(0)} \sigma _z +2i B \alpha k \sigma_y, \\
    A_k^{(-)} &= \text{adj} \, [a_k^{(+)}  a_k^{(-)} -\Delta^{2}_{0} ] \nonumber \\
    &= z^2-\Delta_0^2 + B^2 -(h_k^{(0)})^2 -\alpha^2 k^2 \nonumber \\
    &- 2  z B \sigma_x +2 \alpha k  h_k^{(0)} \sigma _z-2i B \alpha k \sigma_y 
\end{align}
and
\begin{align}
    Q_k &= \det \,  [a_k^{(+)}  a_k^{(-)} -\Delta^{2}_{0} ] \nonumber \\
    &= \det \,  [a_k^{(-)}  a_k^{(+)} -\Delta^{2}_{0} ] \nonumber \\
    &= [ z^2 -\Delta_0^2+ B^2 -(h_k^{(0)})^2 -\alpha^2 k^2 ]^2\nonumber \\
    &- 4 z^2 B^2 - 4 \alpha^2 k^{2}(h_k^{(0)})^2 +4 \alpha^{2} B^{2} k^{2}.
\end{align}
Setting $z=\omega + i 0^+$ we find on the basis of \eqref{G0m_app}
\begin{widetext}
\begin{align}
  g (x)&=  i  \Theta (x) \sum_{s=1}^{4} \frac{ (2m)^4 e^{i k_{s,+} x}}{\prod_{s'\neq s} (k_{s,+}-k_{s',+}) \prod_{s'}(k_{s,+}-k_{s',-})}  \left(\begin{array}{cc} a_k^{(-)} A_k^{(+)} & \Delta_{0} A_k^{(-)} \\ \Delta_{0}   A_k^{(+)} & a_k^{(+)} A_k^{(-)}   \end{array} \right)   \bigg|_{k = k_{s,+}} \nonumber \\
    &-i  \Theta(-x) \sum_{s=1}^{4} \frac{(2m)^4 e^{i k_{s,-} x}}{ \prod_{s'}(k_{s,-}-k_{s',+}) \prod_{s' \neq s}(k_{s,-}-k_{s',-})}  \left(\begin{array}{cc} a_k^{(-)} A_k^{(+)} & \Delta_{0} A_k^{(-)} \\ \Delta_{0}   A_k^{(+)} & a_k^{(+)} A_k^{(-)}   \end{array} \right)  \bigg|_{k = k_{s,-}}, \label{gx_Majorana_final}
\end{align}
where $k_{s,+}$ and $k_{s,-}$ are the the roots of the equation $Q_k =0$ with positive and imaginary parts, respectively.

The expression \eqref{gx_Majorana_final} allows one to easily establish
\begin{align}
    g' (0^+) & =  -   \sum_{s=1}^{4} \frac{ (2m)^4  k_{s,+} }{\det  \prod_{s'\neq s} (k_{s,+}-k_{s',+}) \prod_{s'}(k_{s,+}-k_{s',-})} 
     \left(\begin{array}{cc} a_k^{(-)} A_k^{(+)} & \Delta_{0} A_k^{(-)} \\ \Delta_{0}   A_k^{(+)} & a_k^{(+)} A_k^{(-)}   \end{array} \right)   \bigg|_{k = k_{s,+}} ,  \\ 
    g' (0^-) & =   \sum_{s=1}^{4} \frac{  (2m)^4 k_{s,-} }{ \prod_{s'}(k_{s,-}-k_{s',+}) \prod_{s' \neq s}(k_{s,-}-k_{s',-})} \left(\begin{array}{cc} a_k^{(-)} A_k^{(+)} & \Delta_{0} A_k^{(-)} \\ \Delta_{0}   A_k^{(+)} & a_k^{(+)} A_k^{(-)}   \end{array} \right)  \bigg|_{k = k_{s,-}} .
\end{align}
\end{widetext}

\subsection{Expansion in the spin-orbit dominated regime}
\label{Glazman_result}

To reproduce the results of Ref. [\onlinecite{PhysRevB.101.224501}] in the spin-orbit dominated regime, we perform the expansion of  \eqref{gx_Majorana_final} to the leading order $1/\alpha$. This gives
\begin{align}
& g (x)  \approx  \frac{e^{i C_0 |x|/\alpha} e^{-i  2 m \alpha   x \sigma_{z} }}{2 i \alpha} \left[ \frac{z+ \Delta_0 \tau_{x}}{C_0} - \text{sgn} \, (x) \tau_{z}\sigma_{z} \right]  \nonumber \\
&  + \frac{ e^{i C_{++} |x|/\alpha}}{ 2 i \alpha}  \left[\frac{z  - (B-\Delta_0) \sigma_x}{C_{++}}  + \text{sgn} \, (x)  \tau_z \sigma_z \right]  \frac{1+\tau_x \sigma_x}{2} \nonumber \\
&  + \frac{ e^{i C_{+-} |x|/\alpha}}{2 i \alpha} \left[\frac{z  - (B+\Delta_0) \sigma_x}{C_{+-}}  + \text{sgn} \, (x) \tau_z \sigma_z \right] \frac{1 - \tau_x \sigma_x }{2}  ,
\label{g_SO_dom}
\end{align}
where 
\begin{align}
      C_0 &= z \sqrt{1- \frac{\Delta_0^2}{z^2}} , \\
      C_{\tau \sigma} &= z \sqrt{1- \frac{(\sigma B- \tau \Delta_0)^2}{z^2}} , \quad \tau, \sigma = \pm .  
\end{align}
Next, we establish the quantities relevant for the calculation of the $d$-matrix:
\begin{align}
    g (0^{\pm}) & = g (0) = \frac{1}{2 i \alpha} \sum_{\tau = \pm} \sum_{\sigma = \pm} \bar{g}_{\tau \sigma} \frac{1+ \tau \tau_x}{2} \frac{1+ \sigma \sigma_x}{2} , \\
 \frac{1}{m} g' (0^{\pm}) &= \pm \tau_{z} -  \frac{z + \tau_x \Delta_0}{C_0} \sigma_z ,
\end{align}
where
\begin{align}
    \bar{g}_{\tau \sigma} = \frac{z + \tau \Delta_0}{C_0} + \frac{z - \sigma B + \tau \Delta_0}{C_{\tau \sigma}} .
\end{align}

Inserting these results at $z=\omega + i 0^+$ into \eqref{d_Maj_single}, after a simple albeit tedious calculation we recover the equations 
for the sub-gap bound states at the absent contact potential ($V_0 =0$) 
\begin{align}
    \cos^{2}\frac{\varphi}{2} &= \frac{\omega^2 + \Delta_0^2 - B^2 + C_{++} C_{+-}}{\omega^2 + \Delta_0^2 - B^2 - C_{++} C_{+-}} , 
    \label{ABS_Majorana1} \\
    \cos^{2}\frac{\varphi}{2} &= \frac{\omega^2}{\Delta_0^2} , 
    \label{ABS_Majorana2}
\end{align}
previously reported in Ref.~[\onlinecite{PhysRevB.101.224501}].

We note in passing that for an isolated Majorana wire a boundary Green's function of the form discussed in Appendix \ref{bound_props_aps} suggests the following equation for the bound states
\begin{align}
\det g (0) = \frac{1}{(2 \alpha)^4} \prod_{\tau,\sigma} \bar{g}_{\tau \sigma} =0,
\end{align}
which factorizes into the four equations $\bar{g}_{\tau \sigma} =0$ for $\tau,\sigma = \pm$. The Majorana zero mode $\omega =0$ appears at $B > \Delta_0$ as a solution satisfying the two equations $\bar{g}_{++}=\bar{g}_{--}=0$.

\subsection{Long junction}
\label{Ap_bulk_Norm_majorana_model}
In the case of the long junction we, in addition, need the bulk position space Green's function of the normal central region. Setting $\Delta_0 =0$ in \eqref{g_k_Maj} we represent it as
\begin{align}
    g_k^C = \sum_{\tau= \pm} \frac{\text{adj}\, a_k^{(\tau)}}{\det a_k^{(\tau)}} \frac{1+ \tau \tau_z}{2} =   \sum_{\tau= \pm} g_k^{(\tau)} \frac{1+ \tau \tau_z}{2} ,
\end{align}
where
\begin{align}
    \text{adj}\, a_k^{(\tau)} &= z - \tau h_k^{(0)} + \tau \alpha k \sigma_z - B \sigma_x, \\
    \det a_k^{(\tau)} &= [ z - \tau h_k^{(0)}]^2 - \alpha^2 k^2 - B^2.
\end{align}
It follows
\begin{align}
    g^C (x) =  \sum_{\tau= \pm} g^{(\tau)} (x) \frac{1+ \tau \tau_z}{2} , \quad  g^{(\tau)} (x) = \int \frac{dk}{2 \pi} e^{i k x} g_k^{(\tau)}.
\end{align}
By the virtue of  \eqref{G0m_app} we find for $z=\omega + i 0^+$
\begin{align}
   &  g^{(\tau)}(x) = -\frac{ (2m \alpha)^2}{k^2_{1,+} -k_{2,+}^2} \\
    & \times \sum_{s=1}^{2} (-1)^{s+1} \frac{e^{i k_{s,+} |x|}}{2 i \alpha } \left[ \frac{  \omega - \tau h_{k_{s,+}}^{(0)}  - B \sigma_x }{  \alpha k_{s,+}} + \text{sgn} (x) \tau  \sigma_z \right]  ,  \nonumber
\end{align}    
where $k_{s,+}= - k_{s,-}$ are the roots of the bi-quadratic equation $\det a_k^{(\tau)} =0$ with $\text{Im} \, k_{s,+} > 0$. 

We consequently find
\begin{align}
    &  g^{(\tau)}(0^{\pm}) = -\frac{ (2m \alpha)^2}{k^2_{1,+} -k_{2,+}^2} \\
    & \times \sum_{s=1}^{2} \frac{ (-1)^{s+1} }{2 i \alpha }  \frac{  \omega - \tau h_{k_{s,+}}^{(0)}  - B \sigma_x }{  \alpha k_{s,+}}   
    \nonumber 
\end{align}
and
\begin{align}
    &  g^{(\tau)\, '}(0^{\pm}) =  m \tau  \left[    \pm 1  -  \frac{2 m \alpha  \sigma_z}{ k_{1,+} + k_{2,+}} \right] ,
\end{align}
from which the $\mathcal{L}_{R/L\rightarrow C}$ functions in Eqs. \eqref{L_LtoC} and \eqref{L_RtoC} may be deduced. Hence the $d$ matrix for the model of Section \ref{sec:majorana_thick} may be constructed on the basis of Eqs.  \eqref{d_two_barrier}, \eqref{p0simple}, and \eqref{pWsimple}.  

It is worth noting the useful formula
\begin{align}
    \partial_\varphi d=\frac{i}{4}\left( \begin{array}{cc}\left[\tau_{z}, \mathcal{L}_{L}\right] & 0 \\ 0 &\left[\tau_{z}, \mathcal{L}_{R}\right]  \end{array}\right), \label{Der_spec_mat_long}
\end{align}
which is applicable for the JC calculation by means of Eqs.~\eqref{key_current formula_1}-\eqref{key_current formula_3}.

\end{appendix}

\bibliography{citations}

\end{document}